\documentclass[a4paper,11pt]{article}
\usepackage{jheppub,esint,shuffle,psfrag}
 \usepackage[utf8]{inputenc}
\newcommand\tr{\mathop{\mathrm{tr}}\nolimits}

\newcommand\lr[1]{{\left({#1}\right)}}
\newcommand\e{\mathrm{e}}
\newcommand\MS{\ensuremath{\overline{\text{MS}}}}
\newcommand\DR{\ensuremath{\overline{\text{DR}}}}

\newcommand\re[1]{(\ref{#1})}

\title{The three-loop cusp anomalous dimension in QCD and its supersymmetric extensions}
\author[a,b]{Andrey G.~Grozin,}
\author[c]{Johannes M.~Henn,}
\author[d]{Gregory P.~Korchemsky,}
\author[e]{Peter Marquard}
\affiliation[a]{Budker Institute of Nuclear Physics SB RAS, Novosibirsk 630090, Russia}
\affiliation[b]{Novosibirsk State University, Novosibirsk 630090, Russia}
\affiliation[c]{PRISMA Cluster of Excellence, Johannes Gutenberg University, 55099 Mainz,
Germany}
\affiliation[d]{Institut de Physique Th\'eorique\footnote{Unit\'e Mixte de Recherche 3681 du CNRS}, CEA Saclay, 91191 Gif-sur-Yvette Cedex, France}
\affiliation[e]{Deutsches Elektronen-Synchrotron, DESY, Platanenallee 6, D15738 Zeuthen, Germany}
\emailAdd{A.G.Grozin@inp.nsk.su}
\emailAdd{henn@uni-mainz.de}
\emailAdd{Gregory.Korchemsky@cea.fr}
\emailAdd{peter.marquard@desy.de}
\preprint{  \parbox[t]{25mm}{DESY 15-186\\ IPhT-T15/174 \\ MITP/15-088}}

\abstract{
We present the details of the analytic calculation of the three-loop angle-dependent cusp anomalous dimension in QCD and its supersymmetric extensions, including the maximally supersymmetric $\mathcal{N}=4$ super Yang-Mills theory. The three-loop 
result in the latter theory is new and confirms a conjecture made in our previous paper. We study various physical limits of the cusp anomalous dimension
and discuss its relation to the quark-antiquark potential including the effects of broken conformal symmetry in QCD. 
We find that the cusp anomalous dimension viewed as a function of the cusp angle and the new effective coupling given by 
light-like cusp anomalous dimension reveals a remarkable
universality property -- it takes the same form in QCD and its supersymmetric extensions, to three loops at least.
We exploit this universality property and make use of the known result for the three-loop quark-antiquark potential to predict  
the special class of nonplanar corrections to the cusp anomalous dimensions at four loops.
Finally, we also discuss in detail the computation of all necessary Wilson line  integrals up to three loops using the method of leading singularities and differential equations.  
}

\begin{document}
\maketitle
\flushbottom

\section{Introduction and summary}
\label{S:Intro}
 
The predictive power of QCD as a theory of strong interaction relies on the possibility to predict 
the scale dependence of various observables in terms of anomalous dimensions as a function
of the strong coupling constant and various kinematical invariants.
Well-known examples include the anomalous dimensions of twist-two operators, which govern
the scale violation of structure functions of deep inelastic scattering. In this paper we study another
important anomalous dimension that appears in many physical quantities involving heavy quarks, the
so-called cusp anomalous dimension \cite{Polyakov:1980ca,Gervais:1979fv,Dotsenko:1979wb,Arefeva:1980zd,Brandt:1981kf,Dorn:1986dt,Korchemsky:1987wg}. 

The simplest physical process that leads to the appearance of this anomalous dimension is the scattering of a heavy
quark off an external potential (see e.g. \cite{Neubert:1993mb,Manohar:2000dt,Grozin:2004yc}). In the infinite mass limit, $m_Q\to\infty$, the quark behaves as a classical charged
particle -- it moves with velocity $v_1^\mu$ that changes to $v_2^\mu$ after scattering off the external source with the
momentum transferred  $q^\mu=m_Q(v_1-v_2)^\mu$. Due to the instantaneous acceleration, the heavy quark starts emitting 
gluons  with arbitrary momenta. The gluons with small momenta generate infrared divergences (IR), whereas the gluons
with large momenta introduce a dependence on the ultraviolet (UV) cut-off. As was shown in \cite{Korchemsky:1985xj,Korchemsky:1991zp}, the
dependence of the scattering amplitude on both IR and UV cut-offs is controlled by the cusp anomalous dimension $\Gamma_{\rm cusp}(\phi,\alpha_s)$, which depends
on the Minkowskian recoil angle of the heavy quark, $(v_1v_2)=\cosh\phi_M$, where $\phi=i \phi_M$.  

The heavy quark scattering and its cross channel, the heavy quark production, enter as partonic subprocesses in various 
important physical applications, e.g.  heavy meson form factors in QCD \cite{Falk:1990yz} and top quark pair production \cite{Czakon:2009zw}. In these processes
IR and UV cut-offs are replaced by relevant physical scales leading to the appearance of large perturbative
corrections enhanced by powers of logarithms of the ratios of these scales. Such logarithmic corrections can be resummed to 
all orders in the QCD coupling constant. The cusp anomalous dimension is an important ingredient of the resulting
resummation formulas  which have numerous phenomenological applications (see e.g. \cite{Neubert:1993mb,Czakon:2009zw,Kidonakis:2010ux,Luisoni:2015xha}). 

The cusp anomalous dimension has a simple interpretation in terms
of Wilson loops \cite{Korchemsky:1991zp}. The heavy quark couples to gluons through an eikonal current and, as a consequence,
the heavy quark scattering amplitude reduces to an eikonal phase. This phase is given by an
expectation value of a path-ordered exponential of the gauge field integrated along the classical trajectory of heavy quark.
The latter consists of two semi-infinite rays separated by a relative angle $\phi$, i.e. a Wilson loop with a cusp. Due to 
the presence of the cusp on the integration contour, the vacuum expectation value of the Wilson loop develops specific ultraviolet divergences.
The cusp anomalous dimension $\Gamma_{\rm cusp}(\phi,\alpha_s)$ governs its dependence  
on the ultraviolet cut-off. 

The two-loop result for the cusp anomalous dimension has been known for more than 25 years \cite{Korchemsky:1987wg} (see also
\cite{Kidonakis:2009ev}).
In this paper, we describe details of the three-loop calculation of  this fundamental quantity in QCD. 
The result was previously reported in \cite{Grozin:2014axa,Grozin:2014hna}.
The obtained expression for the cusp anomalous dimension has an interesting dependence on the cusp angle. The following three limits are of physical importance:
\begin{itemize}
\item 
In the small angle limit $\phi \to 0$ the cusped Wilson loop reduces to a straight Wilson line.  In this limit the cusp divergences
disappear and the cusp anomalous dimension vanishes as $-B(\alpha_s)\phi^2$, with $B(\alpha_s)$ being a positively definite
function of the coupling constant. 
\item 
In the large (Minkowskian) angle limit, $\phi=i\phi_M$, with $\phi_M \to \infty$, the cusp anomalous dimension scales linearly with the angle, $ K(\alpha_s)\phi_M$, with $K(\alpha_s)$ being the light-like cusp anomalous dimension, which also governs the large-spin asymptotics of the anomalous dimension of twist-two operators \cite{Korchemsky:1988si}.
\item 
In the limit of a backtracking Wilson line, $\phi\to\pi$, the three-loop cusp anomalous dimension develops a pole $V(\alpha_s)/(\pi-\phi)$.
In a gauge theory with exact conformal symmetry, $V(\alpha_s)$ coincides with the analogous function defining quantum
corrections to the static quark-antiquark potential. We demonstrate that  this relation holds in QCD to up to conformal symmetry breaking corrections proportional to the beta function.
\end{itemize}
 
In addition to QCD, we also compute $\Gamma_{\rm cusp}(\phi,\alpha_s)$ in supersymmetric extensions of QCD. There
are several reasons for doing this. The cusp anomalous dimension depends on the particle content of the theory. We show that,
surprisingly enough, the latter dependence can be eliminated by expressing $\Gamma_{\rm cusp}(\phi,\alpha_s)$ 
in terms of an effective coupling constant $a\sim K(\alpha_s)$ closely related to light-like cusp anomalous dimension mentioned above.  We find that the 
cusp anomalous dimension viewed as a function of the cusp angle $\phi$ and the new coupling $a$ reveals a remarkable
universality property -- it takes the same form in QCD and its supersymmetric extensions, to three loops at least.
Among various supersymmetric gauge theories, the maximally supersymmetric $\mathcal N=4$ Yang-Mills theory plays a special role. This theory is believed to be integrable in the planar limit (see e.g. \cite{Beisert:2010jr}), which opens up the possibility of determining the above-mentioned universal function for an arbitrary coupling constant (in the planar limit at least). 

The coefficients of the expansion of the cusp anomalous dimension in powers of the coupling 
constant depend on various color factors of the $SU(N)$ gauge group. Up to three loops, 
the latter are given by quadratic Casimirs of the $SU(N)$ gauge group, whereas starting from four loops new color factors proportional to higher Casimirs can appear \cite{Gatheral:1983cz,Frenkel:1984pz}. A distinguished feature of such color factors is that they generate non-planar corrections to $\Gamma_{\rm cusp}(\phi,\alpha_s)$. We exploit the above mentioned
universality property of the cusp anomalous dimension and make use of the known result for the three-loop quark-antiquark potential to predict (conjecturally) this special class of nonplanar corrections to $\Gamma_{\rm cusp}(\phi,\alpha_s)$ at four loops.
 
In order to compute the cusp anomalous dimension, we need to separate the IR and UV divergences of the cusped Wilson loop mentioned above.
We use an infrared suppression factor to remove the IR divergences coming from the integration region at large distances, and 
employ dimensional regularization (dimensional reduction in the supersymmetric case) to regulate the UV divergences. The cusp anomalous dimension is obtained from the latter in the usual way via a renormalization group equation. 

We carry out the calculation in momentum space, where the Wilson lines are replaced by eikonal propagators. 
As a technical trick, we use eikonal identities to relate all non-planar integrals appearing in our calculation to (sums and products of) planar integrals.
We classify all planar three-loop vertex diagrams of this type, and relate them to
master integrals using integral reduction programs. All Feynman diagrams are generated in an automatic way, in an arbitrary covariant gauge, and expressed in terms of the master integrals. 

We compute the master integrals applying the differential equations method \cite{Kotikov:1990kg,Kotikov:1991pm,Remiddi:1997ny,Gehrmann:1999as,Argeri:2007up} and using the new ideas of \cite{Henn:2013pwa}, recently reviewed in \cite{Henn:2014qga}. It was proposed in that paper that the differential equations can be cast into a canonical form that makes properties of the answer manifest, and that can be easily solved. 
The canonical form is achieved by writing the differential in a certain basis that can be found systematically using the criteria described in \cite{Henn:2013pwa}. In particular, integrals having constant leading singularities \cite{ArkaniHamed:2010gh} play an important role. The leading singularities \cite{Cachazo:2008vp} of a Feynman integral essentially correspond to residues at certain poles of the integrand, and hence are easily computed.\footnote{Leading singularities also play an important role in the study of multi-loop integrands of ${\mathcal N}=4$ SYM, see e.g. \cite{ArkaniHamed:2012nw}.}
We give a pedagogical introduction to this method, presenting the two-loop computation in full detail, and giving three-loop examples.

We present the analytic result for the three-loop cusp anomalous dimension in terms of harmonic polylogarithms \cite{Remiddi:1999ew}. The latter can be readily evaluated numerically \cite{Maitre:2005uu,Maitre:2007kp}, analytically continued, or expanded \cite{wasow1965asymptotic,Maitre:2005uu,Maitre:2007kp,Ablinger:2010kw,Ablinger:2011te,Ablinger:2012ph,Ablinger:2013cf} around the above-mentioned interesting physical limits. In this way, we reproduce the known result  for $\Gamma_{\rm cusp}(\phi,\alpha_s)$ in  the light-like limit \cite{Vogt:2000ci,Berger:2002sv,Moch:2004pa,Moch:2005tm,Baikov:2009bg} and provide new insights into on the relation to the quark-antiquark potential 
\cite{Kilian:1993nk,Drukker:2011za,Correa:2012nk,Laenen:2014jga}
in the backtracking Wilson line limit. We carry out a number of checks of our results.
At the level of the Feynman integrals, we reproduce correctly previously known results,  
including (the gauge-dependent) heavy quark wave function renormalization \cite{Melnikov:2000zc,Chetyrkin:2003vi}. 
Another important check is the gauge independence of the final result. 

The paper is organized as follows.
In section \ref{S:Cusp} we discuss the main properties of the cusp anomalous dimension, using the one-loop case as an example. In section \ref{sect:setup3loop} we discuss our three-loop Feynman diagram calculation,
while section \ref{section_three_loop_integrals} is devoted to the calculation of the Feynman integrals. 
It contains a detailed discussion of the two-loop case.
In section \ref{sect:results} we summarize our main results, and in section \ref{sect:properties}  we discuss their properties.
Section \ref{last} contains concluding remarks.
There are four appendices. Appendix A summarizes our conventions, Appendix B discusses the heavy quark effective theory
 (HQET) approach to
computing the cusped Wilson loop,
Appendices C and D contain a calculation of certain infinite classes of large $n_f$ terms of the cusp anomalous dimension
 and quark-antiquark potential.

\section{Cusped Wilson loop}
\label{S:Cusp}
 
In a generic four-dimensional Yang--Mills theory a cusped Wilson loop is defined as
\begin{equation}
W = \frac{1}{N_R} \big\langle0|\tr_R P \exp\lr{i g\oint_C dx^\mu A_\mu(x)}|0\big\rangle\,,
\label{def}
\end{equation}
where the gauge field $A_\mu(x)=A_\mu^a(x) T^a$ is integrated
along a closed integration contour $C$. The latter is smooth everywhere,
except for a single point where is has a cusp.
Here $T^a$ are the generators of the $SU(N)$ gauge group in an arbitrary representation $R$
and the normalization factor ${N_R} = \tr_R 1$ is introduced to ensure that $W=1+\mathcal{O}(g^2)$.
The cusped Wilson loop depends on the choice of the representation $R$, which
we take to be an arbitrary irreducible representation of $SU(N)$. Later in the paper we shall discuss the dependence
of the cusp anomalous dimension on  $R$.

\begin{figure}[t!bp]
\centering
\psfrag{v1}[cc][cc]{$v_1$}
\psfrag{v2}[cc][cc]{$v_2$}
\psfrag{phi}[cc][cc]{$\phi$}
\includegraphics[width = 50mm]{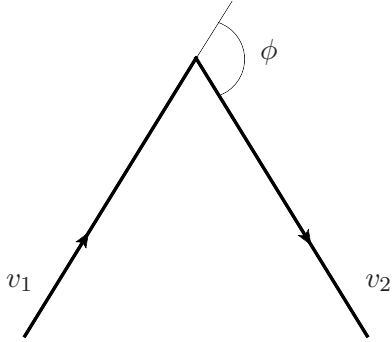}
\caption{The integration contour $C$ entering the definition~(\ref{def}) of the cusped Wilson loop.}
\label{cusp-fig}
\end{figure}

For our purposes we shall choose the integration contour $C$ in~(\ref{def})
to consist of two semi-infinite rays running along two directions $v_1^\mu$ and $v_2^\mu$
(with $v_1^2=v_2^2=1$), with the Euclidean cusp angle $\phi$ (see figure~\ref{cusp-fig})
\begin{equation} 
\cos\phi = v_1\cdot v_2\,.
\label{angle}
\end{equation}
In Minkowski space-time the analogous angle is defined as $\cosh\phi_M=v_1\cdot v_2$
and it is related to (\ref{angle}) as $\phi=i\phi_M$.
The reason for such a choice of the integration contour is twofold.
Firstly, the corresponding cusped Wilson loop $W$ has a clear physical meaning
in the context of heavy quark effective theory
(after analytical continuation from Euclidean to Minkowski space).
Namely, it describes the amplitude for a heavy quark with velocity $v_1$
to undergo the transition into the final state with velocity $v_2$.
We shall make use of this interpretation later in this section.
Secondly, the above choice of the contour facilitates significantly
the evaluation of perturbative corrections to $W$.
In particular, it allows us to apply a powerful technique for computing higher loop Feynman integrals,
as will be discussed in section \ref{section_three_loop_integrals}.

\subsection{Case of study}
\label{sect:case}

Using the definition~(\ref{def}),
we can expand $W$ in powers of the coupling constant
\begin{equation}
W =1 - \frac{1}{2} g^2 C_R \oiint_C dx^\mu d y^\nu D_{\mu\nu}(x-y) + \mathcal{O}(g^4)\,,
\label{W-1loop}
\end{equation}
where  $D_{\mu\nu}(x)$ is the free propagator of the gauge field and 
$C_R=T^a T^a$ is the quadratic Casimir of the $SU(N)$ in the representation $R$. 
As follows from this relation, the lowest order correction to $W$ does not depend on a particle content of Yang--Mills theory.
The latter dependence arises at order $\mathcal{O}(g^4)$.
Going beyond the leading order, we have to specify the underlying gauge theory.
In what follows, we shall consider two special cases:
\begin{itemize}
\item[(i)] gauge field coupled to $n_f$ species of fermions all in the fundamental representation of the $SU(N)$;
\item[(ii)] gauge field coupled to interacting $n_s$ scalars and $n_f$ fermions all in the adjoint representation of the $SU(N)$.
\end{itemize}
The corresponding Lagrangians are specified in Appendix \ref{App:YM}.  
The interaction terms are chosen in such a way that, fine tuning the number of fermions and scalars,
we can use these two cases to describe QCD and supersymmetric Yang--Mills theories, respectively.
In particular, the maximally supersymmetric $\mathcal{N}=4$ SYM theory
corresponds to $n_s=6$ and $n_f=4$.  

For arbitrary number of fermions and scalars the above mentioned theories
are neither conformal, nor supersymmetric.
The $\mathcal{N}=4$ SYM theory plays a special role since both symmetries are exact
for any value of the coupling constant.
In addition, we can define in this theory a supersymmetric extension  
of the cusped Wilson loop \cite{Maldacena:1998im,Rey:1998ik}
\begin{equation}
\mathcal W = {1\over  N_R} \big\langle0|\tr_R   P \exp\lr{i g\int dt \Big[\, \dot x^\mu A_\mu(x)+ \sqrt{\dot x^2}\, n^I(t) \phi_I(x) \Big]} |0\big\rangle\,,
\label{superW}
\end{equation}
where we introduced the parameterisation of the integration contour $x^\mu=x^\mu(t)$
with $\dot x_\mu=\partial_t x_\mu(t)$.
As compared with~(\ref{def}), it has an additional coupling to six scalars $\phi_I$ that
depends on a unit vector $n^I=n^I(t)$ in the internal $S^5$ space.
In analogy with the previous case, we take this vector to be constant
along two semi-infinite rays, $n_1^I$ and $n_2^I$,
except the cusp point where it forms an additional internal cusp angle $\cos\theta=\sum_I n_1^I n_2^I$.
The vacuum expectation value of this Wilson loop operator has been studied in many papers.
Perturbative results are available up to four loops in the planar case (and in part in the non-planar case) \cite{Drukker:1999zq,Drukker:2011za,Correa:2012nk,Henn:2013wfa}; results at strong coupling are available via the AdS/CFT correspondence \cite{Drukker:1999zq,Drukker:2011za,Forini:2010ek}; the small angle asymptotics is known exactly \cite{Correa:2012at}. Finally, the system is governed by integrability \cite{Correa:2012hh,Drukker:2012de}, for further work in this direction see e.g. \cite{Bajnok:2013sya}.

\subsection{Cusp anomalous dimension}
\label{S:1L}

To explain the general framework of our analysis,
let us revisit the one-loop calculation of the cusped Wilson loop~(\ref{W-1loop}).
Anticipating the appearance of divergences in $W$,
we introduce the dimensional regularization with $D=4-2\epsilon$.
Then, we use gauge invariance of $W$ to choose the Feynman gauge, $D_{\mu\nu}(x) = g_{\mu\nu} D(x)$, with
\begin{align}
D(x) &= -i \int \frac{d^D k}{(2\pi)^D} \frac{\e^{ikx}}{k^2+i0}
= \frac{\Gamma(1-\epsilon)}{4\pi^{2-\epsilon}} (-x^2+i0)^{-1+\epsilon}\,.
\label{1L:D}
\end{align}
To the lowest order in the coupling,
we find from~(\ref{W-1loop}) that $W$ is given by a gluon propagator
integrated over the position of its end points on the integration contour.
Parameterizing points on two semi-infinite rays as $-v_1^\mu s$ and $v_2^\mu t$
with $0\le s,t< \infty$, we arrive at the following integral
\begin{align}
I(\phi) ={}& \int_0^\infty ds\,dt\, (v_1 v_2) D(v_1 s+ v_2 t)
\nonumber\\
={}& i \int \frac{d^D k}{(2\pi)^D} \frac{(v_1 v_2)}{(k^2+i0)((kv_1)+i0)((kv_2)+i0)}\,,
\label{zero}
\end{align}
where in the second relation we switched to the momentum representation.
Then,
\begin{equation}
\log W = g^2 C_R \bigl[I(\phi) - I(0)\bigr] + \mathcal{O}(g^4)\,,
\label{W-1}
\end{equation}
where the second term inside the square bracket
comes from the contribution of the gluon propagator attached to the same semi-infinite ray.

It is easy to see from the second relation in~(\ref{zero})
that $I(\phi)$ develops poles in $\epsilon$ both in the infrared, $k^\mu \to 0$,
and in the ultraviolet, $k^\mu\to\infty$.
In the configuration space, for $\rho=s+t$,
the same poles arise from integration over $\rho\to\infty$ and $\rho\to0$, respectively.
Moreover, since the integral in~(\ref{zero}) does not involve any scale,
it vanishes in the dimensional regularisation, $I(\phi)=0$,
thus indicating that infrared (IR) and ultraviolet (UV) poles of $W$
are in a one-to-one correspondence with each other~\cite{Korchemsky:1985xj}.

In order to compute the cusp anomalous dimension,
we have to disentangle UV and IR divergences of $W$.
This can be done by introducing inside the first integral in~(\ref{zero})
the additional factor $\exp(-i\delta(s+t))$ with $\rm Im\,\delta<0$.
It suppresses the contribution of large $(s+t)$
and introduces the dependence on the IR cut-off $\delta$.
In this way, we obtain from (\ref{zero})
\begin{align}
I_\delta(\phi) ={}&  \int_0^\infty ds\,dt\, (v_1 v_2) D(v_1 s+ v_2 t)\e^{-i\delta(s+t)}
\nonumber\\
={}& i \int \frac{d^D k}{(2\pi)^D} \frac{(v_1 v_2)}{(k^2+i0)((kv_1)-\delta+i0)((kv_2)-\delta+i0)}\,,
\label{zero1}
\end{align}
where we introduced the subscript $\delta$ to indicate the dependence on this scale.%
\footnote{Notice that we can use simple transformation properties of $I_\delta(\phi)$
under rescaling, $k^\mu \to z k^\mu$ and $\delta\to z \delta$ (with $z>0$),
to choose $\delta$ to our best convenience.}
Changing the integration variables in the first relation to $y=s/(s+t)$ and $\rho=s+t$, we obtain
\begin{equation}
I_\delta(\phi) = - \frac{\Gamma(2\epsilon)}{(2\pi)^2} (\pi \mu^2/\delta^2)^\epsilon
\left[ \phi\cot\phi + \mathcal{O}(\epsilon)\right]\,,
\end{equation}
where the cusp angle $\phi$ was defined in~(\ref{angle}).
As expected, $I_\delta(\phi)$ develops a UV pole $1/\epsilon$.
Together with~(\ref{W-1}) this leads to the well-known result
for one-loop cusp UV divergence~\cite{Polyakov:1980ca}
\begin{equation}
\log W = - \frac{1}{2\epsilon} \frac{g^2 C_R}{(2\pi)^2} \lr{\phi\cot\phi-1} + \mathcal{O}(\epsilon^0)\,.
\label{lnW1}
\end{equation}
The coefficient in front of $1/\epsilon$ is gauge invariant, 
it does not depend on the IR regulator $\delta$ and defines the one-loop correction to the cusp anomalous dimension.

The properties of cusp singularities of Wilson loops are well understood to all loops 
\cite{Polyakov:1980ca,Gervais:1979fv,Dotsenko:1979wb,Arefeva:1980zd,Brandt:1981kf,Dorn:1986dt}.  
The cusped Wilson loop can be made
finite by subtracting UV poles and expressing the resulting quantity $\log W - \log Z$ in terms of renormalized coupling constant.
In the $\overline{\rm MS}$ scheme, the renormalization $Z-$factor has the
following form
\begin{align}\notag\label{lnZ}
\log Z ={}& -{1\over 2\epsilon}  \lr{{\alpha_s \over \pi}}\Gamma^{(1)} +\lr{{\alpha_s \over \pi}}^2 \left[{\beta_0 \Gamma^{(1)}\over 16\epsilon^2}  - {\Gamma^{(2)}\over 4\epsilon}  \right]
\\[2mm]
{}& +\lr{{\alpha_s \over \pi}}^3 \left[-{\beta_0^2 \Gamma^{(1)}\over 96\epsilon^3} + {\beta_1\Gamma^{(1)} +4\beta_0\Gamma^{(2)} \over 96\epsilon^2} -{\Gamma^{(3)}\over 6\epsilon} \right]+ \dots\,,
\end{align}
where $\alpha_s=g_{\rm YM}^2/(4\pi)$ is the renormalized coupling constant  
satisfying
\begin{align}
{d \log \alpha_s \over d\log\mu}  = -2\epsilon   - 2 \beta(\alpha_s) = -2\epsilon   
- 2\left[   \beta_0 {\alpha_s\over 4\pi}  +\beta_1 \lr{\alpha_s\over 4\pi}^{2} +\dots\right] \,.
\end{align}
The QCD beta-function is well known \cite{vanRitbergen:1997va} (we need it to two loops), while for the case of theory (ii) renormalization was discussed in \cite{Machacek:1983tz,Machacek:1983fi,Machacek:1984zw}. 
The expansion coefficients $\Gamma^{(i)}$ carry the dependence on the cusp angle $\phi$ and define the cusp anomalous dimension
\begin{align}\label{cusp-def}
\Gamma_{\rm cusp}(\phi,\alpha_s) = {d \log Z \over d\log \mu} =  {\alpha_s \over \pi}\, \Gamma^{(1)} + \lr{{\alpha_s \over \pi}}^2 \Gamma^{(2)}
+ \lr{{\alpha_s \over \pi}}^3 \Gamma^{(3)} +\dots
\end{align}
Matching \re{lnW1} into \re{lnZ} we obtain the one-loop cusp anomalous dimension
\begin{align}\label{1-loop}
\Gamma^{(1)}  = C_R\lr{\phi\cot\phi-1} = -C_R\lr{\xi \log x +1}\,,
\end{align}
where the notation was introduced for $x=\e^{i\phi}$ and $\xi=(1+x^2)/(1-x^2)$.

\subsection{Regularization}
\label{section:regularizationHQET}

Comparing \re{zero1} with \re{zero} we observe that the net effect of the IR cut-off is to shift the position of poles of the eikonal 
propagators. 
This transformation has a simple interpretation in the context of heavy
quark effective theory (see Appendix~B for more details). 

\begin{figure}[t!h]
\centerline {\psfrag{v1}[cc][cc]{$v_1$}\psfrag{v2}[cc][cc]{$v_2$}  \psfrag{dots}[cc][cc]{$\dots$} 
\psfrag{V}[cc][cc]{$V$}\psfrag{S}[cc][cc]{$\Sigma$}
  \includegraphics[width = 65mm]{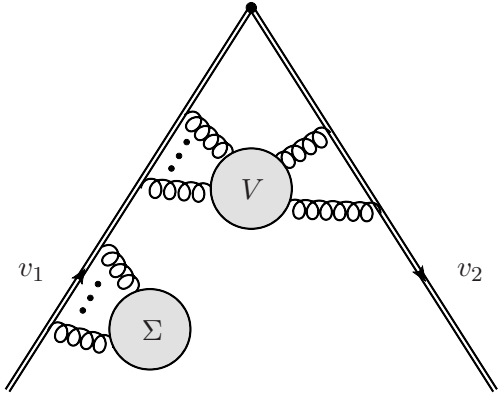}
 }
  \caption{The cusped Wilson loop as a heavy quark transition amplitude. Double line represents a heavy quark propagator, $V$ and $\Sigma$ denote vertex and self-energy corrections, respectively.}
  \label{HQET-fig1}
\end{figure}

As was mentioned earlier, the cusped Wilson loop \re{def} can be interpreted as an amplitude 
for a heavy quark with the velocity $v_1$ to undergo the transition into the state with the velocity $v_2$ (see figure \ref{HQET-fig1}). 
Indeed, the heavy quark propagates along the straight line in the direction of its velocity and the effects of its interaction with gauge fields is
described by a Wilson line evaluated along the classical trajectory of a heavy quark. 

\begin{figure}[h!]
 {\psfrag{v}[cc][cc]{$v$} \psfrag{dots}[cc][cc]{$\dots$} 
 \psfrag{k1}[cc][cc]{$k_1$}\psfrag{k2}[cc][cc]{$k_2$}\psfrag{kn}[cc][cc]{$k_n$}
 \psfrag{rules}[lc][cc]{$\displaystyle = \  g^n {v^{\mu_1} T^{a_1}\over (k_1 v)-\frac12}{v^{\mu_2} T^{a_2}\over  ((k_1+k_2) v) -\frac12} 
 \dots {v^{\mu_n} T^{a_n}\over  \sum_{i=1}^n (k_i v) -\frac12}$}
  \includegraphics[width = 60mm]{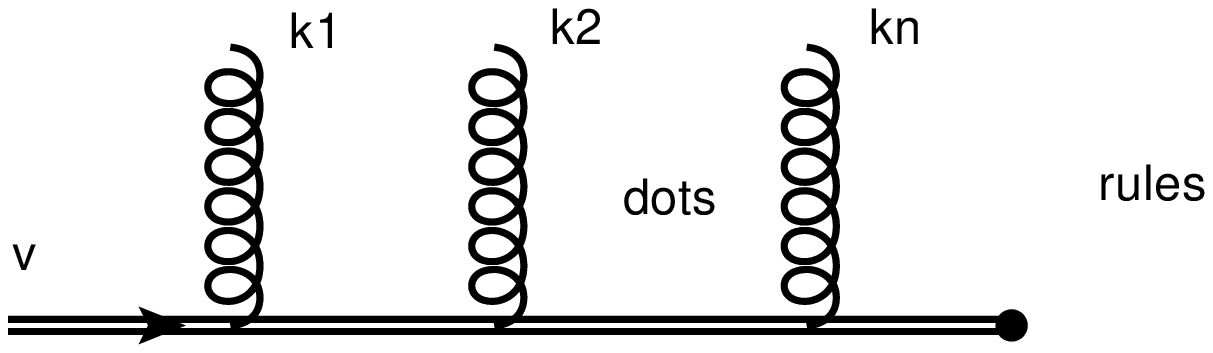}
 }
  \caption{Feynman rules for computing Wilson loop in the momentum space. Double lines stand  for heavy quark propagators
  and wavy lines denotes gluons with outgoing momenta. It is tacitly assumed that all propagators have the same `$+i0$' prescription 
  as in (\ref{delta}).}
  \label{HQET-fig}
\end{figure}

The heavy quark transition amplitude suffers from both IR and UV divergences. The former can be regularized in the momentum
representation by slightly shifting the heavy quark from its mass shell  
\begin{align}\label{delta}
\frac1{(kv_i)+i0}\ \to \ \frac1{(kv_i)-\delta+i0}
\end{align}
with the IR cut-off $\delta$ having the meaning of the residual energy of heavy quark. As was already mentioned in the 
previous subsection, the corresponding Feynman integrals are homogenous functions of loop momenta $k$. This fact allows us to assign to 
$\delta$ an arbitrary real value. It proves convenient to choose $\delta=1/2$.
Applying the regularization (\ref{delta}) with $\delta=1/2$, 
we can make use of a very efficient diagram technique for computing $W$ in the momentum representation beyond the leading 
order (see figure~\ref{HQET-fig} for the corresponding Feynman rules).\footnote{The Feynman integrals obtained in this way are the same that appear in heavy quark effective theory. See appendix B for more details.}

\bigskip
\medskip

To regularize UV divergences we employ dimensional regularization. The cusp divergences come both from the one-particle
irreducible vertex corrections $V(\phi)$  and from self-energy corrections $\Sigma$ to the heavy quark propagators (see figure \ref{HQET-fig1}). 
In virtue of Ward identities, the latter contribution is related to the vertex correction at zero recoil angle $V(0)$ leading to \cite{Korchemsky:1987wg}
\begin{align}\label{lnV}
\log W = \log V(\phi) - \log V(0) = \log Z + O(\epsilon^0)\,.
\end{align}
This relation allows us to compute $\log Z$ from the subset of Feynman diagrams corresponding to vertex corrections $V(\phi)$,
i.e. with non-trivial angular dependence.

\subsection{Nonabelian exponentiation}
\label{sect:NE}

The calculation of the cusp anomalous dimension can be significantly simplified
by making use of the nonabelian exponentiation property of the Wilson loop~\cite{Sterman:1981jc,Gatheral:1983cz,Frenkel:1984pz}.
It allows us to express a logarithm of the Wilson loop, $\log W$,
in terms of a special class of `maximally nonabelian' diagrams, the so-called webs.

\begin{figure}[tbp]
\centering
\psfrag{a}[cc][cc]{(a)}\psfrag{b}[cc][cc]{(b)}\psfrag{c}[cc][cc]{(c)}
\includegraphics[width = .9 \textwidth]{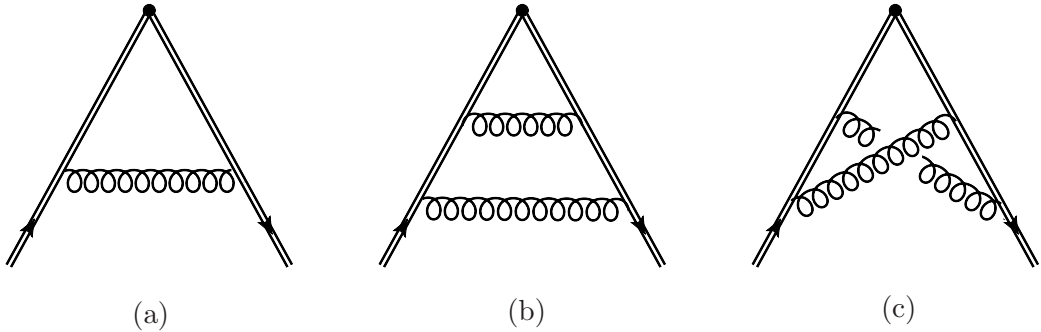}
\caption{Examples of Feynman diagrams contributing to the vertex function $V(\phi)$ at one loop (a)
and two loops, (b) and (c).
The diagram (b) does not contribute to the right-hand of~(\ref{nonab}).}
\label{exp-fig}
\end{figure}

In the special case of gauge theories in which all fields are defined
in the adjoint representation of $SU(N)$,
this leads to the following general expression
\begin{equation}
\log W = C_R  \sum_{n=1}^3 \lr{\frac{\alpha_s}{\pi}}^n C_A^{n-1}
\left[ V_n(\phi) - V_n(0)\right] + \mathcal{O}(\alpha_s^4)\,,
\label{nonab}
\end{equation}
where $C_A=N$ is the quadratic Casimir operator of $SU(N)$ in the adjoint representation,
$f^{abc} f^{and} = C_A \delta^{cd}$,
and $V_n(\phi)$ stands for the sum of certain Feynman integrals
defining $n-$loop corrections to the (one-particle irreducible) vertex function
(see figure~\ref{exp-fig}).
Notice that the expression on the right-hand side of~(\ref{nonab})
only depends on the quadratic Casimirs.  In addition,
it is proportional to $C_R$ that depends on the representation
in which the Wilson loop~(\ref{def}) is defined, the so-called Casimir scaling.
It is expected that both properties are violated at four loops
since the color factors start to depend on higher Casimirs of $SU(N)$.

The power of the nonabelian exponentiation~(\ref{nonab}) is
that it allows us to discard the diagrams whose color factor does not contain terms
of the maximally nonabelian form.
Moreover, we can use~(\ref{nonab}) to express their contribution
in terms of Feynman integrals $V_n$ that appear on the right-hand side of~(\ref{nonab}).
To illustrate this point consider the Feynman diagrams shown in figure~\ref{exp-fig}.
The one-loop diagram shown in figure~\ref{exp-fig}(a) has the color factor $C_R$
and the corresponding Feynman integral defines $V_1(\phi)$.
The two-loop diagrams shown in figures~\ref{exp-fig}(b) and~(c)
have the color factors $C_R^2$ and $C_R(C_R-C_A/2)$, respectively.
Since the second color factor contains the maximally nonabelian term $C_R C_A$,
only diagram shown in  figure~\ref{exp-fig}(c) contributes to~(\ref{nonab}) at two loops.
At the same time, the nonabelian exponentiation implies that the two-loop contribution
to $W$ proportional to $C_R^2$ should be related to one-half of the square
of one-loop contribution. This leads to the following relation between the Feynman integrals
corresponding to diagrams shown in figure~\ref{exp-fig}:\footnote{The relation \re{half} holds up to terms proportional to the 
IR cut-off $\delta$. Such terms do not produce UV divergences and, therefore, do not contribute to the cusp anomalous dimension.}
\begin{equation}
\frac{1}{2} \left[I_{\text{\ref{exp-fig}(a)}}\right]^2 =
I_{\text{\ref{exp-fig}(b)}} + I_{\text{\ref{exp-fig}(c)}}\,.
\label{half}
\end{equation}
Indeed, in configuration space the diagrams shown in figure~\ref{exp-fig}(b) and (c) only differ in the ordering of gluons
attached to two semi-infinite rays and the relation \re{half} follows from the identity $\theta(t_1-t_2) +\theta(t_2-t_1)=1$.

Notice that the diagram shown in figure~\ref{exp-fig}(c) is nonplanar.
We can then use~(\ref{half}) to turn the logic around
and express the contribution of this diagram to~(\ref{nonab})
in terms of planar integrals only.
The same is true at higher loops.
Namely, up to three loops, the vertex function $V_n(\phi)$
on the right-hand side of~(\ref{nonab})
can be expressed in terms of planar Feynman integrals only.
To see this we observe that the sum in~(\ref{nonab}) only depends on $C_A=N$
and  does not contain nonplanar corrections.
Therefore, computing $\log W$ in the planar limit
we can unambiguously determine $V_n(\phi)$ up to three loops.
Starting from four loops, $\log W$ depends on higher $SU(N)$ Casimirs
that generate subleading (nonplanar) corrections suppressed by powers of $1/N^2$ (see section \ref{sect:casimir} below).
They are accompanied by Feynman integrals that are not necessarily planar.


The fact that only planar Feynman integrals are needed up to three loops is a technical simplification
that will be helpful (but not essential) in the calculation described in section \ref{section_three_loop_integrals}.
The main advantage of planar integrals is that  we can define canonical region (or dual) coordinates that
make it easy to deal with the loop integrand, without having the ambiguity of redefinitions
of the loop momenta. This also makes the classification of all required integrals rather straightforward. 

\begin{figure}[tbp]
\centering
\psfrag{v1}[cc][cc]{$v_1$}
\psfrag{v2}[cc][cc]{$v_2$}
\psfrag{phi}[cc][cc]{$\phi$}
\includegraphics[width = \textwidth]{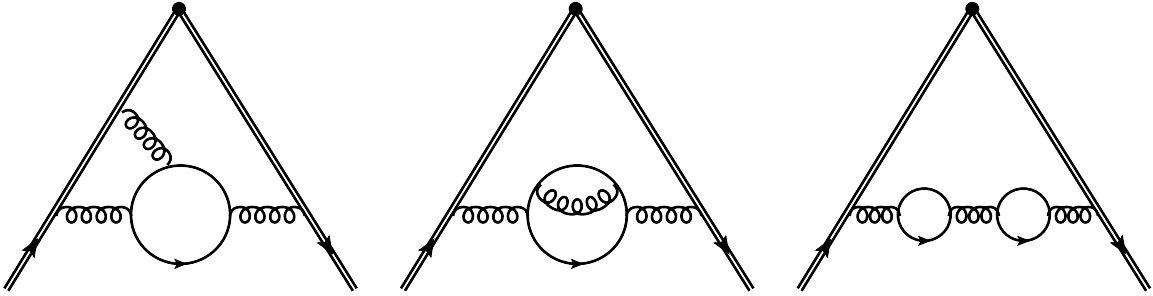}
\caption{Sample diagrams for different $n_f$-dependent color structures appearing at three loops. They contribute to $C_A T_F n_f$,  $C_{F} T_{F} n_{f}$ and $(T_{F} n_{f})^2$ terms, respectively.}
\label{figure-nf-contributions}
\end{figure}

An immediate consequence of nonabelian exponentiation~(\ref{nonab}) is
that the cusp anomalous dimension~(\ref{cusp-def}) has a similar dependence on the $SU(N)$ Casimirs,
\begin{equation}\label{cusp-color}
\Gamma_{\rm cusp}(\phi,\alpha_s)  = C_R \bigg[ \frac{\alpha_s}{\pi} \gamma +\lr{\frac{\alpha_s}{\pi}}^2 C_A \gamma_{A} +\lr{\frac{\alpha_s}{\pi}}^3  C_A^2\gamma_{AA} \bigg]+ \mathcal{O}(\alpha_s^4)\,,
\end{equation}
with $\gamma$, $\gamma_A$ and $\gamma_{AA}$ depending on the cusp angle $\phi$.
We recall that this relation only holds in gauge theories with all fields defined in the adjoint representation.
If some of the fields are defined in the fundamental representation, as it happens in QCD,
the color factors of maximally nonabelian diagrams have more complicated form
and depend on quadratic Casimir in the fundamental representation $C_F=(N^2-1)/(2N)$.
(Examples of three-loop diagrams producing new color factors are shown in figure~\ref{figure-nf-contributions}.)
Nevertheless, similar to the previous case, up to three loops
the color factors that appear in the expansion of $\Gamma_{\rm cusp}(\phi,\alpha_s)$
only depend on quadratic Casimirs of the $SU(N)$.
 One can show that the cusp anomalous dimension
in QCD with $n_f$ fermions in  the fundamental representation has the following form 
\begin{align}
{}& \Gamma_{\rm cusp, QCD}(\phi,\alpha_s)  =  C_R   \biggl[ \frac{\alpha_s}{\pi} \gamma
+ \lr{\frac{\alpha_s}{\pi}}^2 \left( C_A \gamma_A + T_F n_f \gamma_f \right)  
\nonumber\\
{}&  +\left(\frac{\alpha_s}{\pi}\right)^3 \left( C_A^2 \gamma_{AA} + C_F T_F n_f \gamma_{Ff} + C_A T_F n_f \gamma_{Af} + \left( T_F n_f \right)^2 \gamma_{ff} \right)
    \biggr]+ \mathcal{O}(\alpha_s^4)\,,
\label{R:NA}
\end{align}
where $T_F$ defines the normalization of the $SU(N)$ generators in the fundamental representation,
$\tr_F(T^a T^b) = T_F \delta^{ab}$, and the coefficient functions are different, in general, from those in (\ref{cusp-color}). 
As compared with (\ref{cusp-color}), the cusp anomalous dimension in QCD contains the
additional terms proportional to powers of $T_F n_f$. They come from diagrams involving fermion loops (see figure~\ref{figure-nf-contributions}).
 
\subsection{Dependence on the cusp angle}

In order to discuss the dependence of $\Gamma_{\rm cusp}(\phi,\alpha_s)$ on the cusp angle,
it proves convenient to introduce auxiliary (complex) variables
\begin{align}\notag
& x = \e^{i\phi}\,,  && x+x^{-1} = 2 \cos\phi\,,
\\[2mm]
& \xi = \frac{1+x^2}{1-x^2} = i\cot\phi\,,
&& \chi={1-x^2\over x}=-2i\sin\phi\,.
\label{x}
\end{align}
In Euclidean space, for $0\le \phi \le \pi$, we have $|x|=1$.
In Minkowski space, for $\phi=i\phi_M$ with $\phi_M$ real,
the variable $x=\e^{-\phi_M}$ can take arbitrary nonnegative values.
Moreover, due to the symmetry of the definition~(\ref{x}) under $x\to 1/x$
we can assume $0<x<1$ without loss of generality.

We can use the one-loop result~(\ref{1-loop}) to illustrate interesting asymptotic behaviour
of the cusp anomalous dimension in three different limits.
For $\phi\to0$, or $x\to 1$, the integration contour in figure~\ref{cusp-fig}
transforms into a straight line leading to the vanishing of the cusp anomaly
\begin{equation}
\Gamma_{\rm cusp}(\phi,\alpha_s) \stackrel{\phi\to 0}{\sim} - \phi^2 B(\alpha_s)
\label{br}
\end{equation}
with $B = C_R\,\alpha_s /(3\pi) + \mathcal{O}(\alpha_s^2)$ the so-called bremsstrahlung function.
For $\phi\to\pi$, or $x\to-1$, the integration contour degenerates into two antiparallel lines
and the cusp anomalous dimension develops a pole
\begin{equation}
\Gamma_{\rm cusp}(\phi,\alpha_s) \stackrel{\phi\to\pi}{\sim} - \frac{V(\alpha_s)}{\pi-\phi}
\end{equation}
with $V(\alpha_s)=C_R \alpha_s + \mathcal{O}(\alpha_s^2)$ being closely related
to the heavy quark-antiquark potential (we shall clarify this relation in section~\ref{sect:pot}).
In Minkowski space, for large cusp angle, $\phi_M\to\infty$, or $x\to 0$,
the cusp anomaly scales logarithmically
\begin{equation}
\Gamma_{\rm cusp}(\alpha_s, i \phi_M) \stackrel{\phi_M\to\infty}{\sim} K(\alpha_s) \,\phi_M\,,
\end{equation}
with $K(\alpha_s) = C_R \alpha_s/\pi + \mathcal{O}(\alpha_s^2)$
being the light-like cusp anomalous dimension.

Finally, the Wilson line integrals are naively invariant under the crossing transformation $v_{2} \to -v_{2}$,  or equivalently $x\to -x$ (see e.g. one-loop integral (\ref{zero})). This invariance is broken by the Feynman `$+i0$' prescription and, therefore, we expect it to be valid only up to terms picked up from crossing the branch cut on the negative real axis,  
\begin{equation}\label{Disc}
\Gamma(-x+i0) =\Gamma(x)  +  \frac{1}{2} {\rm Disc}\, \Gamma(-x)  \,,
\end{equation}
where $0<x<1$ and $ {\rm Disc}\,\Gamma(-x) := \Gamma(-x + i0) -  \Gamma(-x - i0)$ 
denotes the contribution originating from crossing the branch cut.
For example, at one loop we have from (\ref{1-loop}),  ${\rm Disc}\, \Gamma^{(1)}(-x) = - 2  \pi i\, C_{R} \xi$.
 
\section{Setup of the three-loop calculation}
\label{sect:setup3loop}
 
As explained in section \ref{section:regularizationHQET},
the cusp anomalous dimension can be calculated within
the framework of heavy quark effective theory (HQET). More precisely, we have to compute
three-loop corrections to the vertex function $V(\phi)$  (see figure~\ref{HQET-fig1}) and, then,
apply (\ref{lnV}) and (\ref{lnZ}). 
The corresponding HQET diagrams  contributing to $V(\phi)$ contain two external heavy quarks with velocities $v_1$ and $v_2$.
Note that, by definition, the heavy quarks do not propagate within loops and
therefore, we only have to consider massless particles (gluons, fermions and scalars) inside the
diagrams. The interaction between massless particles is described by the Lagrangians specified in Appendix~\ref{App:YM}.

We use {\tt QGRAF}~\cite{Nogueira:1991ex} to generate all
(one-heavy-quark-irreducible) vertex diagrams in HQET. In total there are 315
three-loop diagrams involving gluons and fermions plus 100 additional diagrams involving scalars. As
explained in Section \ref{sect:NE}, due to nonabelian exponentiation we
only need to calculate the planar diagrams. We find that in the planar limit  there are only 120 diagrams,
plus 32 diagrams with scalars. 

\begin{figure}[h!tb]
\centering
\begin{tabular}{cccc}
  \includegraphics[width=0.22\textwidth]{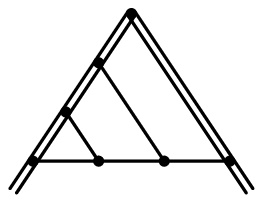}&\includegraphics[width=0.22\textwidth]{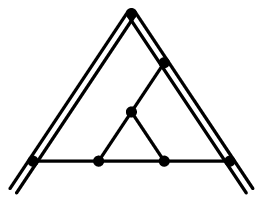}&
  \includegraphics[width=0.22\textwidth]{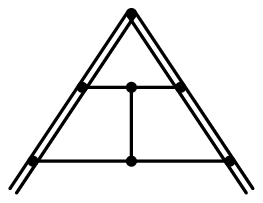}&\includegraphics[width=0.22\textwidth]{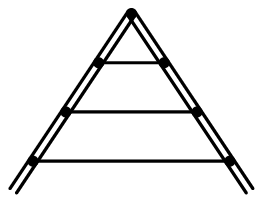}\\[4mm]
  \includegraphics[width=0.22\textwidth]{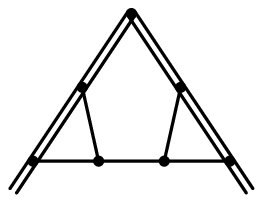}&\includegraphics[width=0.22\textwidth]{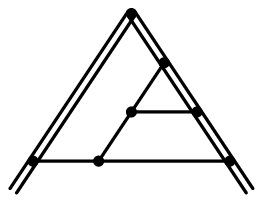}&
  \includegraphics[width=0.22\textwidth]{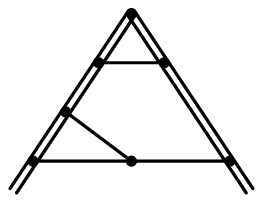}&\includegraphics[width=0.22\textwidth]{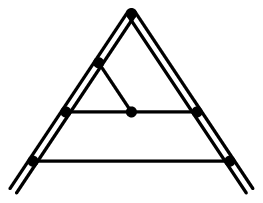}
\end{tabular}
\caption{Generic topologies contributing to the vertex function $V(\phi)$ at three loops in the planar limit. Double lines
denote heavy quarks and solid lines stand for massless gluons, fermions or scalars.
}
\label{F:topo}
\end{figure}

Computing the contribution of three-loop planar diagrams to the vertex function, we
performed the numerator algebra using {\tt Form}~\cite{Vermaseren:2000nd}, {\tt TForm}~\cite{Tentyukov:2007mu} or
{\tt Reduce}~\cite{reduce}. In this way, we obtained scalar HQET integrals which can be
mapped onto the 8 generic topologies
\footnote{Counting the number of needed topologies, we made use of
the symmetry of $V(\phi)$ under exchange of the heavy quarks, $v_1\leftrightarrow v_2$.}
 shown in Fig.~\ref{F:topo} either
manually or using {\tt q2e} and {\tt exp}~\cite{Harlander:1997zb,Seidensticker:1999bb}.

Applying integration-by-parts identities~\cite{Chetyrkin:1981qh}, the three-loop integrals are then reduced to a set of 71 master integrals 
with the help of {\tt Crusher}~\cite{crusher}, {\tt  Fire}~\cite{Smirnov:2008iw,Smirnov:2013dia,Smirnov:2014hma} or 
{\tt LiteRed}~\cite{Lee:2012cn,Lee:2013mka}. The evaluation of the master integrals, which plays a central role in the
calculation, is described in detail in Section \ref{section_three_loop_integrals}. 
Matching the divergent part of the vertex function to the expected form (\ref{lnV}) and (\ref{lnZ}), we obtain the three-loop cusp anomalous
dimension given in Section \ref{sect:results}. Its properties, which also serve as indispensable checks of our calculation, are discussed 
in Section \ref{sect:properties}. 

In addition, in the case of QCD, we compute corrections to the cusp anomalous dimension enhanced by the number
of fermions, $C_R (T_F n_f)^{L-1} \alpha_s^L$ and  $C_RC_F (T_F n_f)^{L-2}\alpha_s^L$. The details of the calculation 
can be found in Appendices C and D.

\section{Three-loop calculation of HQET integrals}
\label{section_three_loop_integrals}

In this section, we describe our choice of the basis of Feynman integrals that contribute to the cusp anomalous dimension at three loops
and present their calculation. An unusual feature of these integrals as compared with the conventional Feynman integrals is that they involve
eikonal or heavy quark propagators (see figure \ref{HQET-fig}).  In what follows we refer to them as HQET integrals.

In section~\ref{section_iterated_integrals}, we start by introducing the generalized polylogarithm functions required in our calculation.
In section~\ref{section:purefunctions} we discuss their weight properties and relation to Feynman integrals.
To explain the procedure, we first explain our method for computing 
the master integrals using differential equations. The two-loop case is reviewed as a pedagogical example in section~\ref{section:hqet2loopexample},
Next, in sections~\ref{choice_mi} and~\ref{section:cuts}, we explain in detail our choice of integral basis.
We give there two complementary points of view, the first being based on analyzing the Wilson line integrals in position space, and the second analyzing generalized cut properties of the same integrals in the momentum-space.
Finally, in section~\ref{sec:DE3loops}, we perform the three-loop calculation of the master integrals.

\subsection{Iterated integrals}
\label{section_iterated_integrals}

We will see below that the HQET integrals required in our calculation can be expressed in terms of certain iterated integrals
studied in the mathematical literature~\cite{Goncharov1,Brown1}. More precisely,  a particular subclass of such integrals,
known in the physics literature as harmonic polylogarithms (HPL)~\cite{Remiddi:1999ew,Gehrmann:2001pz},
is sufficient to express all results.

The harmonic polylogarithms $H_{a_1 , a_2 , \ldots, a_n}(x)$ depend on the set of indices $a_1,\dots,a_n$ taking values $\{-1,0,+1\}$.
They are defined iteratively with respect to their weight $n$. 
The iteration  starts with the weight-one functions
\begin{equation}
H_{1}(x) = - \log(1-x)\,,\qquad H_{0}(x) = \log(x)\,,\qquad H_{-1}(x) = \log(1+x)\,.
\end{equation}
For all indices being different from zero, $a_i\neq 0$ for $i=1,\dots,n$, higher weight functions are defined as 
\begin{equation}
H_{a_1 , a_2 , \ldots, a_n}(x) = \int_0^x f_{a_1}(t) H_{a_2 , \ldots, a_n}(t) dt\,, 
\end{equation}
with the integration kernels
\begin{equation}
f_1(x) = (1-x)^{-1}\,,\qquad f_0(x) = x^{-1}\,,\qquad f_{-1}(x)=(1+x)^{-1}\,.
\end{equation}
In the case of all indices being zero,
we have 
\begin{align}
H_{\underbrace{\scriptstyle 0\,\ldots 0}_{n}}(x) = {1\over n! }(\log x)^n \,.
\end{align}
The weight of $H_{a_1 , a_2 , \ldots, a_n}(x)$ refers to the number of integrations
with logarithmic kernels $dx/x$, $dx/(x+1)$, $dx/(x-1)$
and equals the length of the index vector $\vec a=(a_1 , a_2 , \ldots, a_n)$.
In the physics literature, it is sometimes colloquially referred to as ``transcendentality''.

Iterated integrals satisfy a shuffle algebra,
which expresses the product of a weight $n$ and a weight $m$ function
as a sum over weight $k=n+m$ functions,
\begin{equation}
H_{\vec{a}}(x) H_{\vec{b}}(x) = \sum_{\vec{c}\,\in\,\vec{a}\,\shuffle\,\vec{b}} H_{\vec{c}}(x)\,,
\end{equation}
 where the list $\vec c$ of length $n+m$ arises from ``shuffling'' the lists $\vec{a}=(a_1,\dots,a_{n})$ and $\vec{b}=(b_1,\dots,b_{m})$,
like a deck of cards.

Special values of harmonic polylogarithms at $x=1$ and $x=-1$ are related to nested sums, called Euler sums.
The latter satisfy additional relations, see e.g.~\cite{Broadhurst:1998rz,Remiddi:1999ew},
that allow us to reduce them to a minimal number of constants.
It turns out that in our calculation, only zeta values $\zeta_n =  \sum_{k\ge 1} 1/k^{n}$ are needed.

\subsection{Pure functions of uniform weight}
\label{section:purefunctions}

In our calculation, functions of uniform weight play an important role.
The latter are defined as linear combinations of iterated integrals of the same weight  with coefficients rational
in $x$.
For example,
\begin{equation}
\frac{1}{1-x} H_{1,0,1}(x) +  2 \frac{x}{1+x} H_{0,0,0}(x)
\end{equation}
is a function of uniform weight $3$.
We can go further and define so-called {\em pure functions},
which are  linear combinations of uniform weight functions with rational coefficients, e.g.
\begin{equation}
H_{1,0,1}(x) +  2 H_{0,0,0}(x)\,.
\end{equation}
%
%

Pure functions have nice properties that make them easy to compute.
Most importantly, their differential has a simple form.
If $f^{(k)}(x)$ is a pure function of weight $k$, then
\begin{equation}
d f^{(k)}(x)  = \sum_{i} c_{i} \,  d\log \alpha_i(x) \, g^{(k-1)}_i(x)\,,
\label{d_purefunction}
\end{equation}
where $c_{i} \in \mathbb{Q}$, the $\alpha_i(x)$ are at most algebraic functions,
and $g_i^{(k-1)}(x)$ are certain pure functions of weight $(k-1)$.
For example,
\begin{equation}
d \left[ H_{1,0}(x) + \frac{1}{2} H_{0,-1}(x) \right]
= - d\log(1-x) H_{0}(x) + \frac{1}{2} d\log(x) H_{-1}(x)\,.
\end{equation}
For $k=1$, the expression on the right-hand side of (\ref{d_purefunction}) contains only one term,
since there is only one (independent) weight zero function  $f^{(0)}(x)=1$. As a consequence,
the relation (\ref{d_purefunction}) allows for a simple recursive way of defining a weight $k$ function,
through differential equations.
This is precisely the route that we take in section~\ref{section:hqet2loopexample} below.

Let us imagine we have a set of Feynman integrals that can be evaluated in terms of iterated integrals.
Taking certain linear combinations of these integrals, we may try to express them in terms of pure functions.
But is there a way to tell in advance which Feynman integrals will evaluate to pure functions?

A proposal in this direction was made in~\cite{ArkaniHamed:2010gh},
based on ideas related to generalized unitarity~\cite{Bern:1994zx,Cachazo:2008vp},
and relying on a large body of evidence from computations in ${\mathcal N}=4$ super Yang-Mills.
To understand this, let us imagine a Feynman integral depending on many kinematic variables.
Iterated integrals are multivalued functions in these variables.
The idea is that taking generalized unitarity cuts of Feynman integrals
should in some way correspond to taking discontinuities of these functions
(the precise correspondence between the two objects is an open problem).
Then, taking different discontinuities should project onto different terms
in the expression of an integral in terms of iterated integrals.
The conjecture of~\cite{ArkaniHamed:2010gh} is that if all leading singularities
(corresponding to a series of cuts that completely localize a Feynman integral)
are rational numbers, the answer is a pure function. 

This conjecture was verified in a number of non-trivial examples, see
e.g.~\cite{Drummond:2010cz,Dixon:2011nj}.
Moreover, it turned out that choosing such integrals as a basis
rendered the physical answer much simpler already at the integrand level,
before carrying out the integrations.
(This is in part due to the close relationship between certain unitarity cuts
and infrared divergences, whose appearance is clearer in the new basis choice.)

The understanding of the relationship between Feynman integrals and uniform weight functions
was put on a firmer footing in~\cite{Henn:2013pwa}, by providing a way of proving the conjecture with the help of differential equations.
It is known that Feynman integrals viewed as functions of kinematical invariants satisfy a system
of first-order partial differential equations (see e.g.~\cite{Argeri:2007up,Henn:2014qga} for reviews).
Denoting the set of such integrals by $\boldsymbol{f}(x)$ we have in complete generality
\begin{equation}\label{noncanon}
d {\boldsymbol{f}}( {x} ) = d \tilde{A}({x},\epsilon) {\boldsymbol{f}}( {x} )\,,
\end{equation}
where $\tilde{A}({x},\epsilon)$ is a nontrivial matrix depending on the dimensional regularization 
parameter $\epsilon$ and the differential on both sides is taken with respect to $x$.
In \cite{Henn:2013pwa} it was suggested that by changing the basis 
${\boldsymbol{f}} \to T( {x},\epsilon) {\boldsymbol{f}}$ to uniform weight functions, the differential equation 
(\ref{noncanon}) should take a simple canonical form,
\begin{equation}\label{example_canonicalDE}
d {\boldsymbol{f}}( {x}) = \epsilon \, d\tilde{A}( {x}) {\boldsymbol{f}}( {x}) \,,
\end{equation}
with the matrix $\tilde{A}(x)$ being $\epsilon$ independent and given by a linear combination of logarithms with rational coefficients.
We can write a formal solution to (\ref{example_canonicalDE}) as a path-ordered exponential  
\begin{equation}
\boldsymbol{f}(x) = P \,\e^{\epsilon \int_{\mathcal C} d \, \tilde{A} } \boldsymbol{g}(\epsilon)\,,
\label{pathintegral}
\end{equation}
with some contour $\mathcal{C}$ connecting the base point at $x=1$ with the function argument,
and $\boldsymbol{g}(\epsilon) = \boldsymbol{f}(1,\epsilon)$ being the boundary values.

In fact, in the canonical form~(\ref{example_canonicalDE}), each term in the $\epsilon-$expansion of ${\boldsymbol{f}} = \sum \epsilon^k {\boldsymbol{f}}^{(k)}$ satisfies an equation of the form~(\ref{d_purefunction}), 
which allows one to prove that $ {\boldsymbol{f}}^{(k)}$ has uniform weight $k$.
\footnote{In principle, transcendental constants could enter through the boundary conditions. Experience shows that this does not happen if the basis is chosen according to the criteria explained below.}
Moreover, assigning weight $(-1)$ to $\epsilon$, each ${\boldsymbol{f}}$ has uniform weight, in the sense of the $\epsilon$ expansion.

How does one find an appropriate basis ${\boldsymbol{f}}$? Given the conjecture of \cite{ArkaniHamed:2010gh}, integrals having constant leading singularities in the sense of that paper are a natural choice.
The differential equations then allow one to prove the uniform weight properties of those functions.
We remark that generalized unitarity cuts are also very natural in the context of differential equations.
The reason is that the cut integrals satisfy the same differential equations as the original
integrals, but with different boundary conditions. The cuts allow one to focus on a subset
of integrals that share a common propagator structure. This can be used e.g. for making consistency  checks before the whole system of  differential equations is considered.

Another strategy put forward in  \cite{Henn:2013pwa} is to try to deduce the weight properties from an integral representation, e.g. in Feynman parametrization. This works particularly
well in cases with few propagators, and for Wilson line integrals. We will present various examples below.\footnote{A well-known case where the weight properties of the answer could be deduced from an integral representation is \cite{Kotikov:2004er}. Based on properties of the BFKL equation, the authors conjectured that the leading weight pieces (the ``most complicated part'') of twist-two anomalous dimensions in QCD and supersymmetric Yang--Mills theories should coincide.}

In subsections \ref{choice_mi} and~\ref{section:cuts} we will see various examples of analyzing weight properties of Feynman integrals before integration, either based on parametric representations, or based on generalized unitary cuts.
 
\subsection{Two-loop master integrals and differential equations}
\label{section:hqet2loopexample}

Let us present as an example the full set of differential equations (\ref{example_canonicalDE}) for the two-loop case.
The analysis at three loops will be almost identical, except that the system will be much larger.
Going through the simple two-loop example therefore allows us to be more explicit.

\begin{figure}[tbp]
\centering
\psfrag{V1}[cc][cc]{$$}
\psfrag{V2}[cc][cc]{$$}
\psfrag{k2}[cc][cc]{$a_7$}
\psfrag{k1}[cc][cc]{$a_6$}
\psfrag{k12}[cc][cc]{$a_3$}
\psfrag{a1}[cc][cc]{$a_1$}
\psfrag{a2}[cc][cc]{$a_2$}
\psfrag{a3}[cc][cc]{$a_3$}
\psfrag{a4}[cc][cc]{$a_4$}
\psfrag{a5}[cc][cc]{$a_5$}
\psfrag{a6}[cc][cc]{$a_6$}
\psfrag{a7}[cc][cc]{$a_7$}
\psfrag{a}[cc][cc]{(a)}
\psfrag{b}[cc][cc]{(b)}
\includegraphics[width=0.3 \textwidth]{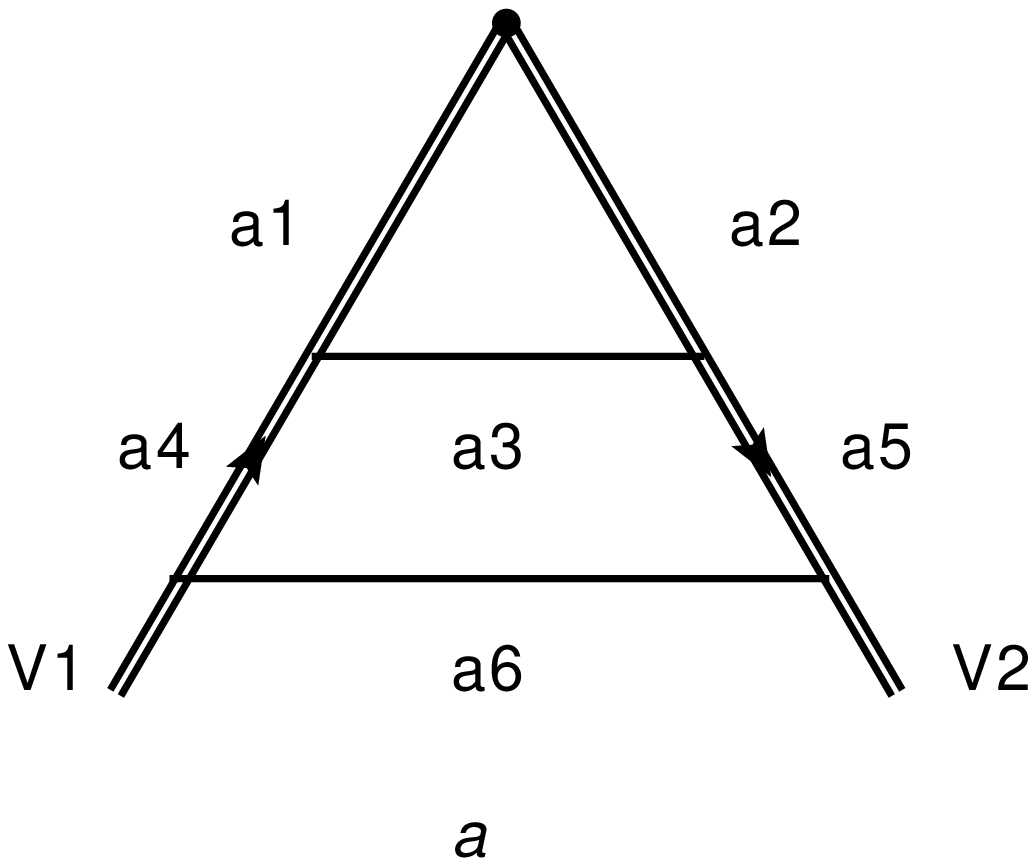}
\quad
\quad
\quad
\includegraphics[width=0.3 \textwidth]{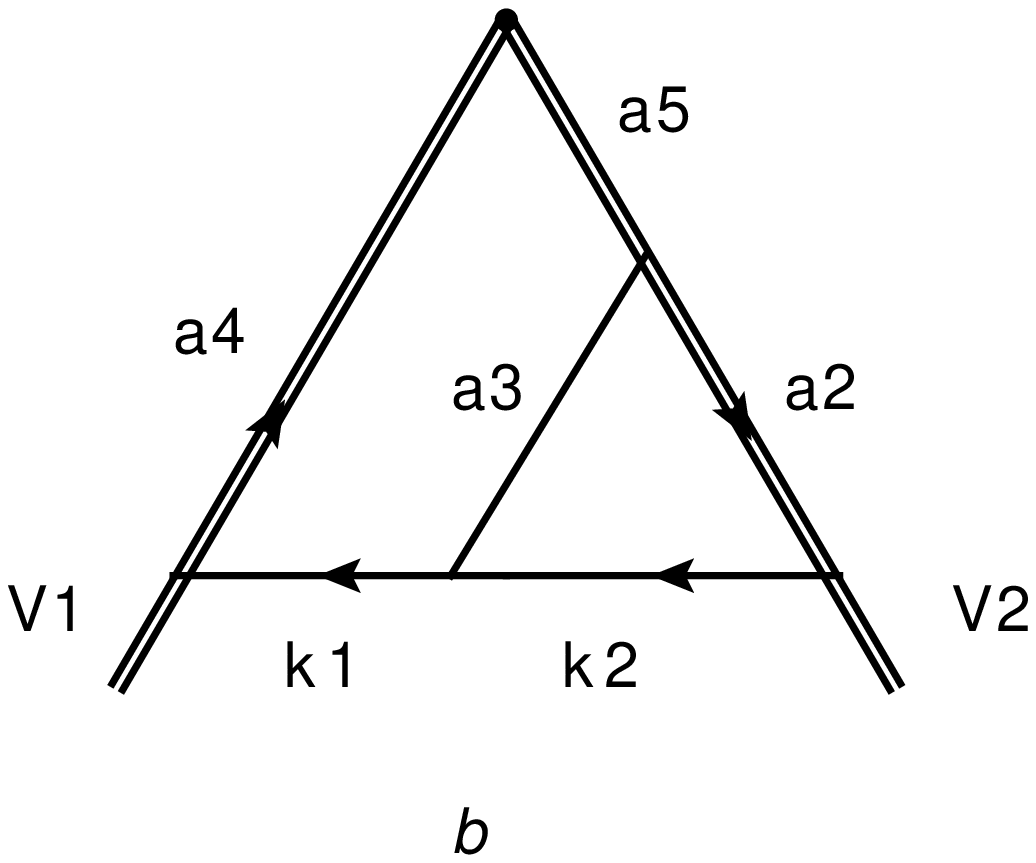}
\caption{Two-loop integral families. Double and solid lines with index $a_i$ stand for eikonal and scalar propagators, respectively, raised to power $a_i$.}
\label{fig:hqet2loopfamilies}
\end{figure}

In order to discuss all planar two-loop integrals
(recall that non-planar integrals can be obtained via eikonal identities),
we introduce the following notation 
\begin{equation}\label{G7}
G_{a_1 ,\ldots a_7} =
\e^{2 \epsilon \gamma_{\rm E} } \int \frac{d^{D}k_{1}  d^{D}k_{2}  }{(i \pi^{D/2})^2} \prod_{i=1}^{7} (Q_{i})^{-a_{i} } \,,
\end{equation}
where $a_1,\dots,a_7$ are arbitrary integer indices and
\begin{align}
& Q_1 = -2 k_2 \cdot v_1 +1 \,,\qquad Q_2 = -2 k_2 \cdot v_2 +1 \,,\qquad Q_3 = -(k_1-k_2)^2\,,
\nonumber\\
& Q_4 = -2 k_1 \cdot v_1 +1 \,,\qquad Q_5 = -2 k_1 \cdot v_2 +1 \,, \qquad Q_6 = -k_1^2 \,,\qquad Q_7 = -k_2^2 \,.
\end{align}
Examining (\ref{G7}) for various values of indices, we identify two families of integrals that match topology of planar 
Feynman diagrams contributing to the cusped Wilson loop at two loops. 
They are shown in figure~\ref{fig:hqet2loopfamilies}(a), (b) and are given by 
\begin{equation}\label{2mas}
 G_{a_1,a_2,a_3,a_4,a_5,a_6,0}\,,\qquad\qquad G_{0,a_2,a_3,a_4,a_5,a_6,a_7}\,,
\end{equation}
respectively.
All other planar two-loop integrals can be obtained by pinching lines (setting some of the $a_{i}$ to zero),
or adding numerators (setting some $a_{i}$ to negative values).

Integral reduction using integration-by-parts (IBP) identities~\cite{Chetyrkin:1981qh}
shows that there are nine master integrals that can come from the two families of integrals shown in (\ref{2mas}). Notice that
integral reduction programs automatically choose a particular integral basis $\boldsymbol{f}$
according to certain criteria.
Such a basis typically does not have the uniform weight properties discussed above,
and hence leads to a complicated form of differential equations (\ref{noncanon}).
In order to bring the differential equations to a simple canonical form (\ref{example_canonicalDE})
we make the following choice of master integrals,
{
\allowdisplaybreaks
\begin{align}\label{f1}
f_1 ={}& \epsilon^4 \chi^2 \, G_{1,1,1,1,1,1,0} \,, \qquad
\raisebox{-10mm}{
\psfrag{V1}[cc][cc]{$$}
\psfrag{V2}[cc][cc]{$$}
\psfrag{k2}[cc][cc]{$$}
\psfrag{k1}[cc][cc]{$$}
\psfrag{k12}[cc][cc]{$$}
\psfrag{a1}[cc][cc]{$$}
\psfrag{a2}[cc][cc]{$$}
\psfrag{a3}[cc][cc]{$$}
\psfrag{a4}[cc][cc]{$$}
\psfrag{a5}[cc][cc]{$$}
\psfrag{a6}[cc][cc]{$$}
\psfrag{a7}[cc][cc]{$$}
\psfrag{x1}[cc][cc]{$$}
\psfrag{x2}[cc][cc]{$$}
\psfrag{x3}[cc][cc]{$$}
\psfrag{x4}[cc][cc]{$$}
\includegraphics[width=0.2 \textwidth]{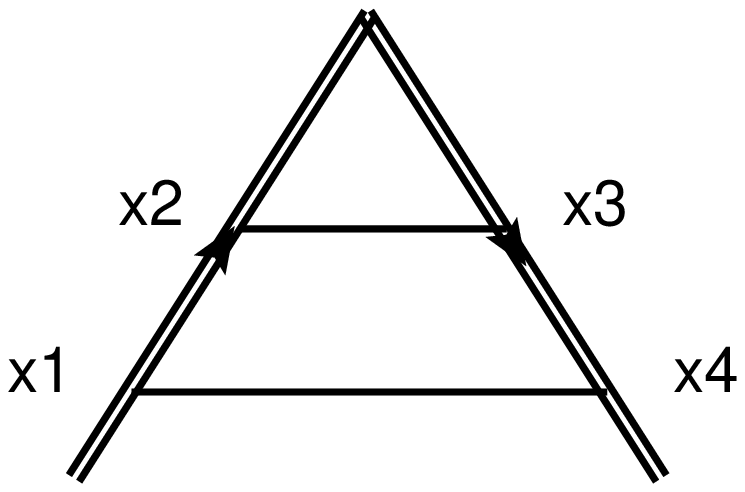}}
\\ \label{f2}
f_2 ={}& \epsilon^3 \chi  \,G_{0,2,1,1,1,1,0} \,,
\qquad
\raisebox{-10mm}{
\includegraphics[width=0.2 \textwidth]{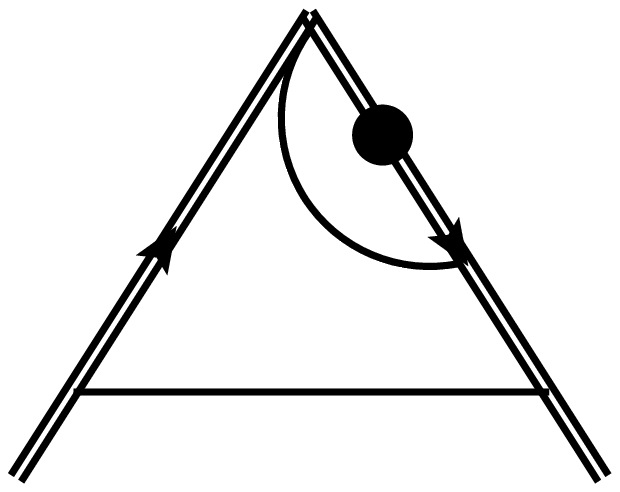}}
\\ \label{f3}
f_3 ={}& \epsilon^3 \chi \, G_{1,1,2,0,0,1,0} \,,
\qquad
\raisebox{-10mm}{
\includegraphics[width=0.2 \textwidth]{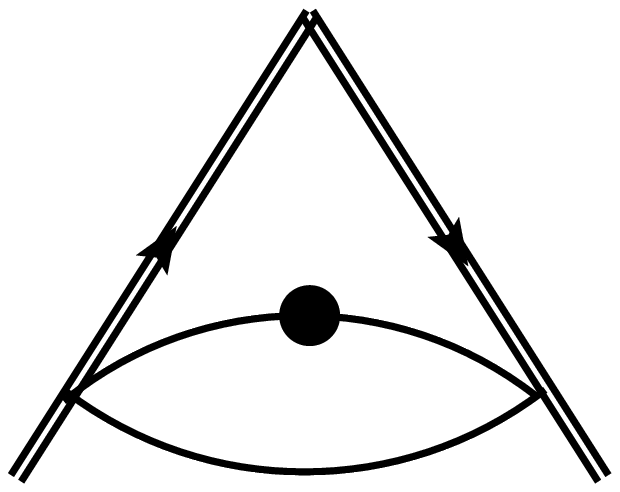}}
\\ \label{f4}
f_4 ={}& \epsilon^2 \, G_{0,1,2,0,0,2,0} \,,
\qquad
\raisebox{-10mm}{
\includegraphics[width=0.2 \textwidth]{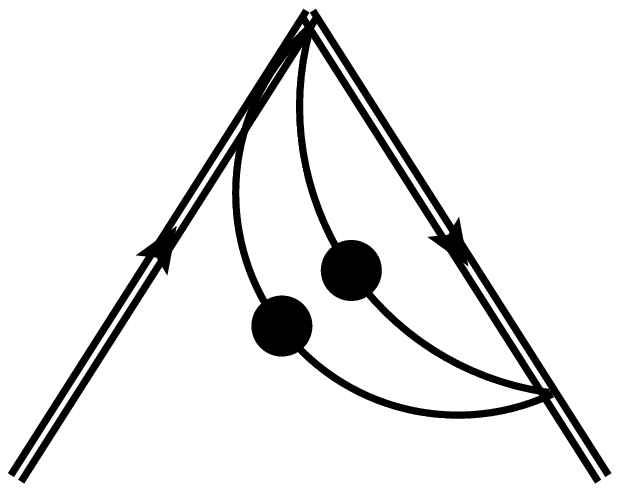}}
\\ \label{f5}
f_5 ={}& \epsilon^2 \,G_{0,2,1,1,0,2,0} \,,
\qquad
\raisebox{-10mm}{
 \includegraphics[width=0.2 \textwidth]{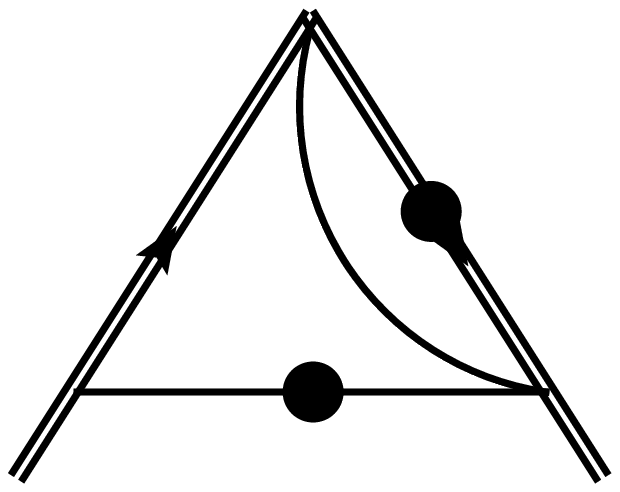}}
\\ \label{f6}
f_6 ={}& \epsilon^3 \chi \,G_{1,1,1,0,1,2,0} \,,
\qquad
\raisebox{-10mm}{
\psfrag{x1}[cc][cc]{$$}
\psfrag{x2}[cc][cc]{$$}
\psfrag{x3}[cc][cc]{$$}
\psfrag{x4}[cc][cc]{$$}
\includegraphics[width=0.2 \textwidth]{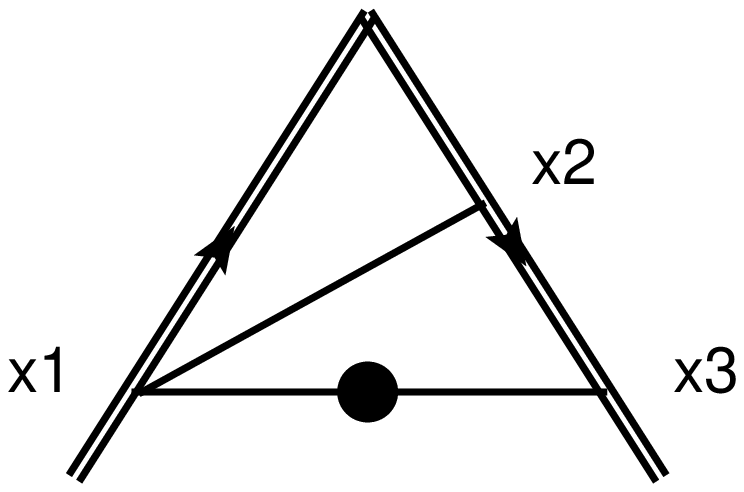}}
\\ \label{f7}
f_7 ={}& \epsilon^4 \chi \, G_{0,1,1,1,0,1,1}\,,
\qquad
\raisebox{-10mm}{
\psfrag{x1}[cc][cc]{$$}
\psfrag{x2}[cc][cc]{$$}
\psfrag{x3}[cc][cc]{$$}
\psfrag{x4}[cc][cc]{$$}
 \includegraphics[width=0.2 \textwidth]{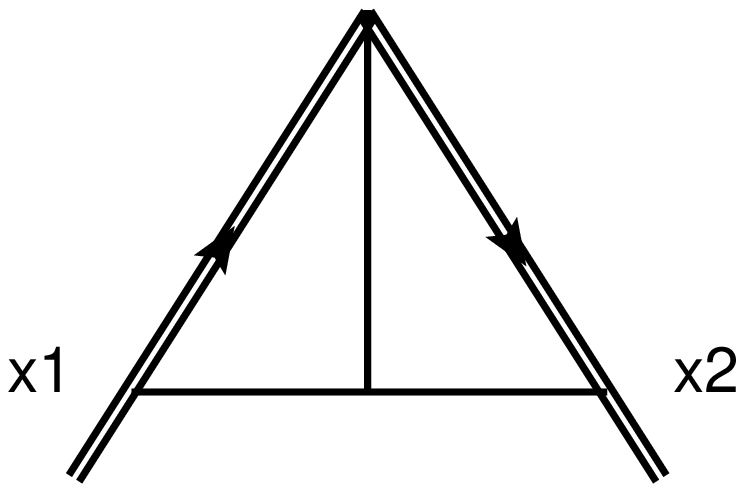}}
\\ \label{f8}
f_8 ={}& \epsilon^2 \, G_{0,1,0,1,0,2,2} \,,
\qquad
\raisebox{-10mm}{
\psfrag{x1}[cc][cc]{$$}
\psfrag{x2}[cc][cc]{$$}
\psfrag{x3}[cc][cc]{$$}
\psfrag{x4}[cc][cc]{$$}
\includegraphics[width=0.2 \textwidth]{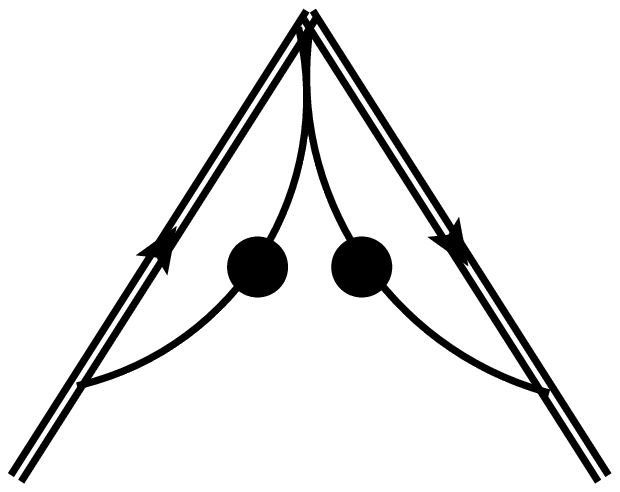}}
\\ \label{f9}
f_9 ={}&\epsilon^3 \chi\, G_{0,1,0,1,1,1,2}
\qquad
\raisebox{-10mm}{
  \includegraphics[width=0.2 \textwidth]{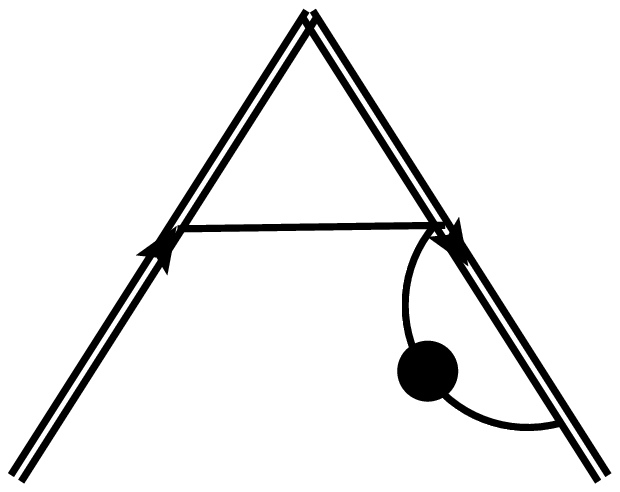}}
\,,
\end{align}
}%
where $\chi = (1-x^2)/x$. Here   dot denotes 
a  propagator squared in momentum space. 
A distinguished feature of this basis is that all functions $f_1,\dots,f_9$ have a uniform weight. This property is by no means obvious and
can be established using the methods discussed in subsections~\ref{choice_mi} and~\ref{section:cuts}.

All integrals depend on dimensionless kinematical variable $x$ 
\begin{equation}\label{defkin}
2 v_1 \cdot v_2 = x+1/x \,,
\end{equation}
with $v_1^2 = v_2^2 = 1$,
and are normalized in such a way that $f_1,\dots,f_9$ are expected to be pure functions of weight zero.
This can be verified by computing their differential with respect to the kinematic variable $x$. Using the definition (\ref{defkin}), we can 
implement this by differentiating the defining Feynman integrals
with respect to $v_1$,
\begin{equation}
\frac{\partial}{\partial x}
= \left[ (v_1 \cdot v_2) v_1^\mu - v_2^\mu \right] \frac{\partial}{\partial v_1^\mu} \,.
\end{equation}
In this way, we find that the set of nine basis integrals ${\boldsymbol{f}}=(f_1,\dots,f_9)$ satisfies the differential equation
(\ref{example_canonicalDE})  
\begin{align}\label{de2loop}
\partial_x {\boldsymbol{f}}( {x}) = \epsilon \, \left({a_2\over x}+{b_2\over x+1}+{c_2\over x-1} \right) {\boldsymbol{f}}( {x})\,,
\end{align}
with $b_2=\text{diag}(4,2,4,0,0,2,2,0,2)$ and 
\begin{align} 
a_2{}&=\left(  
\begin{array}{rrrrrrrrr}
 -2 & -4 & -2 & \ \ 0 & 0 & 1 & 0 & \ \ 0 & 0 \\
 0 & 0 & 0 & 0 & \frac{1}{2} & 0 & 0 & 0 & 0 \\
 0 & 0 & -2 & \frac{1}{2} & 0 & 0 & 0 & 0 & 0 \\
 0 & 0 & 0 & 0 & 0 & 0 & 0 & 0 & 0 \\
 0 & 4 & 0 & \frac{1}{2} & 1 & 0 & 0 & 0 & 0 \\
 2 & 0 & 0 & 0 & -2 & -1 & 0 & 0 & 0 \\
 0 & 0 & 0 & \frac{1}{2} & 2 & 0 & -2 & \frac{1}{2} & 0 \\
 0 & 0 & 0 & 0 & 0 & 0 & 0 & 0 & 0 \\
 0 & 0 & 0 & 0 & 0 & 0 & 0 & 1 & -1 \\
\end{array} 
\right),  \quad 
c_2=\left(
\begin{array}{rrrrrrrrr} 
 0 & 0 & \ 0 & 0 & 0 & 0 & 0 & 0 & 0 \\
 0 & -2 & 0 & 0 & 0 & 0 & 0 & 0 & 0 \\
 0 & 0 & 0 & 0 & 0 & 0 & 0 & 0 & 0 \\
 0 & 0 & 0 & 0 & 0 & 0 & 0 & 0 & 0 \\
 0 & 0 & 0 & -1 & -2 & 0 & 0 & 0 & 0 \\
 0 & 0 & 0 & 0 & 0 & 0 & 0 & 0 & 0 \\
 0 & 0 & 0 & 0 & 0 & 0 & 2 & 0 & 0 \\
 0 & 0 & 0 & 0 & 0 & 0 & 0 & 0 & 0 \\
 0 & 0 & 0 & 0 & 0 & 0 & 0 & 0 & 0 \\
\end{array}
\right).
\end{align}
In order to solve the differential equation (\ref{de2loop}) we also need boundary conditions. The latter can be easily fixed for $x=1$, or equivalently
$v_1^\mu=v_2^\mu$, where no singularities are expected from the Feynman integrals. Since $\chi=0$ in this limit,
the only non-vanishing integrals  in (\ref{f1}) --  (\ref{f9}) are $f_4$, $f_{5}$, and $f_{8}$. For $x=1$ they are reduced to integrals with bubble insertions
and can be easily evaluated (see relations  (\ref{bubbleB2}) below).
In this way, we find
\begin{align}\notag
f_{4}(x=1) ={}& \e^{2 \epsilon \gamma_{\rm E}} \Gamma^2(1-\epsilon) \Gamma(1+4 \epsilon)\,,
\\[1.5mm]\notag
f_{5}(x=1) ={}& -\frac12  f_{4}(x=1)\,,
\\[1.5mm]
f_{8}(x=1) ={}& \e^{2 \epsilon \gamma_{\rm E}} \Gamma^2(1-\epsilon) \Gamma^2(1+2 \epsilon)\,,
\end{align}
with all other integrals vanishing at $x=1$. Making use of  
\begin{equation}
\log  \Gamma(1+ \epsilon)  = - \epsilon \gamma_{\rm E} + \sum_{k\ge 2} \zeta_k \frac{(-\epsilon)^k}{k}\,,
\end{equation}
it is easy to see that the above expressions give rise to uniform weight $\epsilon$ expansions.%
\footnote{This formula also explains why we have chosen the particular normalization factor
$\e^{L \epsilon \gamma_{\rm E}}$ for $L$-loop integrals, namely to avoid the appearance of $\gamma_{\rm E}$ in our results.}

Returning to the differential equation~(\ref{de2loop}), we can write its solution as
\begin{equation}\label{f-exp}
{\boldsymbol{f}}(x) =  \sum_{k \ge 0} \epsilon^k {\boldsymbol{f}}^{(k)}(x) \,.
\end{equation}
Matching the coefficients in front of powers of $\epsilon$ on the both sides of (\ref{de2loop}), we find
that  ${\boldsymbol{f}}^{(k)}(x)$ are given by a $\mathbb{Q}$-linear combination
of harmonic polylogarithms of weight $k$ defined in section~\ref{section_iterated_integrals}.
It is then straightforward to expand ${\boldsymbol{f}}(x)$ to any desired order in $\epsilon$.
For instance, we have
\begin{equation}
f_1(x) = \epsilon^2 H_{0,0}(x)
+ \epsilon^3 \left[ \frac{\pi^2}{3} H_{0}(x) +4 H_{-1,0,0}(x) - H_{0,0,0}(x)  + 2 H_{0,1,0}(x)  + \zeta_3 \right]
+ \mathcal{O}(\epsilon^4)\,.
\end{equation}
In agreement with our expectations, the coefficients in front of powers of $\epsilon$ are pure functions.

We should mention that the basis choice of ${\boldsymbol{f}}(x)$ is not unique.  As we show in the next subsection, we can introduce two other integrals
\begin{align} \label{g1}
g_1(x) ={}& \epsilon^3 \, \chi^2 \, G_{2,1,1,0,1,1,0}
\qquad\raisebox{-10mm}{
\psfrag{x1}[cc][cc]{$$}
\psfrag{x2}[cc][cc]{$$}
\psfrag{x3}[cc][cc]{$$}
\psfrag{x4}[cc][cc]{$$}
\includegraphics[width=0.2 \textwidth]{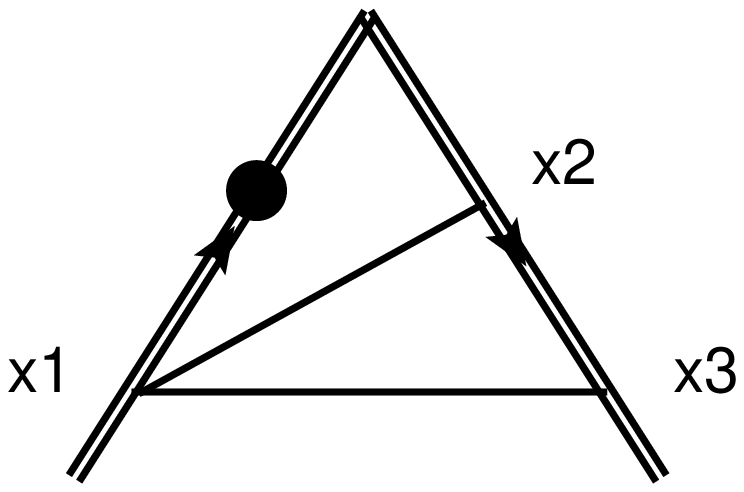}}
\\ \label{g2}
g_2(x)={}& \epsilon^4\, \chi\, G_{0,1,1,1,1,1,1}
\qquad\raisebox{-10mm}{
\psfrag{x1}[cc][cc]{$$}
\psfrag{x2}[cc][cc]{$$}
\psfrag{x3}[cc][cc]{$$}
\psfrag{x4}[cc][cc]{$$}
\includegraphics[width=0.2 \textwidth]{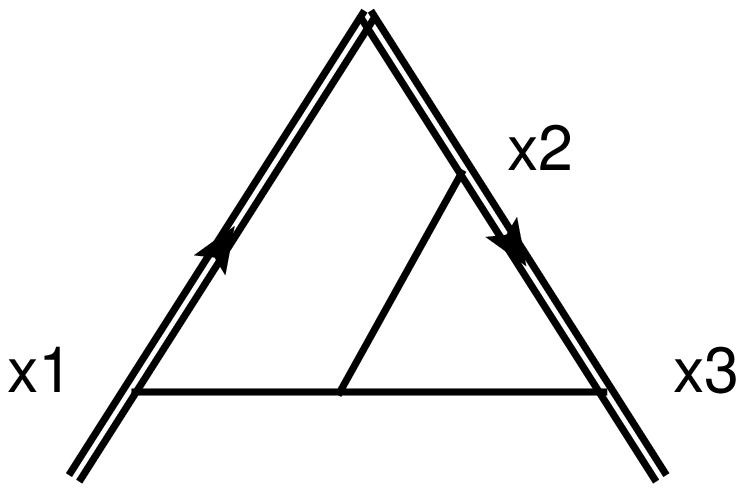}}
\end{align}
that are also pure functions of weight zero. They are related however, via IBP identities,
to the nine basis functions ${\boldsymbol{f}}(x)$. In order for $g_1$ and $g_2$ to be pure functions,
they should be given by a $\mathbb{Q}$-linear combination (independent of $x$ and $\epsilon$)
of the basis integrals.
Indeed, we find that
\begin{equation}
g_1 = f_1\,,\qquad
g_2 = \frac{1}{2} f_6 + f_7 - \frac{1}{2} f_9\,.
\end{equation}
This also means that, replacing e.g. $f_7$ by $g_2$
would have lead to an equally nice set of differential equations.

Of course, not all integrals have such nice properties. As an example, consider the following 
integral that can appear in the Feynman diagram calculation
\begin{align}
G_{-1,1,1,1,1,1,1} =  \frac{1}{2 \epsilon^3 (1-2 \epsilon)} \Big[
f_1 + \frac{1+x^2}{1-x^2} f_3 + \frac{1-\epsilon}{1-2 \epsilon} f_4 + 2 f_5 + \frac{1-x}{1+x} f_6
\nonumber\\
  - 4 \frac{x}{1-x^2} f_7 - \frac{\epsilon}{1-2 \epsilon} f_8 - \frac{1-x}{1+x} f_9  \Big]\,.
\end{align}
This integral is obviously not a pure function. Choosing it as a basis integral would lead to an unnecessarily complicated
dependence of the differential equations on $\epsilon$ and $x$ .

\subsection{Wilson line diagrams in position space and uniform weight integrals}
\label{choice_mi}

In this and in the following subsection we explain the method that we use to identify integrals that
can be evaluated in terms of pure functions. 

\begin{figure}[tbp]
\centering
\psfrag{V1}[cc][cc]{$-s\,v_1^\mu$}
\psfrag{V2}[cc][cc]{$t\,v_2^\mu$}
\psfrag{x1}[cc][cc]{$-s_1 \,v_1^\mu$}
\psfrag{x2}[cc][cc]{$t_1 \,v_2^\mu$}
\psfrag{x3}[cc][cc]{$t_2 \,v_2^\mu$}
\psfrag{a}[cc][cc]{(a)}
\psfrag{b}[cc][cc]{(b)}
\psfrag{c}[cc][cc]{(c)}
\includegraphics[width=0.25 \textwidth]{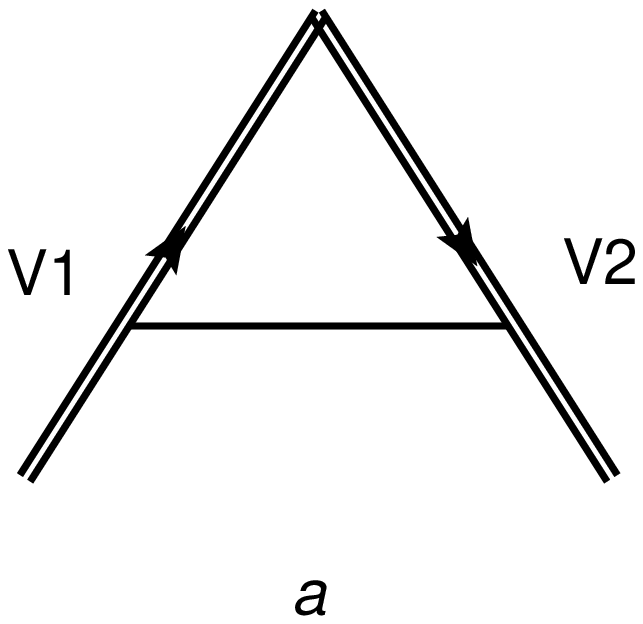}
\quad\quad
\includegraphics[width=0.3 \textwidth]{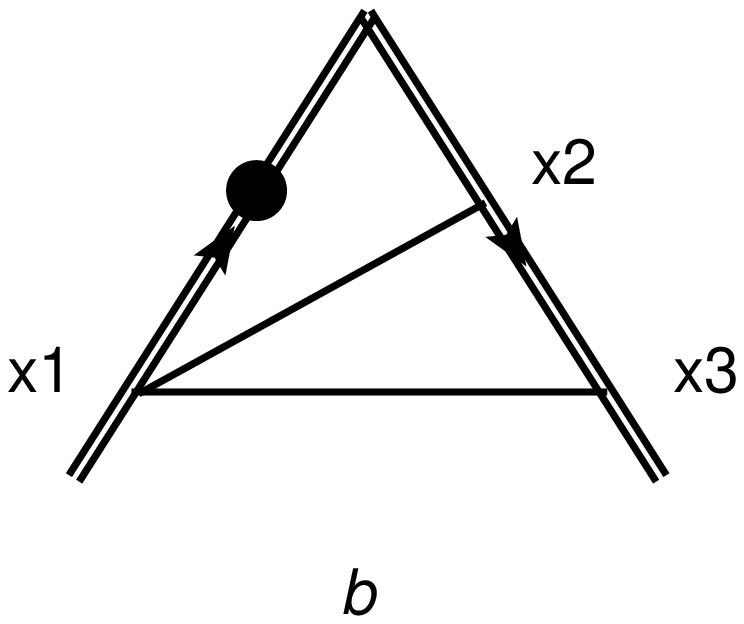}
\quad\quad
\includegraphics[width=0.3 \textwidth]{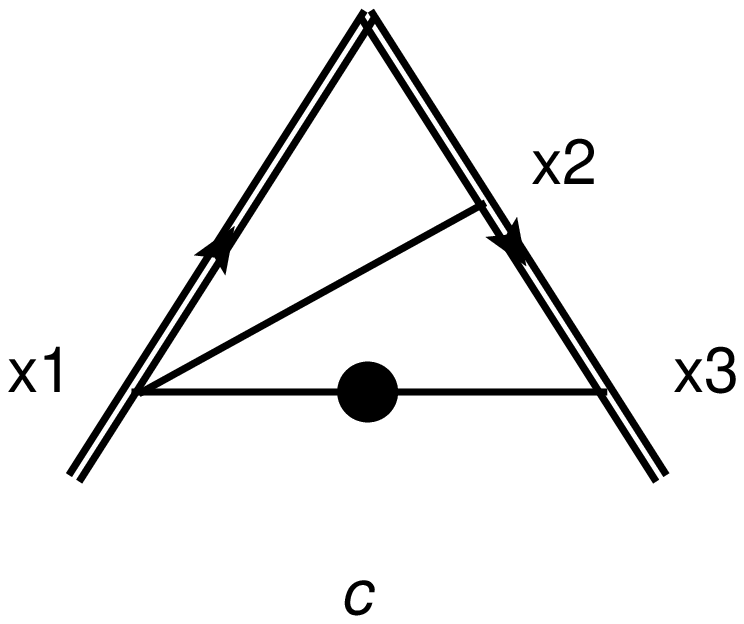}
\caption{Wilson line integrals in position space}
\label{fig:hqetpositionspace1}
\end{figure}

As a warm up example we revisit the calculation of Wilson line integral (see figure~\ref{fig:hqetpositionspace1}(a)) that contributes to the one-loop 
cusp anomalous dimension. It is given in position space by an integral of a scalar propagator connecting two points $-s v_1^\mu$ and $t v_2^\mu$, 
with $s$ and $t$ being the line integration parameters,
\begin{equation}
\frac{1}{(v_1 s + v_2 t)^2} 
= \frac{x}{(s x+t)(s+x)} \,.
\label{identity2d1}
\end{equation}
Using this identity, the integrand can be written in the so-called ``d-log''' form~\cite{Henn:2013wfa}
\footnote{For the time being, we perform the analysis in $D=4$ dimensions,
and do not yet specify the range of integration for $s$ and $t$.}
\begin{equation}\label{dlog}
\int \frac{d s  \wedge d t  }{(v_1 s  + v_2 t )^2} = \frac{x}{1-x^2} \int d\log(s x +t ) \wedge d\log(t x +s )\,.
\end{equation}
In this form, it is manifest that the integral, multiplied by $(1-x^2)/x$,
has a differential of the form (\ref{d_purefunction}), with $n=2$,
and is hence a pure function of weight two.
 Likewise, Wilson line integrals with more propagators stretched  between two (or more) semi-infinite rays are seen to be pure functions of higher weight.
(An algorithm for computing all such contributions was given in ref. \cite{Henn:2013wfa}.)
 
The above analysis is rather formal since the Wilson integral is divergent and requires regularization both in UV and IR. 
As we will see in a moment, regularization does not affect the uniform weight properties of the integral. 
For example, at one loop the regularized Wilson line integral (\ref{zero1}) 
is given, up to overall factor 
\begin{equation}
\int_0^\infty ds dt \,\e^{-i(s+t)/2} \left[ \frac{x}{(s x+t )(s +t  x)}\right]^{1-\epsilon}
= \frac{x}{1-x^2} \int_0^\infty \frac{d\rho}{\rho^{1-\epsilon}} \e^{-i \rho/2}  I_1(x,\epsilon)\,.
\label{oneloopparam1}
\end{equation}
Here we changed variables according to
$s = \rho z$, $t = \rho \bar{z}$ and introduced notation for
\begin{equation}\label{oneloopparam2}
I_1(x,\epsilon) = \int_0^1 d\log \left( \frac{z x + \bar{z}}{z+ \bar{z} x} \right)\;
(z x + \bar{z})^{-\epsilon} (z+ \bar{z}/x)^{-\epsilon}\,,
\end{equation}
with $\bar{z}=1-z$.
The $\rho$ integral in (\ref{oneloopparam1}) gives UV divergence $1/\epsilon$.
More generally, introducing $\rho$ as an overall scale in a given Wilson line integral,
we can always separate the $\rho$ integration from the rest of the calculation.
Moreover, the $\rho$ integral can always be evaluated in terms of gamma function,
typically $\Gamma(L\epsilon)$  at $L$ loops, which does not change the weight properties of the answer, except for an overall offset.
Therefore, in the examples below, we will not discuss further the $\rho$ integration.

Let us examine the integral  $I_1(x,\epsilon)$.
The integrand in (\ref{oneloopparam2}) can be Taylor expanded in $\epsilon$.
At $\epsilon=0$, the integral  $I_1(x,0)$ is obviously evaluates to a logarithm, i.e.\ a pure weight-one function.
It is easy to see that expanding (\ref{oneloopparam2}) at higher orders in $\epsilon$
will increase the weight of the resulting function accordingly.
With the convention that $\epsilon$ has weight $(-1)$, we can therefore see
that $I_{1}(x,\epsilon)$ has uniform weight one.
We could proceed along the lines of section~\ref{section:purefunctions}
(see also refs.~\cite{CaronHuot:2011kk,Henn:2013wfa}) and evaluate the integral,
at a given order in $\epsilon$.
Instead, in this paper, we evaluate all such integrals using differential equations,
with $\epsilon$ as a parameter.
 
We can show in a similar manner that the integrals with $L$ propagators attached to two (or more) semi-infinite rays are expressed in
terms of pure functions of weight $2L$. Indeed, 
we can apply the identity 
(\ref{dlog}) to each propagator to deduce that the integral is given by $(1/\chi)^L$ (with $\chi=(1-x^2)/x$) times a pure function of 
weight $2L$. As in the one-loop case, introducing regularization does not affect this result.~\footnote{There is a small subtlety that the double ladder diagram has a subdivergence,
so that strictly speaking we are not allowed to Taylor expand under the integral sign.
However, we can avoid this problem by performing the same analysis for the crossed ladder diagram,
which is equivalent to the double ladder, up to the one-loop ladder integral squared.}
For example, at two loops, the ladder integral that enters into the definition (\ref{f1}) of the basis function $f_1$,   is given by 
the product of $\chi^{-2}$ and a pure function of weight $4$. Multiplying the ladder integral by $\epsilon^4\chi^2$ we therefore
obtain a pure function of weight $0$. This explains the origin of the normalization factors in the definition of $f_1$.

The above analysis can be generalized to integrals where propagators are raised to some power.
For example, consider the integral of figure~\ref{fig:hqetpositionspace1}(b) where a dot denotes 
a  propagator squared (in momentum space). 
Parametrizing the end-point of propagators according to $s_1 = \rho z$, $t_1 = \rho \bar{z} y$ and $t_2 = \rho \bar{z}$
(with $\bar z=1-z$ and similar for $\bar y$), this leads to (up to an inessential  overall factor and terms suppressed by powers of $\epsilon$), 
\begin{equation}
{1\over \epsilon}\int_0^1 dy\wedge dz z (1-z) P( z, \bar{z} y )P( z, \bar{z} )
={1\over \epsilon\chi^{2}} \int_0^1  d\log\left[ \frac{x y \bar{z} + z }{y \bar{z} + x z }\right] \wedge  d\log \left[ \frac{ x \bar{z} + z }{\bar{z} + x z }\right]\,,
\end{equation}
where we denoted $P(s,t) = [s^2+t^2+s t (x+1/x)]^{-1}$. Here the UV pole $1/\epsilon$ comes from $\rho-$integration and
the additional factor of $z(1-z)$ on the left-hand side comes from the Jacobian of the change of variables and  the doubled eikonal propagator.
This shows that the integral is given by a function of weight three. We can convert it into a pure function of weight zero by multiplying
the integral by the normalization factor $\epsilon^3 \chi^2$. The resulting function coincides with $g_1(x)$ defined in (\ref{g1}).

Another example is the integral shown in figure~\ref{fig:hqetpositionspace1}(c).
The Fourier transform of the doubled propagator gives $\sim  (-x^2)^{-\epsilon}/\epsilon$,
so that this factor is irrelevant at the level of the integrand and can be replaced with $1/\epsilon$.
Parametrizing the line integrals as in the previous example, we obtain
\begin{equation}
{1\over \epsilon}\int_0^1 dy\wedge dz \, \bar{z} P(z,y \bar{z})  = {1\over\epsilon\chi} \int_0^1  d\log\left[ \frac{ y \bar{z} + x z}{x y \bar{z} + z }\right] \wedge  d\log z\,.
\end{equation}
We conclude that this integral multiplied by $\epsilon^3  \chi$ yields a pure function of weight zero. It coincides with the basis function
$f_6$ defined in (\ref{f6}).

These examples might mislead the reader in thinking
that the uniform weight property is rather trivial.
However, this is not the case. For instance, just moving  the dot in the above examples to another propagator destroys this property.
In our analysis, it would result in the impossibility of rewriting the integrand
in a  ``d-log'' form.

For integrals with fewer propagators, bubble subintegrals can appear.
Whenever this happens, the latter can be integrated out,
leaving one with an integral that effectively has one loop less,
up to some gamma functions coming from the integration.
This means that many integrals can be chosen based on the knowledge
of pure functions at the lower loop order.
The relevant formulas are obtained by elementary integrations in Feynman parameter space.
For a bubble on an eikonal line (see (\ref{f2})) and  for a scalar bubble (see (\ref{f3})) we have, respectively,
\begin{align}\notag
{}& \int \frac{d^{D}k_1}{i \pi^{D/2}} \frac{1}{ (-k_1^2)^{a_{1}} [ -2 (k_1+k_2)\cdot v_1 +1]^{a_{2}}}
= (-2 k_2 \cdot v_1 +1)^{D-2 a_1-a_2}  I(a_1 , a_2)\,,
\\
{}& \int \frac{d^{D}k_1}{i \pi^{D/2}} \frac{1}{ (- (k_1+k_2)^2 )^{a_1 } (-k_1^2)^{a_2}}
= (-k_2^2)^{D/2-a} G(a_1 ,a_2 )\,,
\label{bubbleB2}
\end{align}
where $a=a_1+a_2$ and
\begin{align}\notag
I(a_1 ,a_2) ={}& \frac{\Gamma(2 a_1 + a_2 -D) \Gamma(D/2-a_1)}{\Gamma(a_1)\Gamma(a_2)}\,,\\
G(a_1 ,a_2) ={}& \frac{\Gamma(a-D/2) \Gamma(D/2-a_1 )\Gamma(D/2 - a_2)}{\Gamma(a_1) \Gamma(a_2 ) \Gamma(D-a)}\,.
\end{align}
The momentum dependence of these integrals that is important for the present analysis can be simply obtained by power counting.

We can use the relations (\ref{bubbleB2}) to express the two-loop integrals entering the definition of basis functions (\ref{f2}), (\ref{f3}), (\ref{f5}) and  (\ref{f9}) in terms of one-loop integrals. Moreover, the bubble-type integrals entering (\ref{f4}) and (\ref{f8})
can be entirely evaluated in terms of $\Gamma-$functions.
In this way, we verify that $f_2$, $f_3$, $f_4$, $f_5$, $f_8$ and $f_9$ are indeed pure functions
of weight zero.~\footnote{
We remark that, in general, whenever bubble integrals are present,
one may choose further integrals thanks to possibility of adding a numerator,
so that the lower-loop integral has propagators raised to power $\mathcal{O}(\epsilon)$.
Examples of this can be found in~\cite{Henn:2013tua,Henn:2014qga}.
In a certain sense, this phenomenon appears in $f_5$,
since integrating out the sub-integral gives a triangle with one eikonal propagator
raised to power $\mathcal{O}(\epsilon)$.}
 
Let us now discuss the two-loop master integrals with an internal interaction vertex,
cf.\ (\ref{f7}) and (\ref{g2}). It is convenient to analyze them in position space as well.
For simplicity, we will carry out the analysis in four dimensions.
Let us begin with the integral in (\ref{g2})
and denote by $x_1$, $x_2$, $x_3$ the points the three-point vertex is attached to,
with $x_2$, $x_3$ lying on the same Wilson line segment.
(For the integral in (\ref{f7}) we can set $x_2=0$.)
These points can be parametrized by
\begin{equation}
x_1^\mu = - s_1 v_1^\mu\,,\qquad x_2^\mu = t_1 v_2^\mu \,,\qquad x_3^\mu = t_2  v_2^\mu\,,
\end{equation}
with $s_1>0$ and $t_2 > t_1>0$.
Consider carrying out the integration over the internal vertex. The integral involves three scalar 
propagators attached to this vertex and it gives rise in four dimensions to~\cite{Usyukina:1993ch}
\begin{equation}\label{Phi1-def}
{1\over i\pi^2}\int {d^4 x_0\over x_{10}^2 x_{20}^2 x_{30}^2}=\frac{1}{x_{23}^2 \sqrt{\Delta}} \tilde{\Phi}^{(1)}(u,v) \,,
\end{equation}
where $x_{ij}^2=(x_i-x_j)^2$, $u =x_{12}^2/x_{23}^2$, $v=x_{13}^2/ x_{23}^2$, $\Delta= (1-u-v)^2-4 u v$, and
$\tilde{\Phi}^{(1)}$ is a  known pure function of weight two.
Its explicit expression is not relevant for the present analysis.
The latter focuses on the question whether the integrand
can be put in ``d-log'' form.

Just as in the case of propagator exchanges,
there are simplifications due to the fact that the Wilson lines lie in a plane, which leads to simplifications.
After some algebra, we find for the integrand (for $x<1$)
\begin{align}\notag\label{Phi1}
\e^{-i(t_2 +s_1)/2} \frac{ds_1 dt_1 dt_2}{x_{23}^2 \sqrt{\Delta}}
{}& = \frac{x}{1-x^2} \e^{-i(t_2 +s_1)/2} \frac{ds_1 dt_1 dt_2 }{( t_2 -t_1) s_1 } 
\\
{}& = \frac{x}{1-x^2}\, {d\rho}\, \e^{-i\rho/2} \, \frac{dy}{1-y}\frac{dz}{z}
\,,
\end{align}
where in the last relation we changed variables as $t_1 = \rho (1-z) y$, $t_2 = \rho (1-z)$, $s_1= \rho z$.
The $\rho$ integration just gives an overall normalization,
while the remaining integrand can be put into a ``d-log'' form.
Remembering the weight-two function $\tilde{\Phi}^{(1)}$,
we expect that the integral, normalized by $(1-x^2)/x$, gives a pure weight four function.
Then, we multiply it by $\epsilon^4$ to obtain a pure function (\ref{g2}) of weight zero.
 
We can use the calculation above in order to also analyze the integral (\ref{f7}) where $x_{2}=0$.
This is simply achieved by setting $t_1=0$ and dropping the $t_1$ integration in (\ref{Phi1}).
In this case, after changing variables according to $t_2 = \rho (1-z)$, $s_1 = \rho z$ we obtain
\begin{equation}\label{div}
\frac{x}{1-x^2} \frac{d\rho}{\rho}\, \e^{-i\rho/2} \, \frac{dz}{z}\,.
\end{equation}
Notice that the $\rho-$integral is divergent  at $\rho=0$. If we introduced the dimensional regularization
from the beginning, the integrand (\ref{div}) would be modified by the factor $\rho^{2\epsilon}$ leading
to a $1/\epsilon$ pole upon integration over $\rho$. Therefore, as in the previous case, we expect the
integral in (\ref{f7}) to be a uniform function of weight four and, as a consequence, the basis function
$f_7$ to be a pure function of weight zero.

It is clear that the method discussed in this subsection does not rely on a particular loop order
and it proves to be very useful in selecting uniform weight integrals at the three-loop order.

The attentive reader may have noticed that the above analysis relied mainly on the properties of the denominator factors
and not those of the function $\tilde{\Phi}^{(1)}$ defined in (\ref{Phi1-def}).
In fact, ignoring this function corresponds to taking a generalized cut,
making contact with the conjecture of~\cite{ArkaniHamed:2010gh}.
Similarly, and perhaps more easily,
we could have taken the maximal cut of this integral in momentum space,
with the conclusion that it has a unique normalization factor ${x}/({1-x^2})$.
As we will demonstrate in the next subsection, the approach based on generalized cuts
is especially useful for more complicated integrals with many propagator factors that can be cut.
 
\subsection{HQET integrals in momentum space and maximal cuts}
\label{section:cuts}

In this subsection, we perform an analysis of maximal cuts of HQET integrals in momentum space.
The objective is to determine whether a given integral has a unique overall normalization factor,
consistent with being a pure function.
We will start by reviewing some of the integrals of the previous subsection,
and then turn to an example occurring in three-loop computation.

Let us start by verifying the normalization factor of the one- and two-loop ladder integrals.
We work in four dimensions but keep IR regularization with $\delta=1/2$.
The maximal cut of the one-loop integral (\ref{zero1}) is given by
(here and in the remainder of this subsection we will neglect inessential $x-$independent normalization factors)
\begin{equation}
I_{\rm cut} = \int {d^{4}{k}} \, \delta(k^2) \delta(2 k\cdot v_1 -1) \delta(2 k \cdot v_2 - 1)\,.
\end{equation}
There are various ways of evaluating this integral.
To solve the massless on-shell condition for the loop momentum $k^2=0$ we make use of spinor variables (see, e.g., \cite{Dixon:1996wi})
\footnote{Another way could be to use Sudakov decomposition $k^\mu = \alpha v_1^\mu + \beta v_2^\mu +  k_{\bot}^{\mu}$ and carry out integration over $\alpha$, $\beta$ and $k_\bot$.}
\begin{equation}
k_{\alpha\dot\alpha} = \sigma^\mu_{\alpha\dot\alpha} k_\mu
 = \rho\,  \lambda_\alpha   \bar{\lambda}_{\dot\alpha}\,,
\end{equation}
or simply $k= \rho  |\lambda \rangle \lbrack \bar{\lambda}|$. Together with $2k\cdot v_i = \langle \lambda| v_i | \bar{\lambda} \rbrack$,
this leads to
\begin{align}
I_{\text{cut}} \sim{}&
\int d\rho\,\rho\,
\langle \lambda\,d\lambda \rangle
\lbrack \bar{\lambda}\,d\bar{\lambda} \rbrack
\delta(\rho \langle \lambda| v_1 | \bar{\lambda} \rbrack -1)
\delta(\rho \langle \lambda| v_2 | \bar{\lambda} \rbrack - 1)
\nonumber\\
={}& \int \langle \lambda\,d\lambda \rangle \lbrack \bar{\lambda}\,d\bar{\lambda} \rbrack
\frac{\delta(\langle \lambda | (v_1 -v_2) | \bar{\lambda} \rbrack )}
{\langle \lambda | v_1 | \bar{\lambda} \rbrack}
\nonumber\\
={}& - \int \frac{ \langle \lambda\,d\lambda \rangle }{ \langle \lambda | v_1 v_2 | {\lambda} \rangle } \sim \frac{x}{1-x^2}\,.
\end{align}
This is indeed the correct normalization factor, cf.\ eq.~(\ref{oneloopparam1}).
Similarly, a short calculation shows that the maximal cut of the double ladder integral is given by
\begin{equation}
\raisebox{-10mm}{
\psfrag{x1}[cc][cc]{$$}
\psfrag{x2}[cc][cc]{$$}
\psfrag{x3}[cc][cc]{$$}
\psfrag{x4}[cc][cc]{$$}
\includegraphics[width=0.2 \textwidth]{hqet_pic4.eps}} \longrightarrow \quad \left(\frac{x}{1-x^2}\right)^2 \,,
\end{equation}
which is consistent with eq.~(\ref{f1}).

\begin{figure}[h!tbp]
\centering
\psfrag{a}[cc][cc]{(a)}
\psfrag{b}[cc][cc]{(b)}
\psfrag{k1}[cc][cc]{$k_1$}
\psfrag{k12}[cc][cc]{$k_2-k_1$}
\psfrag{k23}[cc][cc]{$k_2-k_3$}
\psfrag{k3}[cc][cc]{$k_3$}
\psfrag{V1}[cc][cc]{$v_1$}
\psfrag{V2}[cc][cc]{$v_2$}
\psfrag{num}[cc][cc]{$$}
\includegraphics[width=0.47 \textwidth]{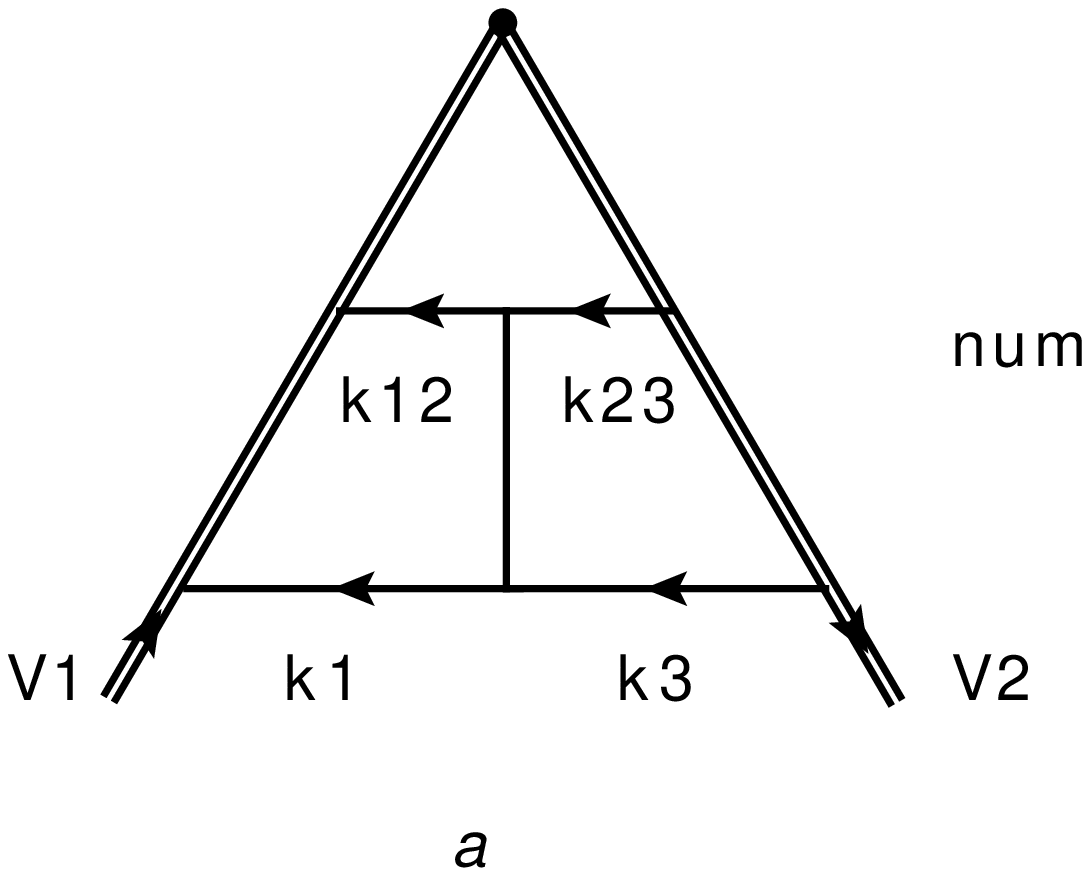}
\qquad
\includegraphics[width=0.29 \textwidth]{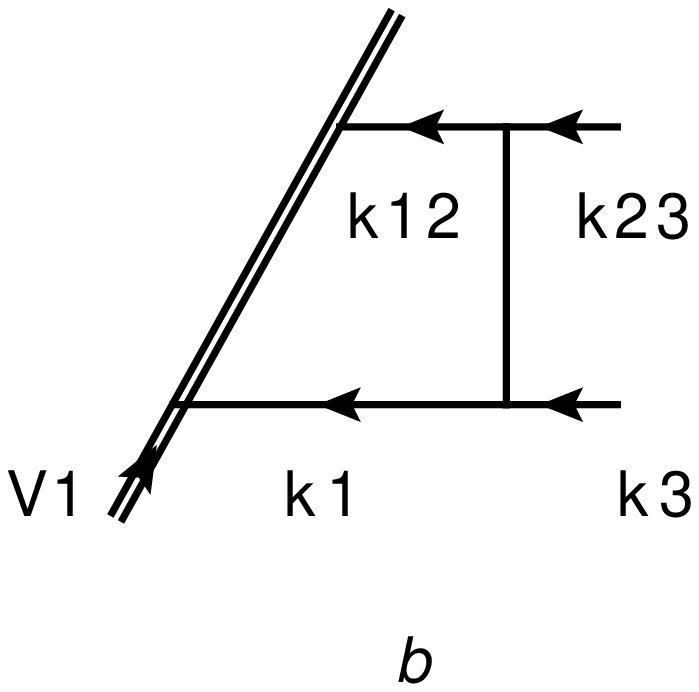}
\caption{Three-loop integral and one-loop subintegral,
whose maximal cuts are considered in the main text.}
\label{fig:hqet3loopcut}
\end{figure}

Let us now consider a more complicated example of three-loop integral containing $9$ propagators 
shown in  figure~\ref{fig:hqet3loopcut}(a).
%
Its maximal cut is best understood by first evaluating the maximal cut
of the one-loop subintegral  
shown in figure~\ref{fig:hqet3loopcut}(b) (with all external legs cut, i.e. $k_3^2=(k_2-k_3)^2=2(k_2 v_1)-1=0$).
The latter is given by
\begin{equation}
\int d^{4}k_1 \delta(k_1^2) \delta( (k_1-k_2)^2) \delta( (k_1-k_3)^2) \delta(2 k_1 \cdot v_1 -1)
\sim \frac{1}{k_2^2 \,(2 k_3 \cdot v_1 -1)}\,,
\label{cutoneloopsubintegal}
\end{equation}
where four delta-functions localize the $k_1-$integral. Applying (\ref{cutoneloopsubintegal}), we effectively reduce the integral 
of figure~\ref{fig:hqet3loopcut}(a) to a two-loop integral. It contains however two additional propagators coming from the right-hand 
side of (\ref{cutoneloopsubintegal}) and does not produce a function of uniform weight.  We can improve the situation by 
inserting into the integral of figure~\ref{fig:hqet3loopcut}(a) a numerator factor depending on loop momenta.
The latter can be chosen, e.g.,
to cancel part of the factors coming from~(\ref{cutoneloopsubintegal}).
In this way we can obtain two-loop integrals that are expected to be of uniform weight
based on the analysis of the previous subsection.

Explicitly, inserting the numerator factors  $(-k_2^2)$ and $(-2k_3\cdot v_1+1)$, we evaluate the maximal cut as
\begin{align}
\raisebox{-10mm}{
\psfrag{k1}[cc][cc]{$k_1$}
\psfrag{k12}[lc][cc]{$k_2-k_1$}
\psfrag{k3}[cc][cc]{$k_3$}
\psfrag{V1}[cc][cc]{$v_1$}
\psfrag{V2}[cc][cc]{$v_2$}
\psfrag{num}[cc][cc]{$\otimes (-k_2^2)$}
\includegraphics[width=0.3 \textwidth]{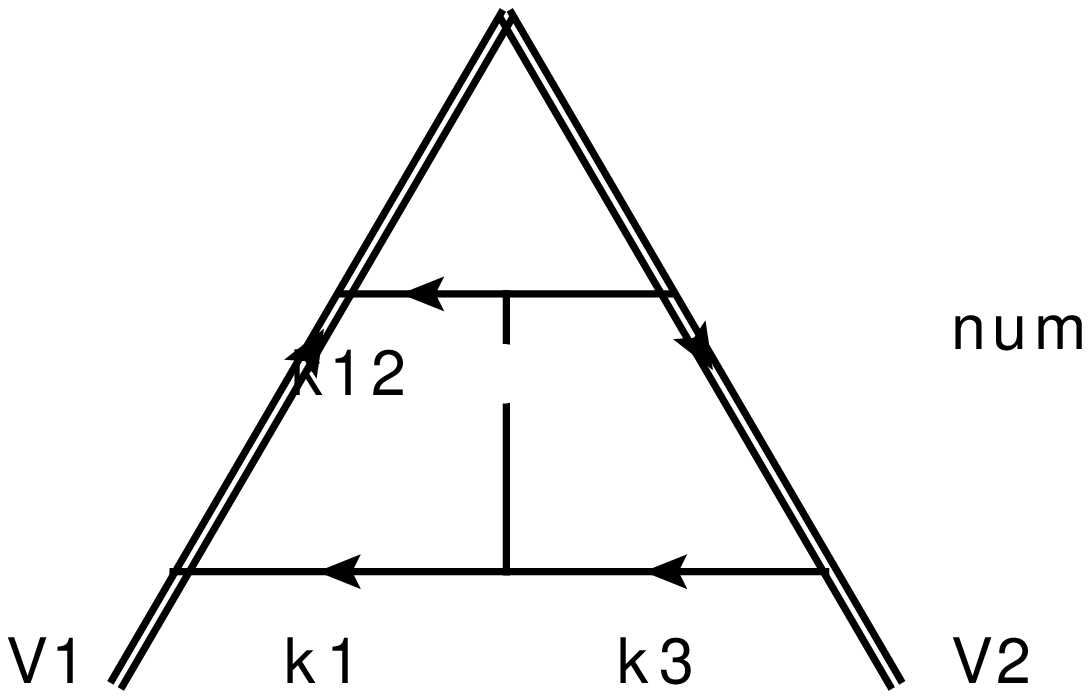}
} &\qquad \longrightarrow&
\raisebox{-10mm}{
\psfrag{x1}[cc][cc]{$$}
\psfrag{x2}[cc][cc]{$$}
\psfrag{x3}[cc][cc]{$$}
\psfrag{x4}[cc][cc]{$$}
\includegraphics[width=0.2 \textwidth]{hqet_pic4.eps}}
& \longrightarrow &&
\left(\frac{x}{1-x^2}\right)^2,
\label{tennis1}\\
\raisebox{-10mm}{
\psfrag{k1}[cc][cc]{$k_1$}
\psfrag{k12}[lc][cc]{$k_2-k_1$}
\psfrag{k3}[cc][cc]{$k_3$}
\psfrag{V1}[cc][cc]{$v_1$}
\psfrag{V2}[cc][cc]{$v_2$}
\psfrag{num}[cc][cc]{$\quad\otimes (-2 k_3 \cdot v_1+1)$}
\includegraphics[width=0.3 \textwidth]{hqet_pic13.eps}
} &\qquad \longrightarrow &
\raisebox{-10mm}{
\psfrag{x1}[cc][cc]{$$}
\psfrag{x2}[cc][cc]{$$}
\psfrag{x3}[cc][cc]{$$}
\psfrag{x4}[cc][cc]{$$}
\includegraphics[width=0.2 \textwidth]{hqet_pic8.eps}}
& \longrightarrow && 
\frac{x}{1-x^2}\,.
\label{tennis2}
\end{align}
With this choice of numerator factors, the three-loop integrals have a unique normalization factor and, therefore,
they are good candidates for pure functions.
This result is not too surprising, given that very similar results were obtained
in \cite{Henn:2013tua} for massless two--to--two amplitudes.

The techniques described in this and the previous subsection allow us to easily and quickly
assemble a list of candidate integrals that give rise to pure functions.
In the case of the generalized unitarity cut or leading singularity analysis,
this is expected based on the conjecture of~\cite{ArkaniHamed:2010gh}.
The differential equation method allows us to prove the uniform weight property
in the cases where it was only conjectured.

\subsection{Three-loop master integrals and differential equations
}
\label{sec:DE3loops}

In this subsection we extend the calculation of HQET master integrals to the three-loop level. As discussed in section~\ref{sect:NE}, 
thanks to eikonal identities we need to calculate only planar HQET integrals. To this end, we define all planar integral families at three loops, describe the choice of master integrals and their computation via differential equations.
Due to the size of the matrices involved, unlike the two-loop case, we select not to present the latter in this paper,
but provide them and other results in the form of ancillary text files.

\subsubsection{Definition of master integrals}

\begin{figure}[tbp]
\centering
\psfrag{V1}[cc][cc]{$$}
\psfrag{V2}[cc][cc]{$$}
\psfrag{k2}[cc][cc]{$a_7$}
\psfrag{k1}[cc][cc]{$a_6$}
\psfrag{k12}[cc][cc]{$a_3$}
\psfrag{a1}[cc][cc]{$a_1$}
\psfrag{a2}[cc][cc]{$a_2$}
\psfrag{a3}[cc][cc]{$a_3$}
\psfrag{a4}[cc][cc]{$a_4$}
\psfrag{a5}[cc][cc]{$a_5$}
\psfrag{a6}[cc][cc]{$a_6$}
\psfrag{a7}[cc][cc]{$a_7$}
\psfrag{a8}[cc][cc]{$a_8$}
\psfrag{a9}[cc][cc]{$a_9$}
\psfrag{a10}[cc][cc]{$a_{10}$}
\psfrag{a11}[cc][cc]{$a_{11}$}
\psfrag{a12}[cc][cc]{$a_{12}$}
\psfrag{a}[cc][cc]{(a)}
\psfrag{b}[cc][cc]{(b)}
\psfrag{c}[cc][cc]{(c)}
\psfrag{d}[cc][cc]{(d)}
\psfrag{e}[cc][cc]{(e)}
\psfrag{f}[cc][cc]{(f)}
\psfrag{g}[cc][cc]{(g)}
\psfrag{h}[cc][cc]{(h)}
\includegraphics[width=0.25 \textwidth]{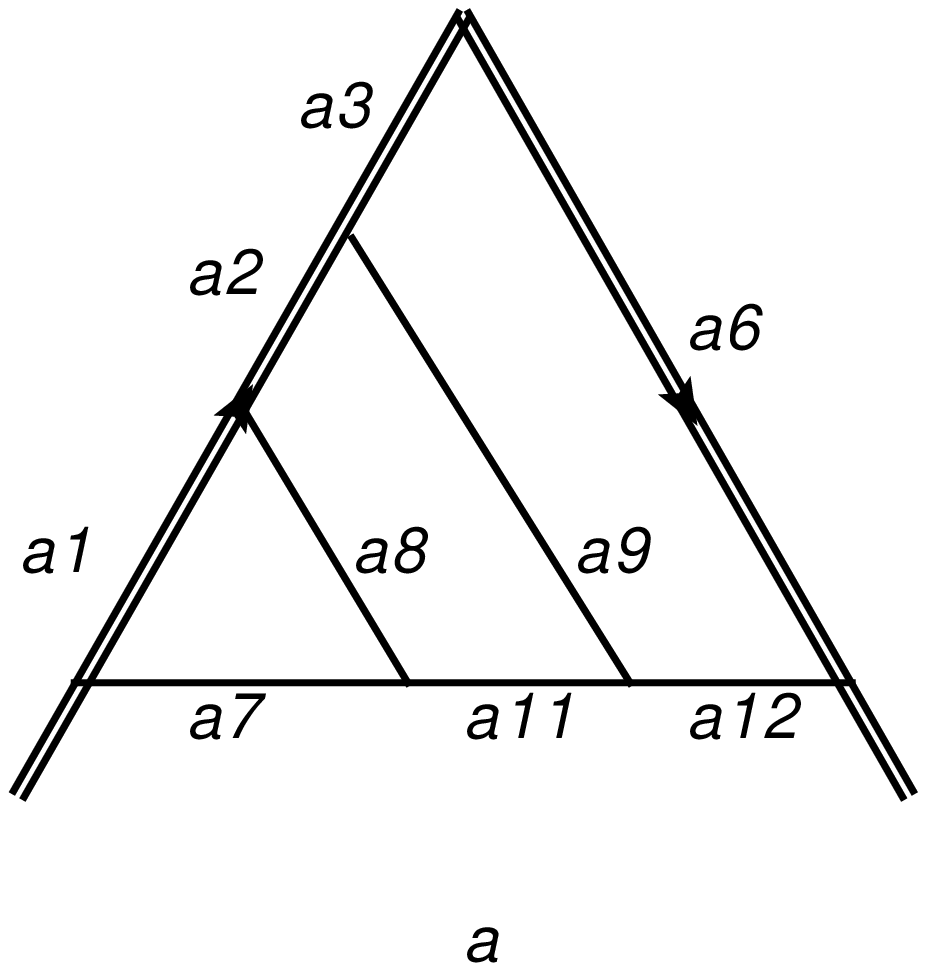}
\quad
\quad
\includegraphics[width=0.25 \textwidth]{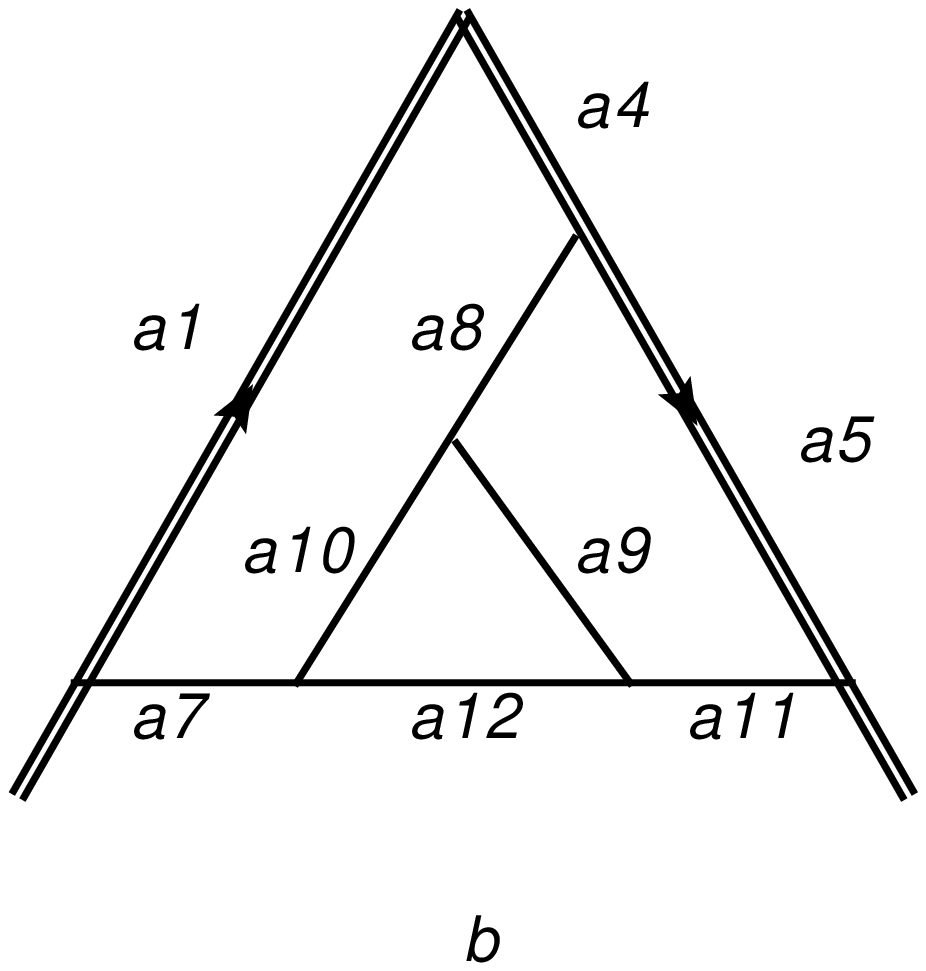}
\vspace{0.4cm}
\\
\includegraphics[width=0.25 \textwidth]{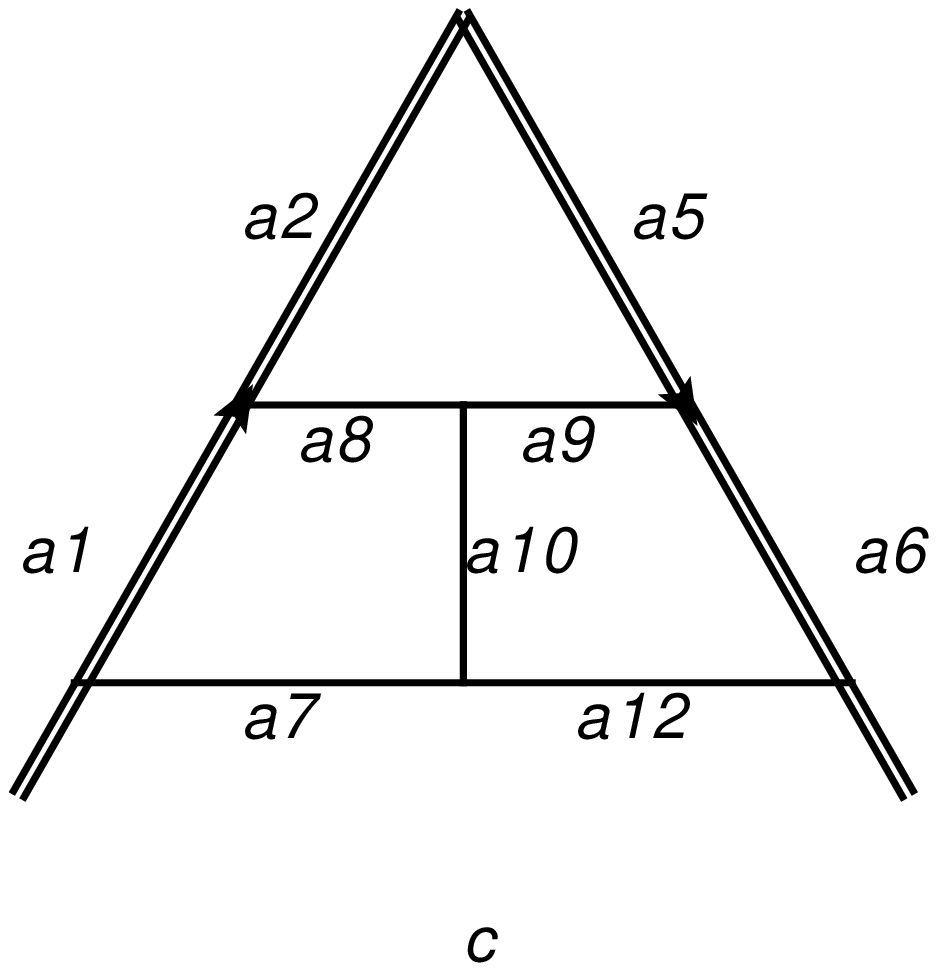}
\quad
\quad
\includegraphics[width=0.25 \textwidth]{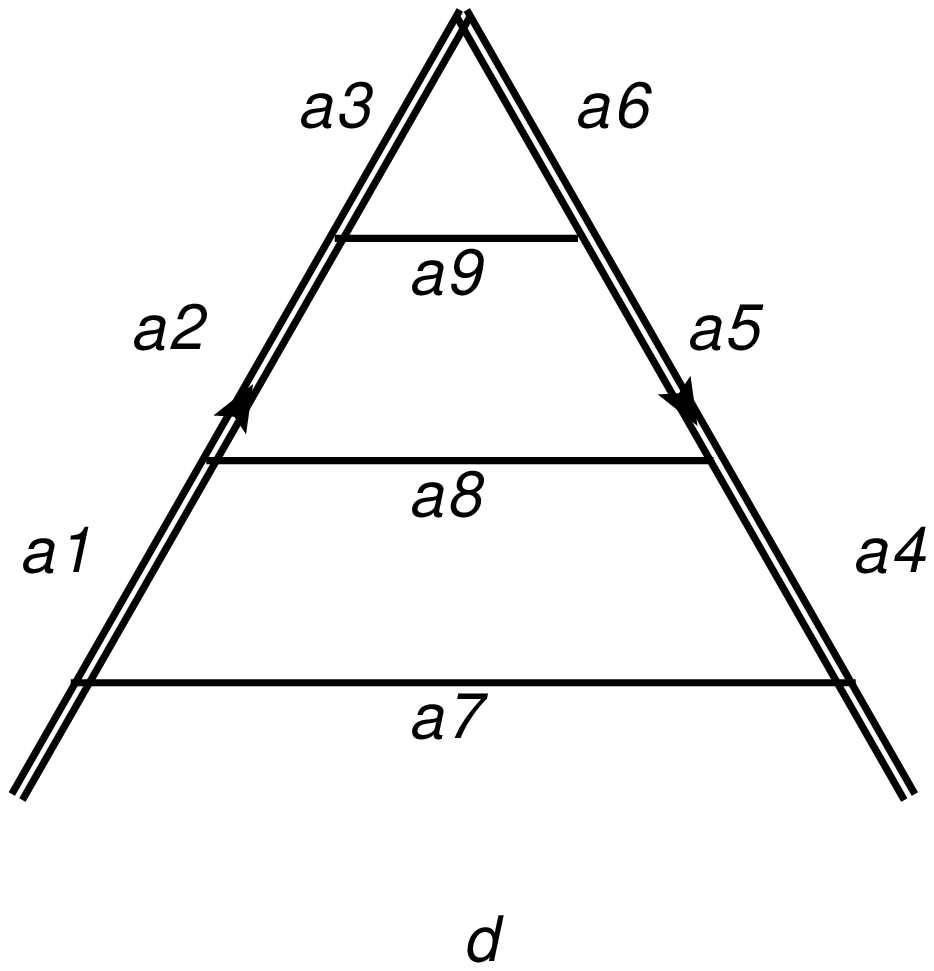}
\vspace{0.4cm}
\\
\includegraphics[width=0.25 \textwidth]{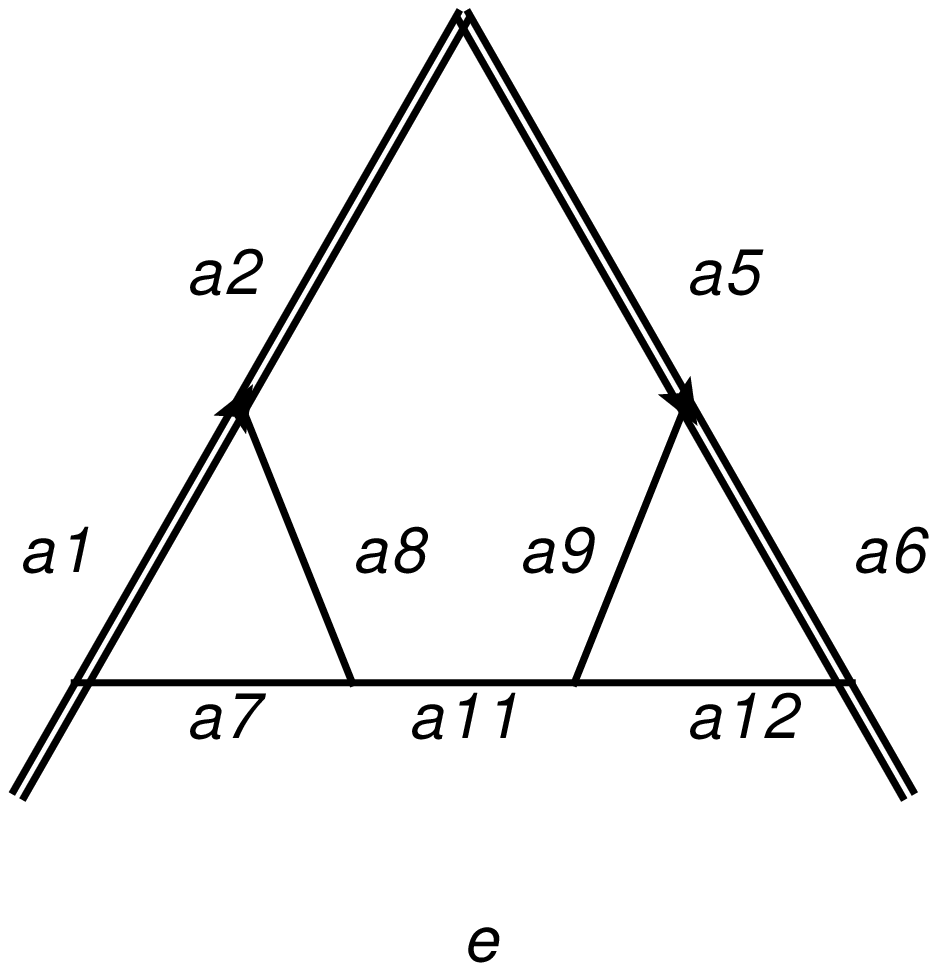}
\quad
\quad
\includegraphics[width=0.25 \textwidth]{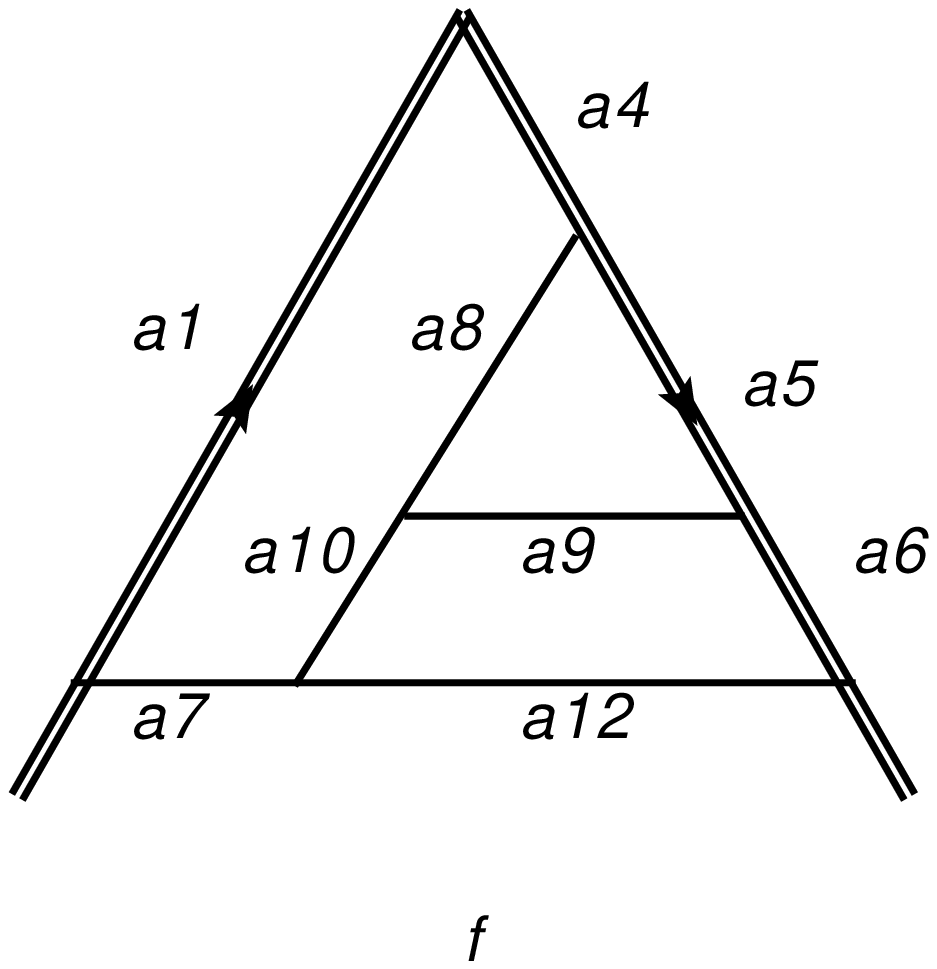}
\vspace{0.4cm}
\\
\includegraphics[width=0.25 \textwidth]{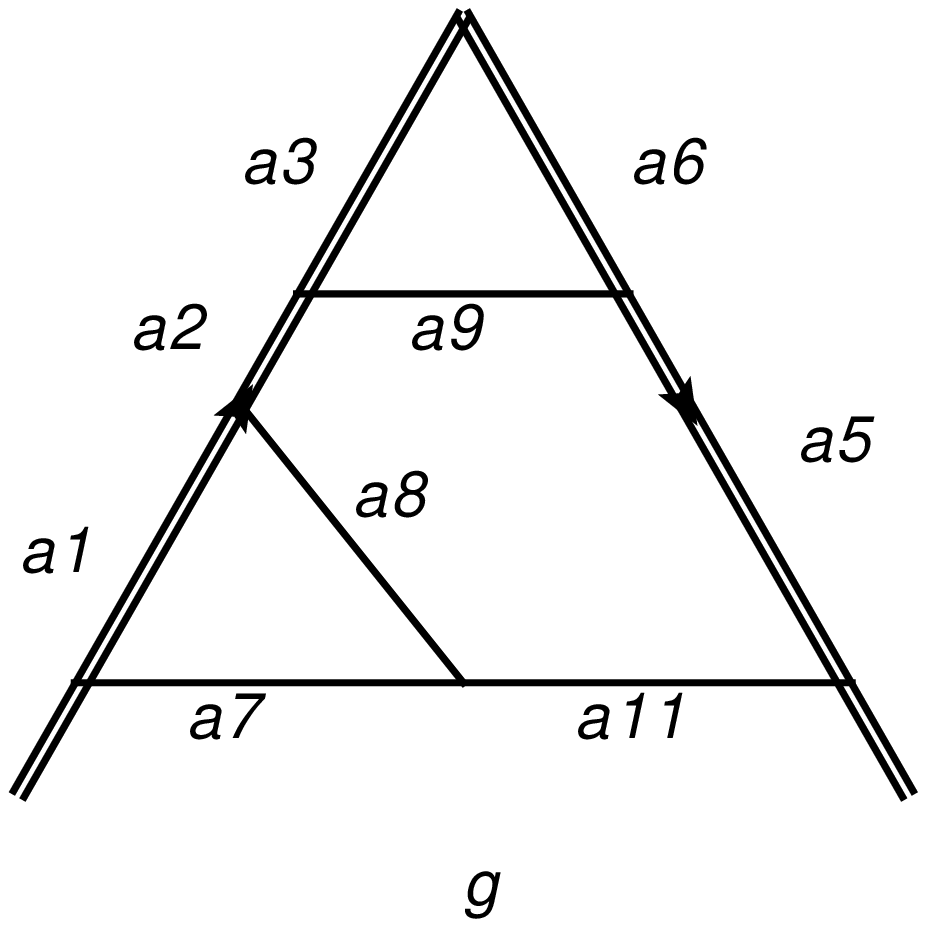}
\quad
\quad
\includegraphics[width=0.25 \textwidth]{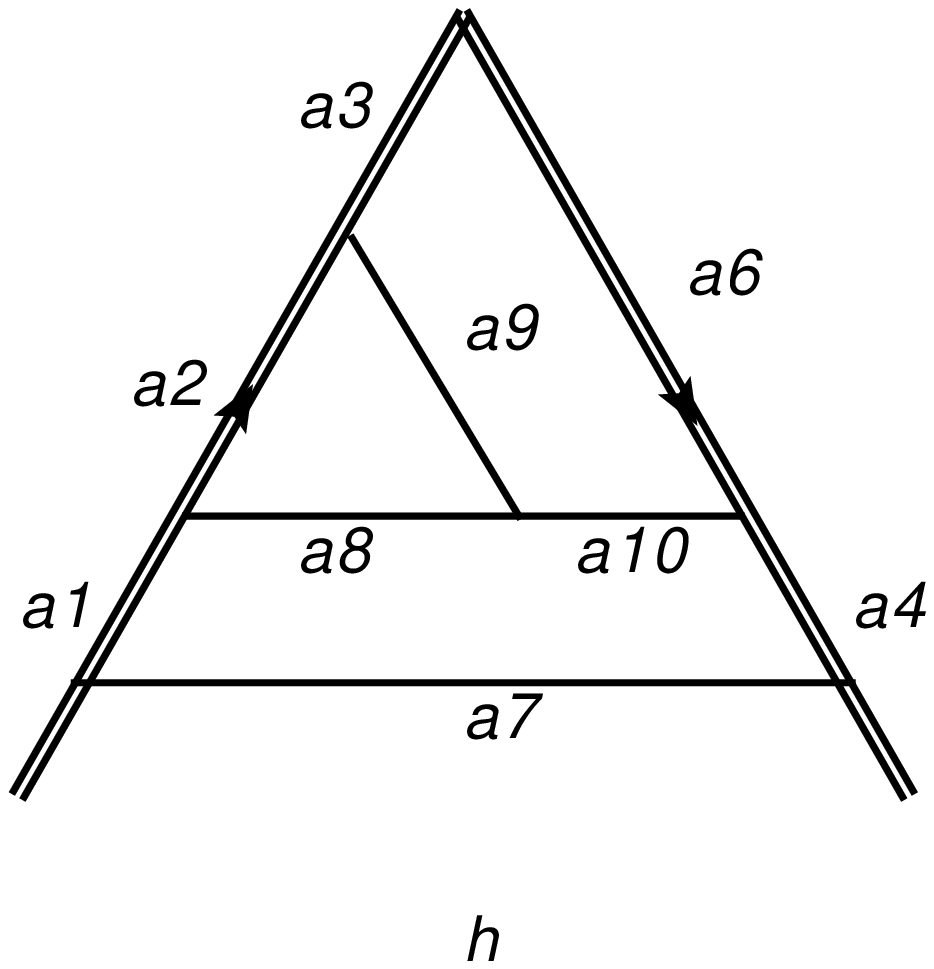}
\caption{The planar three-loop integral families.}
\label{fig:hqet3loopfamilies}
\end{figure}

All planar three-loop HQET integrals can be viewed as some special cases 
of the integral families shown in figure~\ref{fig:hqet3loopfamilies}.
Thanks to planarity it is possible to describe all of them
using a global parametrization of the loop momenta $k_1, k_2, k_3$.
In order to do so, we define the following factors,
\begin{equation}
  \begin{split}
    P_1\ &= -2 k_1\cdot v_1 + 1\,,  \\
    P_4\ &= -2 k_1 \cdot v_2 + 1\,,\\
    P_7\ &= -k_1^2\,,\\
    P_{10} &= -(k_1 - k_3)^2\,,
  \end{split}
\quad\quad
  \begin{split}
   P_2\ &=  -2 k_2 \cdot v_1 + 1\,,\\
    P_5\ &=  -2 k_2  \cdot v_2 + 1\,, \\
    P_8\ &= -(k_1-k_2)^2 \,,\\
    P_{11} &= -k_2^2\,,
  \end{split}\quad\quad
  \begin{split}
   P_3\ &= -2 k_3  \cdot v_1 + 1\,,\\
   P_6\ &=  -2 k_3 \cdot v_2 +
  1\,,  \\
      P_9\ &= -(k_2-k_3)^2\,,\\
  P_{12} &= -k_3^2  \,.
  \end{split}
\end{equation}
We then introduce the following notation for the HQET integrals,
\begin{equation} 
G_{a_1,\ldots a_{12}} =  \e^{3 \epsilon \gamma_{\rm E} } \int \frac{d^{D}k_{1}  d^{D}k_{2} d^{D}k_{3}   }{(i \pi^{D/2})^3} \prod_{i=1}^{12} (P_{i})^{-a_{i} } \,.
\label{deffirenotation}
\end{equation}

\noindent
The integral families shown in figure~\ref{fig:hqet3loopfamilies}
correspond to the following expressions in the notation of (\ref{deffirenotation}) (more generally, the  $a-$indices can of course be different from $1$),
\begin{align}\notag
\begin{split}
& \text{(a):} \quad G_{1,1,1,0,0,1,1,1,1,0,1,1}\,,\\  
& \text{(b):} \quad G_{1,0,0,1,1,0,1,1,1,1,1,1}\,,\\ 
& \text{(c):} \quad G_{1,1,0,0,1,1,1,1,1,1,0,1}\,,\\  
& \text{(d):} \quad G_{1,1,1,1,1,1,1,1,1,0,0,0}\,,
\end{split}
\begin{split}
& \text{(e):} \quad G_{1,1,0,0,1,1,1,1,1,0,1,1}\,,\\  
& \text{(f):} \quad G_{1,0,0,1,1,1,1,1,1,1,0,1}\,,\\  
& \text{(g):} \quad G_{1,1,1,0,1,1,1,1,1,0,1,0}\,,\\  
& \text{(h):} \quad G_{1,1,1,1,0,1,1,1,1,1,0,0}\,.  
\end{split}
\end{align}
Numerator factors can be accommodated by negative values of the indices $a_{i}$.
It is worth pointing out that the labeling in $G$ is not unique,
in the sense that the same integrals can be represented by different index vectors.
This is due to invariance under relabeling of loop momenta,
due to symmetry of some graphs and due to a $v_1 \leftrightarrow v_2$ symmetry of the integrated results.

We remark that with the above setup we can also discuss factorized integrals.
In particular, one-loop integrals multiplying generic two-loop integrals can be treated as a subset of the three-loop integrals.
This is a useful check, and also allows for a convenient calculation of, say,
higher orders of their $\epsilon$ expansion, within the same setup.

Solving the IBP relations for integrals shown in figure~\ref{fig:hqet3loopfamilies}, we find that there 
are $71$ master integrals in total. We choose the master integrals according to the uniform weight criteria explained in detail
in subsections~\ref{choice_mi} and~\ref{section:cuts}, following~\cite{Henn:2013pwa}.
We denote the basis integrals by $\boldsymbol{f}=(f_1,\dots,f_{71})$, 
hoping that using the same letter $f$ that we previously used to denote two-loop basis integrals with
will not lead to confusion. As in the two-loop case, all basis three-loop integrals $\boldsymbol{f}$ are pure functions of $x$ 
of weight zero. All except a handful of integrals could be chosen to be given by a single master integral (\ref{deffirenotation}),
with certain powers of propagators, and normalized appropriately. Only in a few cases it turned out to be necessary to consider 
linear combinations of  integrals  (\ref{deffirenotation}).~\footnote{Expressions for
the basis integrals in terms of master integrals (\ref{deffirenotation}) can be found in the ancillary file {\tt HQET\_3loop\_basis\_f.m}
in the arxiv submission of this paper.}

\subsubsection{Integral subsector at three loops}

As an example of the basis integrals at three loops, let us to return to three-loop integrals discussed in subsection~\ref{section:cuts}
(cf.\ eqs.~(\ref{tennis1}) and~(\ref{tennis2})). In the notation of eq.~(\ref{deffirenotation}) they read
\begin{align}
f_{70} ={}& \epsilon^6 \chi^2\, G_{1,1,0,0,1,1,1,1,1,1,-1,1}\,,
\label{choice70}
\\[2mm]
f_{71} ={}& \epsilon^6 \chi\, G_{1,1,-1,0,1,1,1,1,1,1,0,1}\,.
\label{choice71}
\end{align}
We can use them to illustrate the relationship between generalized cuts
and projections onto sectors of the differential equations. 

Let us consider the maximal cut of these integrals, i.e. replace all scalar and eikonal propagators with their cut version, 
and denote the resulting integrals by $\bar{f}_{70}$ and $\bar{f}_{71}$.
The latter satisfy a closed system of differential equations.
This system of two equations is a subset of the full system of $71$ equations.
This follows from the fact that cut integrals satisfy the same IBP relations as standard ones~\cite{Anastasiou:2002yz}.
Another way of saying this is that cut integrals satisfy the same differential equations as the standard integrals,
but with different boundary conditions (in particular, the remaining basis integrals vanish upon taking the
above mentioned cut, $\bar{f}_{i}=0$ for $i<70$). This means that the subsystem of basis integrals (\ref{choice70}) and (\ref{choice71}) is relevant for the full calculation.
In particular, it can serve as a check
of whether the choice~(\ref{choice70}) and (\ref{choice71}) is consistent
with the canonical form~(\ref{example_canonicalDE}) of the differential equations.

Indeed, we find that the integrals (\ref{choice70}) and (\ref{choice71}) satisfy the system of differential equations,
\begin{equation}
\partial_x
\left(
\begin{array}{c}
 \bar{f}_{70} \\
 \bar{f}_{71} \\
\end{array}
\right)
= \epsilon \left[ \frac{1}{x} \left(
\begin{array}{rr}
 -1 & \frac{2}{3} \\
 3 & -2 \\
\end{array}
\right)
+\frac{1}{x-1}
\left(
\begin{array}{rr}
 -2 & 0 \\
 0 & 2 \\
\end{array}
\right)
+\frac{1}{x+1}
\left(
\begin{array}{cc}
 4 & 0 \\
 0 & 2 \\
\end{array}
\right)
\right]
\left(
\begin{array}{c}
 \bar{f}_{70} \\
 \bar{f}_{71} \\
\end{array}
\right)\,,
\end{equation}
which is consistent with~(\ref{example_canonicalDE}).
Of course, removing the cut, it could be that terms violating the form~(\ref{example_canonicalDE})
are present in off-diagonal terms.
If this is the case, one can attempt to remove them using the methods
discussed in ref.~\cite{Caron-Huot:2014lda} and more recently in \cite{Henn:2014qga,Lee:2014ioa}.
It turns out that this is not needed for the two integrals under discussion.
In our calculation, we resorted to such ``brute-force'' methods
only in the case of a handful of integrals.

\subsubsection{Full system of differential equations}

With the basis $\boldsymbol{f}$ given in ancillary files included in the arxiv submission of this article,
the differential equations take the form 
\begin{equation}
\partial_x \, \boldsymbol{f} =    \epsilon \, \left({a_3\over x}+{b_3\over x+1}+{c_3\over x-1} \right) \, \boldsymbol{f}\,,
\label{DEhqet}
\end{equation}
with constant $71\times 71$ matrices $a_3$, $b_3$ and $c_3$ given in the ancillary file \texttt{HQET\_3loop\_mAtilde.m}. 

We see that eq.~(\ref{DEhqet}) has four regular singular points, $0$, $1$, $-1$, $\infty$.
Due to the $x \leftrightarrow 1/x$ symmetry of the definition~(\ref{defkin}),
only the first three are independent. They correspond, in turn, to the light-like limit (infinite Minkowskian angle),
zero angle limit and backtracking limit.

As before, we can solve eq.~(\ref{DEhqet}) in a Laurant expansion (\ref{f-exp}).
Then we can express $\boldsymbol{f}(x)$  order by order in $\epsilon$ in terms of harmonic polylogarithms.
We use the value of basis integrals at $x=1$ as boundary condition~\cite{Grozin:2000jv,Chetyrkin:2003vi}
(see~\cite{Grozin:2008tp} for a summary).
  
Most of the basis integrals can be evaluated trivially for $x=1$ in terms of Gamma functions.
Boundary conditions for Feynman integrals can often be obtained without additional work, by imposing physical properties. In reference \cite{Henn:2013nsa} this was used e.g. in a bootstrap approach to compute single-scale integrals from differential equations. In the present case, we can use finiteness of the limit $x \to 1$ as our main condition. It turns out that only one non-trivial integral is needed at $x=1$.
It is known up to weight five~\cite{Czarnecki:2001rh}
(but this is the order we are interested in),
\begin{equation}
G_{1, 1, 1,0,0,0, 1, 1, 1, 0, 1, 1}(x=1) = 12 \zeta_2 \zeta_3 - 5 \zeta_5  + \mathcal{O}(\epsilon) \,,
\end{equation}
which is exactly the order we need for our calculation.
It is likely that also this integral could be obtained by inspecting the differential equations more closely, or applying bootstrap ideas as in \cite{Henn:2013nsa}.

\subsubsection{Solution}

As noted above,
the solution to (\ref{DEhqet}) to any order in $\epsilon$ is expressed in terms of harmonic polylogarithms.
The explicit expressions for basis integrals $\boldsymbol{f}(x)$ up to weight five can be found in the ancillary file {\tt HQET\_3loop\_HPL.m}.%
\footnote{A curious feature is that integral $f_{71}$ is apparently finite as $\epsilon \to 0$
and a weight six function, and therefore appears only at order $\epsilon^6$ in our normalization.}
As an example, we have
\begin{align}
f_{44} ={}& \epsilon^5 \frac{1-x^2}{x} G_{1,0,1,0,1,0,1,1,2,0,1,0}
\nonumber\\
={}&  \epsilon^4 \bigg[ - \frac{1}{6} \pi^2 H_{0,0}(x) - \frac{2}{3}\pi^2 H_{1,0}(x)- 4 H_{0,-1,0,0}(x) +2 H_{0,0,-1,0}(x)
\nonumber\\
&{} \qquad\qquad\quad + 2 H_{0,1,0,0}(x) - 4 H_{1,0,0,0}(x) + 4 \zeta_3 H_{0}(x) - \frac{17 \pi^4}{360} \bigg]\,.
\end{align}

We remark that the differential equation, or equivalently, the path integral~(\ref{pathintegral})
encodes all the information about the symbol~\cite{Goncharov1,Brown1,Brown2,Goncharov:2010jf} of the result
(and all possible symbol related simplifications are already manifest).
The latter can immediately be computed as a corollary.
In order to do this, in addition to the matrix $\tilde{A}$, only the leading term $\boldsymbol{f}^{(0)}$ in $\epsilon-$expansion (\ref{f-exp}) is required.
The latter reads
\begin{align}
\boldsymbol{f}^{(0)} ={}&\big(-1,-1,0,\tfrac{1}{2},-1,0,0,0,0,0,\tfrac{1}{6},0,-\tfrac{1}{6},-\tfrac{1}{3},0,0,\tfrac{1}{12},0,\tfrac{1}{4}, 0,-1,
\nonumber\\
&\hphantom{\{}
0,0,0,0,0,0,0,0,0,-\tfrac{1}{12},0,0,0,\tfrac{1}{4},0,0,0,-1,0,0,0,0,0,0,0,
\nonumber\\
&\hphantom{\{}
0,0,0,0,0,0,0,0,0,0,0,0,0,0,0,0,0,0,0,0,0,0,0,0,0\big)\,.
\end{align}

We have evaluated all basis integrals using \texttt{Fiesta} \cite{Smirnov:2013eza} at the value $x=1/4$
and found perfect agreement with analytical formulas, within the error bars.

There are a number of analytic checks.
Out of   $71$ master integrals,
$7$ are straight-line ones (studied in~\cite{Grozin:2000jv,Chetyrkin:2003vi}),
$8$ can be chosen as products of lower-loop integrals,
and $10$ correspond to the one-loop triangle integral with $\epsilon$-dependent powers of denominators
(studied in~\cite{Grozin:2011rs}).
One non-trivial integral at $x=1$ was obtained in our approach from the finiteness of the $x\to1$ limit for all integrals entering the differential equations. Previously, it was computed in terms of a hypergeometric function in ref.~\cite{Beneke:1994sw} (another hypergeometric representation was derived in~\cite{Grozin:2008nu}).
We can expand it in $\epsilon$ using the \texttt{Mathematica} package \texttt{HypExp}~\cite{Huber:2005yg}.
The result, written up to weight five, is
\begin{align}
G_{1,0,1,0,0,0,1,2,1,0,0,1}(x=1) =  \frac{1}{(1-2 \epsilon) \epsilon^4} \bigg[{}& -\frac{\pi^2}{9} \epsilon^2
+ \frac{14}{3} \zeta_3 \epsilon^3 - \frac{337 \pi^4}{540} \epsilon^4
\nonumber\\[2mm]
{}& + \left( \frac{295}{18} \pi^2 \zeta_3 +\frac{500}{3} \zeta_5  \right) \epsilon^5 +\mathcal{O}(\epsilon^6) \bigg]\,.
\end{align}
We found perfect agreement with our result.

\subsubsection{Check of supersymmetric Wilson loop in $\mathcal{N}=4$ SYM}
\label{checkN4sysyWL}

We can also perform analytic checks of our results by comparing to the supersymmetric cusped 
Wilson loop (\ref{superW}) in $\mathcal{N}=4$ super Yang--Mills theory,\footnote{In this section, for simplicity of presentation, we choose the Wilson loop (\ref{superW})  to be in the fundamental representation, $C_R=C_F=(N^2-1)/(2N)$, and we take the planar limit, corresponding to $C_R=N/2$.}
\begin{equation}
\mathcal W = 1 + \sum_{L=1}^{3} \left( \frac{g^2 N}{8 \pi^2}  \right)^L \mathcal W^{(L)} + \mathcal{O}(g^8) \,.
\end{equation}
It was computed at three loops in ref.~\cite{Correa:2012nk}, using a different method.
This quantity depends on two cusp angles $\phi$ and $\theta$ and the dependence on the latter angle enters through the following
variable
\begin{align}\label{xi0}
\xi_0(\phi,\theta)=i{\cos\phi-\cos\theta\over \sin\phi}\,.
\end{align}
Up to three loops, the perturbative corrections to $\mathcal W$ can be expressed in terms of master  integrals defined in (\ref{G7}) and (\ref{deffirenotation})~\footnote{Strictly speaking, the calculation in ref.~\cite{Correa:2012nk} was performed for $\theta=0$ in which case $\xi_0= (1-x)/(1+x)$.} 
\begin{align} \notag
\mathcal W^{(1)} ={}& - \frac{1}{2} \xi_0 \chi G_{111}\,,
\\\notag
\mathcal W^{(2)} ={}& + \frac{1}{4} \left[ \xi_0\chi G_{0,1,1,1,0,1,1} +  (\xi_0\chi)^2 G_{1,1,1,1,1,1,0} \right]\,,
\\
\mathcal W^{(3)} ={}& - \frac{1}{8} \Bigl[ \xi_0\chi \left( G_{1,0,0,0,0,1,1,1,1,0,1,1} + 2 \, G_{1,0,0,-1,1,1,1,1,1,1,0,1} \right)
\nonumber\\
&\hphantom{- \frac{1}{8} \Bigl[\Bigr.}{}
+ (\xi_0\chi)^2  \left( G_{1,1,0,1,0,1,1,1,1,1,0,0} + G_{1,1,0,0,1,1,1,1,1,1,-1,1} \right)
\nonumber\\
&\hphantom{- \frac{1}{8} \Bigl[\Bigr.}{}
+ (\xi_0\chi)^3  G_{1,1,1,1,1,1,1,1,1,0,0,0} \Bigr]\,.
\end{align}
where $G_{111} = (\epsilon\chi)^{-1} \log x$.

Using the obtained results, we reproduce results of the three-loop computation performed in \cite{Correa:2012nk}.
For example, the following three-loop integral was computed there
(taking into account the conversion between the different regulators),
\begin{equation}
G_{1,0,0,0,0,1,1,1,1,0,1,1} = \frac{1}{\epsilon} \frac{x}{1-x^2} \left[ - \frac{14}{135} \pi^4 H_{0}(x) - \frac{8}{9} \pi^2 H_{0,0,0}(x) - \frac{16}{3} H_{0,0,0,0,0}(x) \right] + \mathcal{O}(\epsilon^0)\,.
\label{check4a}
\end{equation}
In terms of the three-loop basis integrals defined above it reads
\begin{equation}
G_{1,0,0,0,0,1,1,1,1,0,1,1} = \epsilon^{-6} \frac{x}{1-x^2} (f_{32}-f_{30} )\,.
\label{check4paper}
\end{equation}
Using the explicit results for $f_{30}$ and $f_{32}$ we found full agreement with~(\ref{check4a}) (note that the individual results for $f_{32}$ and $f_{30}$ are rather complicated in comparison).
In the similar manner, we reproduce the remaining three-loop integrals
\begin{align} \notag
   G_{1,1,1,1,1,1,1,1,1,0,0,0} \ \ ={}&  \epsilon^{-6} \chi^{-3} f_{56}\,,
\\[2mm] \notag
   G_{1,1,0,0,1,1,1,1,1,1,-1,1} ={}&  \epsilon^{-6} \chi^{-2} f_{70}\,,
\\[2mm] \notag
  G_{1,1,0,1,0,1,1,1,1,1,0,0} \ \ ={}&  \frac{1}{4} \epsilon^{-6} \chi^{-2}(f_{29}-f_{36}+f_{50 })\,,
\\[2mm]
 G_{1,0,0,-1,1,1,1,1,1,1,0,1} ={}&  - \frac{1}{4} \epsilon^{-6} \chi^{-1}
\big(f_3 + 2 f_{12} - f_{20} - 4 f_{30} - 4 f_{32} + 8 f_{33}
\nonumber\\
  & \hphantom{- \frac{1}{4} \epsilon^{-6} \xi_0 (}{}
- f_{37} + f_{38} + 2 f_{44} - 4 f_{49} + 4 f_{60}\big)\,,
\end{align}
where  $\chi = (1-x^2)/x$. 

We can use the above results to compute the three-loop cusp anomalous dimension for the supersymmetric Wilson loop 
in $\mathcal{N}=4$ super Yang--Mills theory
\begin{align}
\log \mathcal W = -\sum_{L\ge 1} \frac{1}{2L \epsilon} \left( \frac{g^2 N}{8 \pi^2} \right)^L
  \mathnormal{\Gamma}^{(L)} (\phi,\theta) + \mathcal{O}(\epsilon^0)\,.
\end{align}
In perfect agreement with findings of ref.~\cite{Correa:2012nk}, we find
\begin{align}\notag\label{susy-Gamma}
\mathnormal{\Gamma}^{(1)}  ={}& \xi_0 \,   \frac{1}{2} H_{1}(y)\,,
\\ \notag
\mathnormal{\Gamma}^{(2)}  ={}&  \xi_0 \, \left[ -\frac{\pi^2}{6} H_{1}(y) - \frac{1}{4} H_{1,1,1}(y)  \right]  
 +  \xi^2_0\, \left[ \frac{1}{2} H_{1,0,1}(y) + \frac{1}{4} H_{1,1,1}(y) \right]\,,
\\ \notag
\mathnormal{\Gamma}^{(3)}  ={}& \xi_0 \left[  \frac{\pi^4}{12} H_{1}(y) + \frac{\pi^2}{4} H_{1,1,1}(y) +\frac{5}{8}  H_{1, 1, 1, 1, 1}(y) \right] +  \xi^2_0\left[ -\frac{\pi^2}{6} H_{1,0,1}(y) -\frac{\pi^2}{3} H_{0,1,1}(y)  \right.  
\\ \notag
& \left. 
-\frac{\pi^2}{4}H_{1,1,1}(y)  -H_{1,1,1,0,1}(y)  -\frac{3}{4} H_{1,0,1,1,1}(y) -H_{0,1,1,1,1}(y) -\frac{11}{8} H_{1,1,1,1,1}(y)    \right.  
\\ \notag
& \left. 
 -\frac{3}{2}\zeta_3 H_{1,1}(y)  \right] 
 + \xi^3_0 \left[ H_{1,1,0,0,1}(y) + H_{1,0,1,0,1}(y) +H_{1,1,1,0,1}(y) + \frac{1}{2} H_{1,1,0,1,1}(y) 
   \right.  
 \\
& \left. 
 +\frac{1}{2} H_{1,0,1,1,1}(y) +\frac{3}{4} H_{1,1,1,1,1}(y)  \right]\,,
\end{align}
with $y=1-x^2$ and $\xi_0$ given by (\ref{xi0}). 
This is a highly nontrivial test of the calculation of the master integrals. 
Within the differential equations method, the calculation of a given integral requires the knowledge
of all integrals appearing in sub-topologies (obtained by removing propagator factors). Since the integrals
needed here have a maximal number of propagator factors, this calculation is also a consistency check
of many other integrals appearing for example in QCD.

\section{Results}
\label{sect:results}

The basis integrals defined in the previous section allow us to compute the cusped Wilson loop (\ref{lnV}). 
For example, we can express the three-loop correction to $W$ as
\begin{align}\label{W3}
W^{(3)} = \sum_{i=1}^{71} C_i  \, f_i(x)\,,
\end{align}
where $C_i$ are coefficient functions rational in $x$ and depending on $\epsilon$. 
Their explicit form is not particularly enlightening.
We can write similar formulas for $W^{(1)}$ and $W^{(2)}$.~\footnote{As was already mentioned, the three-loop integrals computed here can also
be used to express all required one- and two-loop integrals, by writing the latter
as factorized three-loop integrals, where the additional factors are trivial.}
Then, we extract divergent part of $\log W$ and
match it into expected form (\ref{lnZ}) of $\log Z$. In this way, we verify gauge independence of $Z-$factor, reproduce well-known result
for $\beta-$function and extract the three-loop cusp anomalous dimension.

\newpage

\subsection{Coefficient functions} 
 
To express our results for three-loop cusp anomalous dimension we introduce the following functions 
\begin{equation}\label{A-B}
\begin{aligned}  
{A}_1(x) =&  \xi \,   \frac{1}{2} H_{1}(y)\,,\quad\quad\\
%
{A}_2(x)  =& \, \left[ \frac{\pi^2}{3} + \frac{1}{2} H_{1,1}(y) \right]  + \xi \left[- H_{0,1}(y) -\frac{1}{2} H_{1,1}(y) \right] \,, \\
%
{A}_{3}(x) =& \;\;  \xi \, \left[ -\frac{\pi^2}{6} H_{1}(y) - \frac{1}{4} H_{1,1,1}(y)  \right]  
 +  \xi^2\, \left[ \frac{1}{2} H_{1,0,1}(y) + \frac{1}{4} H_{1,1,1}(y) \right]  \,,\\
%
%
{A}_{4}(x) =&
 \, \left[  -\frac{\pi^2}{6} H_{1,1}(y) -\frac{1}{4} H_{1,1,1,1}(y)  \right]  + \xi \left[  \frac{\pi^2}{3} H_{0,1}(y) +\frac{\pi^2}{6} H_{1,1}(y) + 2 H_{1,1,0,1}(y)   \right. \\ & \left.  
 +\frac{3}{2} H_{0,1,1,1}(y) 
 + \frac{7}{4} H_{1,1,1,1}(y) + 3 \zeta_3 H_{1}(y)  \right]  +  \xi^2 \left[ -2 H_{1,0,0,1}(y)-2 H_{0,1,0,1}(y)
  \right. \\
& \left.  -2 H_{1,1,0,1}(y)- H_{1,0,1,1}(y) - H_{0,1,1,1}(y) -\frac{3}{2} H_{1,1,1,1}(y) \right]\,,\\
{A}_{5}(x) =&\, \xi \left[  \frac{\pi^4}{12} H_{1}(y) + \frac{\pi^2}{4} H_{1,1,1}(y) +\frac{5}{8}  H_{1, 1, 1, 1, 1}(y) \right] +  \xi^2\left[ -\frac{\pi^2}{6} H_{1,0,1}(y) -\frac{\pi^2}{3} H_{0,1,1}(y)  \right.  \\
& \left. 
-\frac{\pi^2}{4}H_{1,1,1}(y)  -H_{1,1,1,0,1}(y)  -\frac{3}{4} H_{1,0,1,1,1}(y) -H_{0,1,1,1,1}(y) -\frac{11}{8} H_{1,1,1,1,1}(y)    \right.  \\
& \left. 
 -\frac{3}{2}\zeta_3 H_{1,1}(y)  \right] 
 + \xi^3 \left[ H_{1,1,0,0,1}(y) + H_{1,0,1,0,1}(y) +H_{1,1,1,0,1}(y) + \frac{1}{2} H_{1,1,0,1,1}(y) 
   \right.  \\
& \left. 
 +\frac{1}{2} H_{1,0,1,1,1}(y) +\frac{3}{4} H_{1,1,1,1,1}(y)  \right] \,,
 \\ 
B_{3}(x) =& \left[  - H_{1,0,1}(y) + \frac{1}{2} H_{0,1,1}(y) - \frac{1}{4} H_{1,1,1}(y)\right]   \\& + \xi \left[ 2 H_{0,0,1}(y) + H_{1,0,1}(y) + H_{0,1,1}(y) + \frac{1}{4} H_{1,1,1}(y) \right]\,,  \\
B_{5}(x) =& 
\frac{x}{1-x^2} \left[-\frac{\pi^4}{60} H_{-1}(x) -\frac{\pi^4}{60} H_{1}(x) - 4H_{-1,0,-1,0,0}(x) + 4 H_{-1,0,1,0,0}(x)   \right.  \\
& \left. \phantom{\frac{x}{1-x^2}}
 - 4 H_{1,0,-1,0,0}(x)   
+ 4 H_{1,0,1,0,0}(x) + 4 H_{-1,0,0,0,0}(x) + 4 H_{1,0,0,0,0}(x) \right.  \\ 
&  \phantom{\frac{x}{1-x^2}} + 2 \zeta_3  H_{-1,0}(x) + 2 \zeta_3 H_{1,0}(x)   \bigg] \,, 
\end{aligned}
\end{equation}
where we recall that $\xi = (1+x^2)/(1-x^2)$  
and $y=1-x^2$. 
The subscript of $A_i$ and $B_i$ indicates the (transcendental) weight of the functions.

The three-loop cusp anomalous dimension involves particular linear combinations of these functions
\begin{align}\label{tildeA}
\tilde{A}_{i} = A_{i}(x) - A_{i}(1)\,,\qquad \tilde{B}_{i} = B_{i}(x) - B_{i}(1)\,,
\end{align}
and
 \begin{align}
\gamma_{AA} ={}& \frac{1}{4} \left( \tilde{A}_5  + \tilde{A}_{4} + \tilde{B}_{5} + \tilde{B}_{3} \right)
+ \frac{67}{36} \tilde{A}_{3} + \frac{29}{18} \tilde{A}_{2}
+ \left( \frac{245}{96} + \frac{11}{24} \zeta_3 \right) \tilde{A}_{1}\,,
\nonumber\\
\gamma_{ff} ={}& - \frac{1}{27} \tilde{A}_1\,,\qquad\qquad
 \gamma_{Ff} = \left( \zeta_3 - \frac{55}{48} \right) \tilde{A}_1\,,
\nonumber\\
\gamma_{Af} ={}& - \frac{5}{9} \left( \tilde{A}_2 + \tilde{A}_{3} \right)
- \frac{1}{6} \left( 7 \zeta_{3} + \frac{209}{36} \right) \tilde{A}_{1}\,,
\nonumber\\
\gamma_{ss} ={}&  \frac{1}{432} \tilde{A}_1\,,\qquad\qquad
\gamma_{sf} = \frac{7}{16} \tilde{A}_1\,,
\nonumber\\
\gamma_s \ ={}& - \left(\frac{1039}{1728} + \frac{1}{48} \zeta_3 \right) \tilde{A}_1
- \frac{1}{9} (\tilde{A}_2 + \tilde{A}_3)\,.
\label{resultc4}
\end{align}
As follows from the definition, these functions vanish for zero cusp angle, or equivalently $x=1$.
 
\subsection{Three-loop cusp anomalous dimension}

In QCD with $n_f$ fermion flavours, we obtained the following result for the three-loop  cusp anomalous dimension~(\ref{cusp-def}) in the \MS{} scheme  
\begin{align}\notag \label{QCD-3loops}
\Gamma^{\MS}_{\rm  QCD} {}& =  {\alpha_s \over \pi} C_R\,\tilde{A}_1 + \lr{{\alpha_s \over \pi}}^2 C_R \biggl[ \frac{1}{2} C_A \left(\tilde{A}_2 + \tilde{A}_3\right)
+ \left( \frac{67}{36}  C_A
- \frac{5}{9} T_F n_f \right) \tilde{A}_1 \biggr]
\\[2mm]
{}& + \lr{{\alpha_s \over \pi}}^3 C_R \biggl[C_A^2\, \gamma_{AA}+ (T_F n_f)^2\gamma_{ff} + C_F T_F n_f\gamma_{Ff} + C_A T_F n_f\gamma_{Af}\bigg]\,,
\end{align}
where $C_F=(N^2-1)/(2N)$ and $C_A=N$ are the quadratic Casimir operators
of the $SU(N)$ gauge group in the fundamental and adjoint representation, respectively, and $T_F=1/2$ for fermions in the fundamental
representation. The relation (\ref{QCD-3loops}) involves the coefficient functions defined in (\ref{tildeA}) and (\ref{resultc4}).
 
In the gauge theory with $n_f$ fermions and $n_s$ scalars in the adjoint representation of the $SU(N)$, the three-loop cusp anomalous
dimension is given by
\begin{align}\notag\label{adj-3loops}
\Gamma^{\MS}_{\rm  adj} {}& =  {\alpha_s \over \pi} C_R\,\tilde{A}_1 + \lr{{\alpha_s \over \pi}}^2 C_R C_A \biggl[ \frac{1}{2}  \left(\tilde{A}_2 + \tilde{A}_3\right)
+  \left(\frac{67}{36} - \frac{5}{18} n_f- \frac{1}{9} n_s\right) \tilde{A}_1 \biggr]
\\[2mm]
{}& + \lr{{\alpha_s \over \pi}}^3 C_R C_A^2 \biggl[\gamma_{AA}+\frac14n_f^2\gamma_{ff} +  n_s^2\gamma_{ss} +  \frac12 n_s n_f \gamma_{sf}+ \frac12 n_f(\gamma_{Ff} + \gamma_{Af})+   n_s\gamma_s\bigg],
\end{align}
with the same coefficient functions (\ref{tildeA}) and (\ref{resultc4}). Denoting this expression as $\Gamma_{\rm adj}(n_f, n_s)$, we can get the three-loop cusp anomalous dimension in supersymmetric Yang--Mills theories with different number of supercharges $\mathcal N$ by adjusting the number of fermions and scalars
\begin{align} 
\Gamma _{\mathcal{N}=1} ={}& \Gamma_{\rm  adj}\big(n_f=1,n_s=0 \big) \,,
\nonumber\\[2mm]
\Gamma _{\mathcal{N}=2} ={}& \Gamma_{\rm  adj}\big(n_f=2,n_s=2 \big) \,,
\nonumber\\[2mm]
\Gamma _{\mathcal{N}=4} ={}& \Gamma_{\rm  adj}\big(n_f=4,n_s=6 \big) \,.
\label{rules}
\end{align}
In section \ref{sect:schemechange}, we will also give the result for $\Gamma _{\mathcal{N}=4}$ in the dimensional
reduction scheme.
 
A close examination of~(\ref{adj-3loops}) and~(\ref{resultc4}) shows that
the coefficients $\gamma_{ff}, \dots, \gamma_s$ describing $n_f$ and $n_s$-dependent contribution at three loops,
involve the same functions $\tilde{A}_1$, $\tilde{A}_2$ and $\tilde{A}_3$
that already appeared at two loops.
This suggests that these coefficients are not independent.
Indeed, we show in the next section that the  cusp anomalous dimension has an interesting hidden structure
that allows us to predict all $n_f$ and $n_s$-dependent terms at three loops at least.

Notice that all functions in (\ref{A-B}) except $B_5(x)$ depend on $y=1-x^2$ and, therefore, they are formally invariant
under $x\to -x$. However, due to the presence of the cut that runs along negative $x$, these functions acquire an additional
contribution under $x\to -x$ proportional to their discontinuity across the cut (see (\ref{Disc})). For the function $B_5(x)$ the
situation is slightly different. The linear combination of harmonic polylogarithms inside the brackets in $B_5(x)$ formally changes 
the sign under $x\to -x$. It is compensated however by the odd prefactor $x/(1-x^2)$, so that $B_5(x)$ has the same parity
properties as the other coefficient functions. In this way, we verify that our results for three-loop cusp anomalous dimension  
(\ref{QCD-3loops}) and (\ref{adj-3loops}) satisfy the relation (\ref{Disc}).

\section{Properties of the cusp anomalous dimension}
\label{sect:properties}

\subsection{Casimir scaling}
\label{sect:casimir}

Let us discuss the dependence of the cusp anomalous dimension on the $SU(N)$ color factors.
These factors appear as a result of manipulation with traces involving the $SU(N)$ generators
of various representations.
More precisely,
in the case of QCD we encounter the $SU(N)$ generators of three different representations:
fundamental  for fermions $(F)$,
adjoint for gluons $(A)$ 
and, in addition, some arbitrary representation $R$
that enters into the definition~(\ref{def}) of the cusped Wilson loop.
In the case of $\mathcal{N}=4$ SYM,
the generators in the fundamental representation do not appear
since all fields are defined in the adjoint of the $SU(N)$.

We observe from~(\ref{QCD-3loops})
that the dependence of the three-loop cusp anomalous dimension on the representation $R$
enters through an overall factor given by the quadratic Casimir of this representation $C_R= T^a T^a$,
the so-called Casimir scaling
\begin{equation}
\Gamma_{\rm cusp}(\phi,\alpha_s) = C_R \,\gamma(\phi,\alpha_s) + \mathcal{O}(\alpha_s^4)\,,
\label{C-scaling}
\end{equation}
with $\gamma(\phi,\alpha_s)$ being independent of $R$.
As was already mentioned in section~\ref{sect:NE},
we expect this scaling to be broken at four loops due to appearance of higher Casimirs.

\begin{figure}[tbp]
\centering
\includegraphics[width=\textwidth]{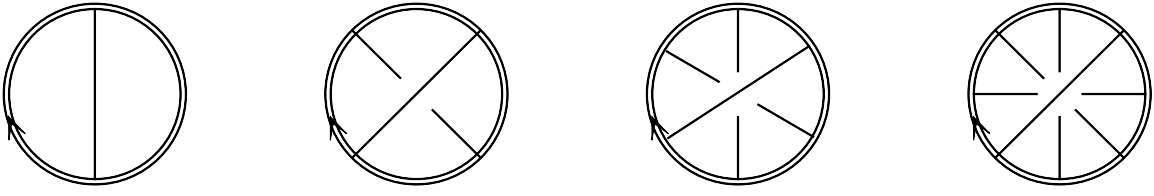}
\caption{Graphical representation of the color factors $C_1$, \dots, $C_4$.
Double line represents $\tr_R[T^{a_1}\cdots T^{a_n}]$,
solid line denotes $\delta^{a_i a_j}$.}
\label{color-fig}
\end{figure}

To understand this property,
let us examine possible color factors
that can appear in the perturbative expansion of Wilson loop~(\ref{def})
up to four loops.
To simplify the analysis we first examine supersymmetric Yang--Mills theory.
The color factor in this case consists of terms having the form $\tr_R[T^{a_1} \cdots T^{a_n}] C_{a_1 \dots a_n}$
with each $T^{a_i}$ corresponding to a gluon attached to the integration contour.
The tensor $C_{a_1 \dots a_n}$ is a product of $\delta_{a_i a_j}$ and $i f^{a_i a_j a_k}$ factors.%
\footnote{
There exists no subset of $i f^{a_i a_j a_k}$ without external indices
(such a subset would correspond to a vacuum subdiagram).}
So, there always exists a $i f^{a_i a_j a_k}$ factor directly contracted to the trace of $T^{a_i}$.
Substituting $i f^{abc} T^c = [T^a,T^b]$,
we transform such terms into the sum of two terms having  less $i f^{a_i a_j a_k}$ factors
(but more $T^{a_i}$ factors inside the trace).
Applying this procedure recursively, we finally reduce any color factor to a linear combination of terms
of the same form where all $C-$tensors are products of $\delta_{a_i a_j}$ only.
In this way, we obtain the basic color factors shown in figure~\ref{color-fig}.

The remaining color factors can be reduced to products and sums of the basic ones.
Going through the calculation we find
\begin{align}
C_1 ={}& \tr_R[T^a T^a]/N_R = C_R\,,
\nonumber\\
C_2 ={}& \tr_R[T^a T^b T^a T^b]/N_R = C_R (C_R-C_A/2)\,,
\nonumber\\
C_3 ={}& \tr_R[T^a T^b T^c T^a T^b T^c]/N_R = C_R (C_R-C_A/2) (C_R-C_A)\,,
\label{Cs}
\end{align}
where $N_R=\tr_R 1$ is the dimension of the representation and $f^{abc} f^{abd} = C_A\delta^{cd}$
with $C_A=N$ being the quadratic Casimir of the adjoint representation of $SU(N)$.
An important difference of $C_4$ compared to~(\ref{Cs}) is that
it cannot be expressed in terms of quadratic Casimirs only.
More precisely, its takes the form
\begin{equation}
C_4 = \tr_R[T^a T^b T^c T^d T^a T^b T^c T^d]/N_R
= \frac{d_R^{abcd} d_A^{abcd}}{N_R} + \dots\,,
\label{C4}
\end{equation}
where the ellipsis denotes terms involving quadratic Casimirs $C_R$ and $C_A$. Here $d_R^{abcd}$ and $d_A^{abcd}$ are fully symmetric tensors
\begin{align}\notag
d^{abcd} = \frac{1}{6} \tr\big[ {}& T^a T^b T^c T^d + T^a T^b T^d T^c + T^a T^c T^b T^d 
\\
+ &{}  T^a T^c T^d T^b + T^a T^d T^b T^c + T^a T^d T^c T^b\big]\,,
\label{d-tensor}
\end{align}
with the generators $T^a$ defined in two different representations.

The color factors $C_n$ appear in the expression for the cusp anomalous dimension starting from $n$ loops.
The very fact that~(\ref{C4}) is not proportional to the quadratic Casimir
for the generic $SU(N)$ representation $R$
implies that the Casimir scaling~(\ref{C-scaling}) should be violated at four loop
unless some miraculous cancellation happens
leading to the vanishing (angle dependent) coefficient function accompanying $C_4$.

Notice that the color factors~(\ref{Cs}) contain higher power of $C_R$.
As we explained in section~\ref{sect:NE}, in virtue of nonabelian exponentiation,
the cusp anomalous dimension should involve maximally nonabelian factors only.
Up to three loops they take the form $C_R$, $C_RC_A$ and $C_RC_A^2$.
This means that the cusp anomalous dimension depends on particular combinations
of the color factors, $C_1$, $C_2-C_1^2$ and $C_3+2 C_1^3-3 C_1 C_2$.
At four loops, the maximally nonabelian color factors are of two kinds,
$C_R C_A^3$ and $d_R^{abcd} d_A^{abcd}/N_R$.
The latter color factor leads to a violation of  the Casimir scaling~(\ref{C-scaling}) at four loops
and induces a nonplanar correction to the cusp anomalous dimension.

Let us now consider the color factors in QCD.
An important difference with the previous case is that the fermions are defined in the fundamental representation.
This leads to the appearance of additional color factors proportional to the number of fermion flavours $n_f$.
Each fermion loop produces a factor of $n_f$ and the maximal power of $n_f$ scales with the loop order.
In particular, the color factors linear in $n_f$ have the form
$n_f \tr_R[T^{a_1} \cdots T^{a_n}] \tr_F[T^{b_1} \dots T^{b_m}]  C_{a_1 \ldots a_n;b_1\ldots b_m}$,
with the $C$ tensor being given by a product of Kronecker symbols.
As in the previous case,
up to three loops $n_f$-dependent color factors can be expressed
in terms of quadratic Casimirs $C_R$, $C_A$ and $C_F$,
where $T^a T^a= C_F$ is the quadratic Casimir of the fundamental representation of $SU(N)$.
Most importantly, the additional $n_f$ dependence
does not affect the Casimir scaling~(\ref{C-scaling}) at three loops
but it modifies the form of the function $\gamma(\phi)$.
At four loops, we encounter the color factor $n_f d_R^{abcd}d_F^{abcd}/N_R$ analogous to~(\ref{C4}),
with the completely symmetric $d_F$ tensor given by~(\ref{d-tensor}) in the fundamental representation.
As before, it is not proportional to $C_R$ and, therefore, leads to violation of the Casimir scaling.

To summarize, the general expression
for the four-loop contribution to the cusp anomalous dimension
violating the Casimir scaling is
\begin{equation}
\Delta \Gamma_{\rm cusp} (\phi,\alpha_s) 
= \lr{\alpha_s\over \pi}^4 \left[
f_A(\phi) \frac{d_R^{abcd} d_A^{abcd}}{2N_R} + f_F(\phi) \,n_f \frac{d_R^{abcd} d_F^{abcd}}{2N_R} \right]
+ \mathcal{O}(\alpha_s^5)\,,
\label{non-planar}
\end{equation}
where $f_A(\phi)$ and $f_F(\phi)$ are some functions of the cusp angle depending on the choice of the gauge theory.
Here the second term inside the brackets is present only if fermions are defined in the fundamental representation,
e.g.\ $f_F(\phi)=0$ in $\mathcal{N}=4$ SYM.
In the special case of $R$ being the fundamental representation of the $SU(N)$, we have
\begin{equation}
\frac{d_F^{abcd} d_A^{abcd}}{2N_F} =  C_F \frac{N (N^2+6)}{48}\,,\qquad\qquad
\frac{d_F^{abcd} d_F^{abcd}}{2N_F} =  C_F\frac{N^4-6N^2+18}{96 N^2}\,,
\label{dd}
\end{equation}
with $C_F=(N^2-1)/(2N)$.
Since these color factors involve various powers of $N$,
the expression on the right-hand side of~(\ref{non-planar})
generates nonplanar corrections to the cusp anomalous dimension.

\subsection{Renormalization scheme change}
\label{sect:schemechange}

We recall that the three-loop calculation of the cusp anomalous dimension
has been performed using dimensional regularization (DREG).
However supersymmetry is broken in DREG
since for $D=4-2\epsilon$ the number of bosonic and fermionic degrees of freedom
do not match for $\epsilon\neq 0$.
To restore the supersymmetry, we can employ dimension reduction (DRED) \cite{Siegel:1979wq}.
In this scheme the gauge fields have four components in $D$ dimensions
and the difference with DREG comes from the contribution
of additional $(4-D)$ components of the gauge field, the so-called $\epsilon$-scalars.
Since the number of scalars $n_s$ is a free parameter in our calculation,
we can easily accommodate the contribution of $\epsilon$-scalars
by replacing $n_s\to n_s+2\epsilon$.

Additional complications arise due to necessity to introduce evanescent coupling constants
describing the self-interaction of $\epsilon$-scalars and their coupling with fermions.
In a generic gauge theory, the renormalization group evolution of the evanescent couplings
differs from that of the gauge coupling and, therefore, they have to be treated differently.
However, in a supersymmetric theory the beta-functions of these two sets of coupling
necessarily coincide allowing us to identify them at any scale.
In this case,
to compute the cusp anomalous dimension in the \DR{} scheme it suffices to replace
$n_s\to n_s+2\epsilon$ in expression~(\ref{lnZ}) for the $Z$ factor in the \MS{} scheme,
identify the residue at the pole $1/\epsilon$
and take into account the relation between the coupling constants in the two schemes~\cite{Harlander:2006rj}
\begin{equation}
\left.\alpha^{\DR}_{s} \right|_{\rm QCD}= \alpha_s^{\MS} \left[1
+ \frac{\alpha_s^{\MS}}{\pi} \frac{C_A}{12}
+ \lr{\frac{\alpha_s^{\MS}}{\pi}}^2
\lr{\frac{11}{72} C_A^2 - \frac{1}{8} C_F T_F n_f} + \mathcal{O}(\alpha_s^3)\right]\,,
\label{coup1}
\end{equation}
for fermions in the fundamental representation, and
\begin{equation}
\left.\alpha^{\DR}_{s}\right|_{\rm adj} = \alpha_s^{\MS} \left[1
+ \frac{\alpha_s^{\MS}}{\pi} \frac{C_A}{12}
+ \lr{\frac{\alpha_s^{\MS}}{\pi}}^2C_A^2
\lr{\frac{11}{72}   - \frac{n_f}{16} } + \mathcal{O}(\alpha_s^3)\right]\,,
\label{coup}
\end{equation}
for fermions in the adjoint representation. 
Notice that scalars do not contribute to (\ref{coup}) at three loops.
This leads to the following relation for the cusp anomalous anomalous dimension in the two schemes
\begin{equation}
\Gamma_{\rm cusp}^{\DR}(\phi,\alpha_s^{\DR}) = \Gamma_{\rm cusp}^{\MS}(\phi,\alpha_s^{\MS})\,.
\label{schemes}
\end{equation}
In the special case of $\mathcal{N}=4$ SYM theory, for $n_s=6$ and $n_f=4$,
we use the relations~(\ref{coup}) and~(\ref{schemes}) together with~(\ref{adj-3loops})
to find the three-loop cusp anomalous dimension in \DR{} scheme
\begin{align}
\Gamma_{\mathcal{N}=4}^{\DR}(\phi,\alpha_s) =  C_R {}&\bigg[\frac{\alpha_s }{\pi} \tilde{A}_1
+ \frac{1}{2} \lr{\frac{\alpha_s  }{\pi}}^2 N (\tilde{A}_2 +\tilde{A}_3)
\nonumber\\
&{} + \frac{1}{4} \lr{\frac{\alpha_s}{\pi}}^3 N^2
(- \tilde{A}_2 + \tilde{A}_4 + \tilde{A}_5 + \tilde{B}_3 + \tilde{B}_5)\bigg]  \,.
\label{DR-3loops}
\end{align}
This confirms a conjecture made in our previous paper \cite{Grozin:2014hna}.
 
\subsection{Asymptotics for large cusp angles}

To examine the limit of large Minkowskian angles, we substitute $\phi=i \phi_{M}$,
or equivalently $x=\e^{-\phi_{M}}$, and put $x\to 0$.
In this limit, the cusp anomalous dimension is expected to have a logarithmic behaviour~\cite{Korchemsky:1985xj,Korchemsky:1991zp}
\begin{equation}
\Gamma_{\rm cusp}(\phi,\alpha_s) = K(\alpha_s) \log(1/x) + \mathcal{O}(x^0)\,,
\label{large}
\end{equation}
with $K(\alpha_s)$ the so-called light-like cusp anomalous dimension.

We use~(\ref{QCD-3loops}) to find at three loops in QCD
\begin{align}\notag \label{K}
K^{\MS}_{\rm QCD}(\alpha_s)  =C_R  \biggl\{ {}&  \frac{\alpha_s}{\pi}+ \lr{\frac{\alpha_s}{\pi}}^2\bigg[C_A  \left( \frac{67}{36} - \frac{\pi^2}{12}  \right)
-\frac{5}{9} T_F n_f\bigg]
\\ \notag
+ {}& \lr{\frac{\alpha_s}{\pi}}^3\bigg[  
 C_A^2 \biggl( \frac{245}{96} - \frac{67 \pi^2}{216} + \frac{11 \pi^4}{720} + \frac{11}{24} \zeta_3 \biggr)- \frac{1}{27} (T_F n_f)^2 
\\ 
+ {}& C_A T_F n_f \left( -\frac{209}{216} + \frac{5 \pi^2}{54} - \frac{7}{6}\zeta_{3} \right)
+ C_F T_F n_f \left( \zeta_3 -\frac{55}{48} \right) \bigg]\bigg\}.
\end{align}
We verify that this expression is in perfect agreement with the known result~\cite{Korchemsky:1987wg,Moch:2004pa}. 

In a similar manner,
we obtain an analogous expression in a supersymmetric Yang-Mills theory, with $n_f$ fermions and $n_s$ scalars in the adjoint
representation, and, then, convert the result into the \DR{} scheme with a help of~(\ref{coup}) to get
\begin{align}\notag \label{K1}
K^{\DR}_{\rm adj}(\alpha_s)  =C_R  \biggl\{ {}&  \frac{\alpha_s}{\pi}+ \lr{\frac{\alpha_s}{\pi}}^2C_A   \left( \frac{16}{9} - \frac{\pi^2}{12}-\frac5{18} n_f -\frac{n_s}9  \right)
\\ \notag
+ {}& \lr{\frac{\alpha_s}{\pi}}^3 C_A^2 \bigg[  
 \frac{1817}{864} - \frac{8 \pi^2}{27} + \frac{11 \pi^4}{720} + \frac{11}{24} \zeta_3  - \frac{n_f ^2}{108} +{n_s^2\over 432} + {7\over 32} n_f n_s
\\ 
{}& \qquad\quad  +   n_f \left( -\frac{91}{96} + \frac{5 \pi^2}{108} - \frac{\zeta_{3}}{12}  \right)
+   n_s \left(-{1007\over 1728} + {\pi^2\over 54}-{\zeta_3\over 48} \right) \bigg]\bigg\}\,.
\end{align}
To obtain from this expression the three-loop light-like cusp anomalous dimension in $\mathcal N=4$ SYM, we
adjust the parameters following~(\ref{rules}),
\begin{equation}
K^{\DR}_{\mathcal{N}=4}(\alpha_s) = C_R \biggl[ \frac{\alpha_s}{\pi}
- \frac{\pi^2}{12} \lr{\frac{\alpha_s}{\pi}}^2 C_A
+ \frac{11}{720} \pi^4 \lr{\frac{\alpha_s}{\pi}}^3 C_A^2 \biggr]
+ \mathcal{O}\lr{\alpha_s^4}\,,
\label{K-N=4}
\end{equation}
in agreement with~\cite{Kotikov:2004er}.

\subsection{Universal scaling function}

We can use the large angle asymptotics of the cusp anomalous dimension~(\ref{large})
to introduce a new effective coupling constant $a$:  \footnote{It is also known in QCD literature as physical coupling constant \cite{Catani:1990rr}.}
\begin{align}\label{a}
a = {\pi\over C_R}  K(\alpha_s) = \alpha_s \left[1 + {\alpha_s\over \pi} K^{(1)} + \lr{\alpha_s\over \pi}^2 K^{(2)} + O(\alpha_s^3) \right].
\end{align}
Inverting this relation we can expand the cusp anomalous dimension in powers of $a$
and define the following function
\begin{equation}\label{Omega}
\Omega(\phi,a) := \Gamma_{\rm cusp}(\phi,\alpha_s)\,,
\end{equation}
where $\Gamma(\phi,\alpha_s)$ and $K(\alpha_s)$ are evaluated in the same scheme.
The expansion coefficients of the two functions are related to each other as
\begin{align}
\Gamma_{\rm cusp}(\phi,\alpha_s) ={}& \frac{\alpha_s}{\pi} \Omega^{(1)}
+ \lr{\frac{\alpha_s}{\pi}}^2 \Bigl(\Omega^{(2)} + K^{(1)} \Omega^{(1)}  \Bigr)
\nonumber\\
&{} + \lr{\frac{\alpha_s}{\pi}}^3
\Bigl(\Omega^{(3)} + 2 K^{(1)} \Omega^{(2)} + K^{(2)} \Omega^{(1)}\Bigr)
+\mathcal{O}(\alpha_s^4)\,,
\label{Gamma-Omega}
\end{align}
with $\Omega^{(i)}(\phi)$ being the coefficients of the expansion of $\Omega(\phi,a)$
in powers of $a/\pi$.
According to~(\ref{schemes}), the change of the renormalization scheme (from \MS{} to \DR)
amounts to a finite renormalization of the coupling constant.
An immediate consequence of~(\ref{schemes}) is that the coefficients $\Omega^{(i)}$
are the same in the two renormalization schemes.
This is not the case however for the expansion coefficients of the cusp anomalous dimension,
$\Gamma^{(i)}$ and $K^{(i)}$.

Let us first compute the function $\Omega(\phi,a)$ in $\mathcal{N}=4$ SYM.
Using~(\ref{DR-3loops}) and~(\ref{K-N=4}), we obtain from~(\ref{Gamma-Omega})
\begin{align}
\Omega(\phi,a) ={}&C_R\bigg[ \frac{a}{\pi}   \tilde{A}_1
+ \lr{\frac{a}{\pi}}^2 \frac{C_A}{2}
\left(\frac{\pi^2}{6} \tilde{A}_1 + \tilde{A}_2 + \tilde{A}_3\right)
\nonumber\\
&{} + \lr{\frac{a}{\pi}}^3 \frac{C_A^2}{4}
\left( - \tilde{A}_2 + \tilde{A}_4 + \tilde{A}_5 + \tilde{B}_3 + \tilde{B}_5
- \frac{\pi^4}{180} \tilde{A}_1 + \frac{\pi^2}{3} (\tilde{A}_2 + \tilde{A}_3) \right)\bigg]\,.
\label{Omega-exp}
\end{align}
By construction, this function takes the same form in \MS{} and \DR{} schemes.

Similarly, we can apply the relations~(\ref{adj-3loops})
 and~(\ref{K})
to compute the corresponding function $\Omega(\phi,a)$ in QCD
and in a generic Yang--Mills theory containing fermions and scalars.
Since the cusp anomalous dimension depends on the particle content of the theory,
we should expect to find different results for $\Omega(\phi,a)$.
Using the obtained results for the cusp anomalous dimension,
we found that the function $\Omega(\phi,a)$ is independent on the number of fermions and scalars!

This remarkable property immediately implies that, at least to three loops,
the function $\Omega(\phi,a)$ is the same in any gauge theory,
\begin{equation}
\Omega_{\mathcal{N}=4}(\phi,a) = \Omega_{\text{QCD}}(\phi,a) = \Omega_{\text{YM}}(\phi,a)\,.
\label{univ}
\end{equation}
Combining this relation with~(\ref{Gamma-Omega}),
we conclude that all $n_f$ and $n_s$ dependent terms in $\Gamma(\phi,\alpha_s)$
are generated from lower-loop terms through expansion of $K(\alpha_s)$ in powers of $\alpha_s$.
It would be interesting to elucidate the origin of the relation~(\ref{univ})
as well as its validity beyond three loops.

We would like to mention that similar phenomenon has been also observed in other supersymmetric Yang-Mills theories. 
In particular, various quantities in three-dimensional $\mathcal N=6$ supersymmetric ABJM theory \cite{Gromov:2008qe} and in
$\mathcal N = 2$ superconformal Yang-Mills theory   \cite{Mitev:2014yba,Mitev:2015oty} can be obtained from their counter 
partners in $\mathcal N=4$ SYM  by replacing the coupling constant by the universal `effective' coupling. 

Let us examine the properties of the function $\Omega(\phi,a)$.

In the large angle limit, for $\phi=-i \log x$ with $x\to 0$,
we combine together~(\ref{large}) and~(\ref{Omega}) to see
that $\Omega(\phi,a)$ has universal asymptotic behavior
\begin{equation}
\Omega(\phi,a) = \frac{a}{\pi} C_R \log (1/x) + \mathcal{O}(x^0)\,,
\label{Omega-large}
\end{equation}
where the coefficient in front of the logarithm does not receive corrections
and is one-loop exact, that is $\Omega^{(i)}=\mathcal{O}(x^0)$ for $i\ge 2$.
Matching this relation into~(\ref{Omega-exp}),
we find that the linear combinations of $\tilde{A}$ and $\tilde{B}$ functions
that appear in the expansion of $\Omega(\phi,a)$ at two and three loops
remain finite in the large angle limit.

In the small angle limit, for $\phi\to 0$,
the integration contour in figure~\ref{cusp-fig} reduces to the straight line
leading to the vanishing of the cusp anomalous dimension.
For small cusp angle $\phi$ we expect that
\begin{equation}
\Omega(\phi,a) = -\phi^2 B_\Omega(a) + \mathcal{O}(\phi^4)\,,
\label{B-Omega}
\end{equation}
where $B_\Omega(a)$ is an analog of the bremsstrahlung function~(\ref{br}).
We use~(\ref{Omega-exp}) to obtain the three-loop result
\begin{align}
B_\Omega(a) ={}& C_R \biggl[ \frac{a}{3\pi}
+ \left(\frac{a}{\pi}\right)^2 \frac{C_A}{4} \left(1 - \frac{\pi^2}{9}\right)
\nonumber\\
&\hphantom{C_R \biggl[\biggr.}{}
+ \left(\frac{a}{\pi}\right)^3 \frac{C_A^2}{12}
\left( - \frac{5}{3} - \frac{\pi^2}{6} + \frac{\pi^4}{20} - \zeta_3 \right) \biggr]
+ \mathcal{O}(a^4)\,.
\label{B3loops}
\end{align}
As before this function takes the same form in any gauge theory (at three loops at least)
and does not depend on the choice of the renormalization scheme.

Substituting~(\ref{B-Omega}) into~(\ref{Omega}) we find for the bremsstrahlung function~(\ref{br})
\begin{equation}
B(\alpha_s) = B_\Omega(a)\,, \qquad C_R \frac{a}{\pi} = K(\alpha_s)\,.
\label{BB}
\end{equation}
Then, we use the obtained three-loop results~(\ref{B3loops}) and~(\ref{K}) to get in QCD
\begin{align}\notag
B_{\rm QCD}^{\MS} (\alpha_s) = C_R {}& \biggl\{ \frac{\alpha_s}{3\pi}
+ \lr{\frac{\alpha_s}{\pi}}^ 2\bigg[ C_A \left(\frac{47}{54} - \frac{\pi^2}{18}  \right)
- \frac{5}{27} T_F n_f\bigg] 
\\
+{}& \lr{\frac{\alpha_s}{\pi}}^3 \bigg[ C_A^2 \lr{\frac{473}{288} - \frac{85}{324} \pi^2 + \frac{\pi^4}{72} + \frac{5}{72} \zeta_3}
- \frac{1}{81} (T_F n_f)^2
\nonumber\\
&{} + C_A T_F n_f \lr{- \frac{389}{648} + \frac{5}{81} \pi^2 - \frac{7}{18} \zeta_3}
+ C_F T_F n_f \lr{- \frac{55}{144} + \frac{\zeta_3}{3}} \bigg]\bigg\}\,.
\end{align}
The two-loop correction to $B_{\rm QCD}^{\MS} (\alpha_s) $ agrees with~\cite{Korchemsky:1987wg,Korchemsky:1991zp},
the three-loop result   is new.
In $\mathcal{N}=4$ SYM we find from~(\ref{BB}) and~(\ref{K-N=4})
\begin{align}
B_{\mathcal N=4}^{\DR}(\alpha_s) =  C_R \biggl[{}& \frac{\alpha_s}{3\pi}
+ \left(\frac{\alpha_s}{\pi}\right)^2 \frac{C_A}{4} \left(1 - \frac{2\pi^2}{9}\right)
\nonumber\\
&{} + \left(\frac{\alpha_s}{\pi}\right)^3 \frac{C_A^2}{12}
\left( - \frac{5}{3} - \frac{2\pi^2}{3} + \frac{\pi^4}{6} -\zeta_3 \right) \biggr]
 \,.
\end{align}

\subsection{The relation to the quark-antiquark potential}\label{sect:pot}

As another check of our results, let us consider the limit $\phi=\pi-\delta$ with $\delta\to 0$, or equivalently
$x=\e^{i(\pi-\delta)}\to -1$. In this limit, the two rays forming the cusp become anti-parallel and the one-cusp
anomalous dimension \re{1-loop} develops a pole $\Gamma^{(1)}\sim - C_R \pi/\delta$. It is expected that
the cusp anomalous dimension should have the same behaviour up to three loops, whereas at four loops it
receives corrections of the form $(\log\delta)/\delta$  \footnote{It is interesting to note that for the locally supersymmetric
Wilson loop similar corrections appear in the cusp anomalous dimension already at two loops \cite{Erickson:1999qv,Pineda:2007kz}.}
\begin{align}\label{V-cusp}
\Gamma_{\rm cusp}(\pi-\delta,\alpha_s) \stackrel{\delta\to 0}{\sim} -C_R {\alpha_s\over \delta}V_{\rm cusp} (\alpha_s) + O(\alpha_s^4 \log\delta /\delta),
\end{align}
with $V_{\rm cusp} (\alpha_s)= 1+ \lr{\alpha_s/\pi} V^{(1)} + \lr{\alpha_s/\pi}^2 V^{(2)}$ depending on the renormalization
scheme.

As before, it is convenient to examine the asymptotic behavior of the universal function $\Omega(\phi,a)$.
Indeed, we find from \re{Omega-exp} that it develops a pole $1/\delta$ at three loops
\begin{align}\label{Omega-pole}
\Omega(\pi-\delta,a) \stackrel{\delta\to 0}{\sim} -C_R {a\over \delta}\left[1- {a\over\pi} C_A\lr{1-{\pi^2\over 12}} + \lr{a\over\pi}^2C_A^2
\lr{{5\over 4} + {\pi^2\over 12} - {49\,\pi^4 \over 2880} }\right] + O(a^4),
\end{align}
We note that this relation  comes about as a result of nontrivial cancellation of more singular contributions coming from 
various terms in \re{Omega-exp}. See discussion in section \ref{sect:results}.

We substitute \re{Omega-pole} into \re{Gamma-Omega} and use the three-loop result for the light-like cusp anomalous dimension \re{K} and \re{K-N=4} to verify that the cusp anomalous dimension  satisfies \re{V-cusp} in QCD and in $\mathcal N=4$ SYM. The
corresponding functions $V_{\rm cusp} (\alpha_s)$  are given  by 
\begin{align} \notag
V_{\rm cusp,\,QCD}^{\overline{\rm MS}} {}&= 1+ {\alpha_s\over\pi} \lr{{31\over 36} C_A - {5\over 9} n_f T_F} + \lr{\alpha_s\over\pi}^2 \bigg[
C_A^2 \lr{{23\over 288}+{\pi^2\over 4} -{\pi^4\over 64}+{11\over 24}\zeta_3}
\\[1.5mm]  \label{qq}
{}& \qquad -{1\over 27} (n_f T_F)^2 +C_F n_f T_F \lr{\zeta_3-{55\over 48}}+ C_An_fT_F\lr{-{7\over 6}\zeta_3+{31\over 216}}\bigg]\,,
\\[3mm]
\label{qq-N4}
V_{\rm cusp,\,\mathcal N=4}^{\overline{\rm DR}} {}&= 1- {\alpha_s\over\pi} C_A  + \lr{\alpha_s\over\pi}^2 C_A^2\lr{\frac54 + { \pi^2 \over 4} - {\pi^4\over 64}} + O(\alpha_s^3)\,.
\end{align}

Let us compare the relation \re{V-cusp} with an analogous expression for color-singlet contribution to the static potential of two heavy color sources carrying 
the $SU(N)$ charge $C_R$
in generic Yang-Mills theory. In the momentum representation, it has the form
\begin{align}
V_R(\boldsymbol{q}) = -C_R {4\pi  \alpha_s(\boldsymbol{q}^2) \over   \boldsymbol{q}^2} V_{Q\bar Q} \left(\alpha_s( \boldsymbol{q}^2) \right)
\end{align}
where the function $V_{Q\bar Q}$ depends on the coupling constant normalized at the scale $\mu^2=  \boldsymbol{q}^2$
\begin{align}\label{V-gen}
V_{Q\bar Q}(\alpha_s) =1+ {\alpha_s\over 4\pi} a_1 +\lr{\alpha_s\over 4\pi}^2 a_2 +O(\alpha_s^3)  \,,
\end{align}
with the expansion coefficients $a_1$ and $a_2$ known both in QCD \cite{Peter:1996ig,Peter:1997me,Schroder:1998vy} and in $\mathcal N=4$ SYM \cite{Prausa:2013qva}. In the coordinate representation, the 
potential is given by 
\begin{align}\label{poten-R}
V_R(r) =\int {d^3\boldsymbol{q} \over (2\pi)^3}\e^{i \boldsymbol{q}\boldsymbol{r}} V_R(\boldsymbol{q}) = -C_R {\bar\alpha_s  \over   r} \big[V_{Q\bar Q}  (\bar\alpha_s  )
+\Delta V(\bar\alpha_s)\big]\,,
\end{align}
with    $\bar\alpha_s=\alpha_s(\mu^2 =  \e^{-2\gamma_{\rm E}}/r^2)$ and $\Delta V(\bar\alpha_s)$ is proportional to the beta-function
\begin{align}\label{DeltaV}
\Delta V(\alpha_s) = {\pi^2\over 3} \lr{\alpha_s\over 4\pi}^2 \beta_0^2 + O(\alpha_s^3)\,.
\end{align}
As was observed in \cite{Kilian:1993nk}, the one-loop correction to \re{qq} coincides with analogous correction to heavy quark-antiquark
static potential  \re{V-gen}  in QCD, i.e. $a_{1,\rm QCD}^{\overline{\rm MS}}={31\over 36} C_A - {5\over 9} n_f T_F$. Of course, the coincidence is not accidental and can be understood in the conformal limit of QCD. 

Namely, for small $\delta$ we can define a conformal transformation $x\to y$ that maps two almost antiparallel semi-infinite rays, shown
in Figure~\ref{cusp-fig} for $\phi=\pi-\delta$, into two (infinite) lines separated by distance $\delta$. To show this, we assume that 
the cusp point is located at the origin and introduce the radial and angular coordinates $x_0=r\cos\phi$, $
\vec x= r \vec n \sin\phi$ with $x_\mu^2=r^2=e^{2\rho}$ and $\phi=\pi-\delta$, so that the metric takes the form
\begin{align}
ds^2 = dx_0^2 + d\vec x^2 = e^{2\rho}\left[d\rho^2 + d\delta^2 + d\vec n^2 (\sin\delta)^2 \right]
\sim e^{2y_0} \left(dy_0^2 + d\vec y^{\,2}\right).
\end{align}
where in the last relation we took $\delta\to 0$ and introduced new coordinates $y_0=\rho$ and $\vec y= \vec n \, \delta$.
As follows from the last relation, the transformation $x\to y$ is conformal at small~$\delta$.
 
If the conformal symmetry were exact, as  it happens in $\mathcal N=4$ SYM theory,
the conformal transformation $x\to y$ would allow us to identify 
the Wilson loops evaluated in two different configurations, thus leading to the expected relation between the cusp
anomalous dimension \re{V-cusp}  and the static potential  \re{poten-R} for $r=\delta$
\begin{align}\label{noanomaly}
 V_{\rm cusp,\,\mathcal N=4}(\alpha_s)= V_{{Q\bar Q},\mathcal N=4}(\alpha_s) \,.
\end{align}
Note that, in virtue of conformal symmetry, the coupling constant does not depend on the renormalization scale, 
$\bar\alpha_s=\alpha_s$, and, in addition, $\Delta V=0$ in \re{poten-R}. 

In the case of QCD, the conformal symmetry is broken by a nonzero beta function. As a consequence, the Wilson
loop receives additional, conformal symmetry breaking corrections under the transformation $x\to y$ which generate
the difference between the cusp anomalous dimension and the static potential in QCD. Since these corrections are 
necessarily proportional to the beta-function, we expect that the difference between \re{V-cusp} and \re{poten-R} (for $r=\delta$)
should be also proportional to $\beta(\alpha_s)$, see e.g. \cite{Braun:2003rp}.~\footnote{The situation here is similar to 
that for the Crewther relation in QCD. The conformal symmetry breaking corrections to this relation have been
studied in \cite{Kataev:2010du,Kataev:2014zha} }

Notice that the expansion of the cusp anomalous dimension \re{V-cusp} and the static potential \re{poten-R}
runs in powers of coupling constant normalized at different scales, $\alpha_s(\mu^2)$ and $\alpha_s(\e^{-2\gamma_{\rm E}}/r^2)$,
respectively. In agreement with our expectations, the difference between the two couplings is proportional to beta-function 
multiplied by logarithms of the ratio of the two scales. Choosing $\mu^2=\e^{-2\gamma_{\rm E}}/r^2$
we can eliminate such logarithms and arrive at the following relation \footnote{We did not include $\Delta V(\alpha_s)$ into this relation since, by definition \re{DeltaV}, this function is proportional to beta-function and, therefore, can be absorbed into 
$C(\alpha_s)$.}
\begin{align}\label{anomaly}
V_{\rm cusp,\,\rm QCD}(\alpha_s)-V_{{Q\bar Q},\rm QCD}(\alpha_s)=  \beta(\alpha_s) C(\alpha_s) \,,
\end{align}
with $\beta(\alpha_s) = (\frac{11}3 C_A-\frac43 T_F n_f)  {\alpha_s/(4\pi)} + O(\alpha_s^2)$ and $C(\alpha_s)$ being some function
of the coupling constant.  

The relations \re{noanomaly} and \re{anomaly} can be tested using the known two-loop result for
 the static potential \re{V-gen} in $\mathcal N=4$ SYM \cite{Prausa:2013qva} and in QCD \cite{Peter:1996ig,Peter:1997me,Schroder:1998vy}. Replacing $V_{\rm cusp}(\alpha_s)$
 by its expressions \re{qq} and \re{qq-N4}, we verified the relations \re{noanomaly} and \re{anomaly} and 
identified the lowest order correction to $C(\alpha_s)$ in the  $\overline{\rm MS}$ scheme
\begin{align}\label{C-fun}
C^{\overline{\rm MS}}(\alpha_s) = {\alpha_s\over\pi} \lr{-{47\over 27}C_A + {28\over 27}\, n_f T_F} + O(\alpha_s^2)\,,
\end{align}
where $O(\alpha_s^2)$ term depends on $\Gamma_{\rm cusp}$ at four loops.
\footnote{The simple form of relation \re{anomaly}  suggests that there should exist another, direct way of computing the
conformal anomaly $C(\alpha_s)$ for the cusped Wilson loop in the $\delta\to 0$ limit.}

It was found in \cite{Smirnov:2008pn,Anzai:2009tm,Smirnov:2009fh} that the three-loop correction to the static potential $V_{\rm QCD}(r)$ involves higher  $SU(N)$
Casimirs defined in \re{dd}. As we argued in section \ref{sect:casimir}, the same happens for the cusp anomalous dimension 
\re{non-planar}  at four loops.
Applying \re{anomaly} we can relate the corresponding terms order-by-order in the coupling. 
In particular, assuming that the two-loop correction to \re{C-fun} does not involve higher Casimirs, we
can use \re{anomaly} to predict the four-loop correction to $\Gamma_{\rm cusp}(\pi-\delta)$ proportional to higher Casimirs
in  $\delta\to 0$ limit. Together with \re{non-planar} this leads to the following asymptotic behavior of the functions 
$f_A(\pi-\delta)$ and $f_F(\pi-\delta)$ for $\delta\to 0$
\begin{align}\label{f-as}
f_A(\pi-\delta) \ \sim \ -{\kappa_A\over 64 \,\delta}\,,\qquad\qquad f_F(\pi-\delta) \ \sim \  -{\kappa_F\over 64 \,\delta}\,,
\end{align}
with the numerical coefficients $\kappa_A$ and $\kappa_F$
obtained in \cite{Smirnov:2008pn,Anzai:2009tm,Smirnov:2009fh} by direct Feynman diagram (numerical) calculation 
\begin{align}\label{kappa}
\kappa_A=-136.39(12)\,,\qquad\qquad
\kappa_F=-56.83(1)\,.
\end{align}

\subsection{Nonplanar corrections at four loops}

We recall that nonplanar corrections first appear in $\Gamma(\phi,\alpha_s)$
at four loops and have the general form~(\ref{non-planar}).
Applying~(\ref{Gamma-Omega}) we can relate them to nonplanar correlations to the function $\Omega(\phi,a)$
and to the light-like cusp anomalous dimension (\ref{a})
\begin{align}\notag
 \Delta\Omega(\phi,\alpha_s) {}& =\lr{\frac{\alpha_s}{\pi}}^4  \Delta\Omega^{(4)} \,,
\\\notag
  \Delta K(\alpha_s)\phantom{\phi} {}&= \lr{\frac{\alpha_s}{\pi}}^4 C_R\Delta  K^{(3)} \,,
\\
 \Delta \Gamma(\phi,\alpha_s) {}& = \lr{\frac{\alpha_s}{\pi}}^4 \Bigl(\Delta\Omega^{(4)}+\Omega^{(1)} \Delta K^{(3)}\Bigr)  \,,
\label{Delta1}
\end{align}
with $\Omega^{(1)}=C_R \tilde{A}_1$.

In general, four-loop nonplanar corrections $\Delta\Omega^{(4)}$ and
$C_R\Delta  K^{(3)}$ have the same form as~(\ref{non-planar}) and are given by a sum of two higher Casimirs.
Notice that one of the Casimirs is accompanied by the factor of $n_f$.
Assuming that~(\ref{univ}) is valid at four loops,
we find that $\Delta\Omega^{(4)}$ should be $n_f$ independent and, therefore,
involve only one Casimir leading to
\begin{align}
 \Delta\Omega^{(4)} = {}& f_\Omega(\phi) \frac{d_R^{abcd} d_A^{abcd}}{2N_R}\,,
\nonumber\\
 \Delta K^{(3)} ={}& K_A \frac{d_R^{abcd} d_A^{abcd}}{2N_R C_R} + K_F n_f \frac{d_R^{abcd} d_F^{abcd}}{2N_R C_R}\,,
\label{f-Omega}
\end{align}
with $K_A$ and $K_F$ independent of the cusp angle as well as of the number of flavours $n_f$.
Substituting these relations into~(\ref{Delta1}) and matching the resulting expression into~(\ref{non-planar}) we obtain
\begin{align}
f_A(\phi) ={}& f_\Omega(\phi) +  K_A \tilde{A}_1(\phi)\,,
\nonumber\\[2mm]
f_F(\phi) ={}& K_F \tilde{A}_1(\phi)\,.
\label{ff}
\end{align}
Since $K_F$ does not depend on $\phi$, we can fix its value by examining
the asymptotic behavior of the both sides of the last relation for $\phi\to\pi$.
Taking into account~(\ref{f-as}) together with $\tilde{A}_1(\pi-\delta)\sim -1/\delta$ we get
\begin{equation}
K_F = \frac{\kappa_F}{64}\,,
\end{equation}
with $\kappa_F$ given by~(\ref{kappa}).
This leads to the following prediction for the $n_f$ dependent part of nonplanar correction~(\ref{non-planar})
to the cusp anomalous dimension
\begin{equation}\label{pred}
f_F(\phi) = \frac{\kappa_F}{64} \tilde{A}_1(\phi)\,.
\end{equation}

In distinction with $f_F(\phi)$, the expression for $f_A(\phi)$ in~(\ref{ff}) involves in addition
the function $f_\Omega(\phi)$ defined in~(\ref{f-Omega}).
Although the explicit form of the function $f_\Omega(\phi)$ is unknown,
we can use~(\ref{f-Omega}) and~(\ref{ff}) to deduce some of its properties.
Namely, examining the asymptotic behavior of both sides of the first relation in~(\ref{ff})
for $\phi=\pi-\delta$ with $\delta\to0$ we find that this function has to satisfy
\begin{equation}
f_\Omega(\pi-\delta) \sim - \frac{1}{\delta} \lr{\frac{\kappa_A}{64} - K_A}\,.
\end{equation}
In addition, in the large angle limit, for $\phi=-i\log x$ with $x\to 0$,
it follows from~(\ref{Omega-large}) and~(\ref{f-Omega}) that $\Delta \Omega^{(4)}$ should stay finite
in this limit leading to $f_\Omega(\phi)=\mathcal{O}(x^0)$.
This property excludes the possibility for $f_\Omega(\phi)$ to be proportional to $\tilde{A}_1$.
 
To summarize,  we demonstrated in this subsection that assuming the validity of~(\ref{univ}) at four loops
leads to a definite prediction  (\ref{pred}) for $n_f$ dependent part of the nonplanar correction to the cusp anomalous dimension
 (\ref{non-planar}).

\subsection{Comparison with the supersymmetric cusp anomalous dimension}

It is instructive to compare (\ref{DR-3loops}) with the analogous result for the supersymmetric Wilson loop (\ref{superW})
\begin{align}
\mathnormal{\Gamma} (\phi,\theta,\alpha_s) =  C_R  \bigg[\frac{\alpha_s }{\pi} \mathnormal{\Gamma}^{(1)} 
+ \frac12 \lr{\frac{\alpha_s}{\pi}}^2 N \mathnormal{\Gamma}^{(2)} + \frac14 \lr{\frac{\alpha_s  }{\pi}}^3 N^2 \mathnormal{\Gamma}^{(3)}\bigg]  ,
\end{align}
with $\mathnormal{\Gamma}^{(1)}$, $\mathnormal{\Gamma}^{(2)}$ and $\mathnormal{\Gamma}^{(3)}$ defined in (\ref{susy-Gamma}).   
In comparison with  (\ref{DR-3loops}), this expression depends on the internal cusp angle $\theta$ on $S^5$. 

The $\theta-$dependence enters into (\ref{susy-Gamma}) through $\xi_0$ given by (\ref{xi0}). For $\theta=\pi/2$ we find 
\begin{align}
\xi_0\left(\phi,{\pi\over 2}\right) =\xi(\phi)= i \cot\phi ={1+x^2\over 1-x^2}\,,
\end{align}
leading to $\mathnormal{\Gamma}^{(1)} =A_1$, $\mathnormal{\Gamma}^{(2)} =A_3$ and $\mathnormal{\Gamma}^{(3)} =A_5$. In this
way, we arrive at
\begin{align}
\mathnormal{\Gamma} (\phi,\pi/2,\alpha_s) =  C_R  \bigg[\frac{\alpha_s }{\pi} A_1
+ \frac12 \lr{\frac{\alpha_s}{\pi}}^2 N A_3 + \frac14 \lr{\frac{\alpha_s  }{\pi}}^3 N^2 A_5\bigg]  \,.
\end{align}
Recalling that $\tilde A_i=A_i(x)-A_i(1)$, we observe that this expression involves the same coefficient functions  as (\ref{DR-3loops}). However,  
in distinction with (\ref{DR-3loops}), it does not vanish for $\phi=0$ but  for $\phi=\theta=\pi/2$, or equivalently $x=i$. Then, defining
\begin{align}
\widetilde {\mathnormal{\Gamma}} (\phi,\alpha_s) = \mathnormal{\Gamma} (\phi,\pi/2,\alpha_s) - \mathnormal{\Gamma} (0,\pi/2,\alpha_s)
\end{align}
we find  
\begin{align}
\Gamma_{\mathcal{N}=4}^{\DR}(\phi,\alpha_s) - \widetilde {\mathnormal{\Gamma}} (\phi,\alpha_s) =  \frac{1}{2}C_R N \lr{\frac{\alpha_s  }{\pi}}^2 \bigg[   \tilde{A}_2  
 +  {\frac{\alpha_s}{2\pi}}  N 
(- \tilde{A}_2 + \tilde{A}_4   + \tilde{B}_3 + \tilde{B}_5)\bigg]  
\end{align}
It is interesting to analyze the properties of the terms on the right-hand side of this equation.

First of all, in the light-like limit $x \to 0$, they have to give a finite limit, since the scalar coupling to the supersymmetric Wilson loop  (\ref{superW}) is suppressed in this limit. Indeed, we observe that $\tilde{A}_{2}$, $\tilde{A}_{4}$, $\tilde{B}_{3}$ and $\tilde{B}_{5}$ all go to constants or vanish in this limit.
Second, by definition, they are also well-behaved in the small angle limit, where they modify the coefficients in the Taylor expansion. Third, the limit of the backtracking Wilson line is more interesting.
At two loops, the function $\tilde{A}_2$ has a term $\propto \log \delta /\delta$, which is required to cancel a corresponding term in $\tilde{A}_{3}$. Such terms are present in the supersymmetric Wilson line operator at two loops due to certain ultrasoft effects, but not in the case of the bosonic Wilson line operator. Likewise, at three loops, the functions $\tilde{A}_{4}$ and $\tilde{B}_{3}$ are required to cancel $1/\delta^2$, $(\log\delta)^2/\delta$, and $\log \delta /\delta$ terms not present in the final result. Finally, the function $\tilde{B}_{5}$ just contributes a term $45 \pi^5/\delta$ in this limit.

\section{Conclusions}
\label{last}

In this paper, we computed the angle-dependent three-loop cusp anomalous dimension in QCD and in a supersymmetric Yang-Mills theories. The obtained expressions are rather compact and are given in terms of harmonic polylogarithmic functions that can be readily evaluated numerically.  We discussed in detail special physical limits of the cusp anomalous dimension and, in particular, placed special emphasis on the backtracking Wilson line limit that is related to the quark-antiquark potential. We showed that this relation holds in QCD up to a conformal symmetry breaking corrections proportional to the beta function and identified the leading contribution to the conformal anomaly. It would be interesting to investigate whether the latter can be computed from the first principles.

We found that, unexpectedly, the results for the different theories considered are very similar. In fact, up to three loops at least, they can be written in terms of a single universal function evaluated at an effective charge given by the light-like cusp anomalous dimension. 
Assuming that this property holds at higher loops, we derived the contribution of the $n_f$-dependent term that violates Casimir scaling
and produces a nonplanar correction to the cusp anomalous dimension at four loops. 

\section*{Acknowledgements}

A.G.'s work was supported by RFBR grant 12-02-00106-a and by the Russian Ministry of Education and
Science. J.M.H. is supported in part by a GFK fellowship and by
the PRISMA cluster of excellence at Mainz university. G.P.K. is supported in part by the French National Agency for Research
(ANR) under contract StrongInt (BLANC-SIMI-4-2011). P.M. was supported in part by the European Commission through contract 
 PITN-GA-2012-316704 (HIGGSTOOLS).

\appendix

\section{Definition of Yang--Mills theories}\label{App:YM}


Throughout the paper, we consider two Yang--Mills theories with different particle content.

In the first case, for gauge fields coupled to $n_f$ species of Dirac fermions, we have
\begin{equation}
\mathcal{L}_{\text{QCD}} = - \frac{1}{2} \tr \left( F_{\mu\nu} F^{\mu\nu} \right) + \sum_{i=1}^{n_f} i \bar\psi_i \gamma^\mu D_\mu \psi_i\,,
\end{equation}
where $F_{\mu\nu}=F_{\mu\nu}^a T^a$ and $D_\mu=\partial_\mu - i g A_\mu^a T^a $
with $T^a$ being the generators of the fundamental representation of the $SU(N)$ normalized as
\begin{equation}
\tr \big({T^a T^b}\big) = T_F \delta^{ab}\,,\qquad
T^a T^a = C_F = \frac{N^2-1}{2N}\,,\qquad T_F=1/2\,.
\label{Ta}
\end{equation}
The fermion fields $\psi_i$ are defined in the fundamental representation of the $SU(N)$
and carry the additional flavour index $i=1$, \dots, $n_f$.

In the second case, for gauge fields coupled to $n_s$ scalars and $n_f$ fermions, we have
\begin{align}
\mathcal{L}_{\text{adj}} ={}& \tr \biggl\{ - \frac{1}{2} F_{\mu\nu} F^{\mu\nu}
+ 2 i \bar{\lambda}_{\dot\alpha A} \sigma_\mu^{\dot\alpha\beta} \mathcal{D}^\mu \lambda_\beta^A
- \mathcal{D}_\mu \phi^I \mathcal{D}^\mu \phi^I
+\frac{1}{2} g^2 [\phi^I,\phi^J] [\phi^I,\phi^J]
\nonumber\\
&{} - \sqrt{2} g \lambda^{\alpha A} (T_I)_{AB} [ \phi^I,\lambda_\alpha^B]
+ \sqrt{2} g \bar\lambda_{\dot\alpha A} (T_I^\dagger)^{AB} [\phi^I,\bar\lambda^{\dot\alpha}_B] \biggr\}\,,
\label{LII}
\end{align}
where $\mathcal{D}^\mu = \partial^\mu - i g [A^\mu,\phantom{,}]$ is the covariant derivative in the adjoint representation
and all fields $\Phi=\{\lambda,\bar\lambda,\phi \}$ are matrix-valued in the $SU(N)$ group,
$\Phi=\Phi^{a}T^a$ with the generators $T^a$ defined in~(\ref{Ta}).
The scalar fields $\phi^I$ and the two-component Weyl fermions,
$(\lambda_\alpha^{A})^\dagger=\bar\lambda_{\dot\alpha A}$ and $\lambda^{\alpha A}=\epsilon^{\alpha\beta} \lambda_\beta^A$
(with $\alpha$, $\dot\alpha=1$, $2$),
carry the additional flavour index $I=1$, \dots, $n_s$ and $A=1$, \dots, $n_f$, respectively.
In the second line of~(\ref{LII}), the Yukawa coupling involves the matrix $T_I = (T_I)_{AB}$.

The reason for choosing the Lagrangian in the form (\ref{LII}) is that, by fine tuning the number of fermions and scalars,
we can use it to describe supersymmetric Yang--Mills theories with different number of supercharges. In particular,
the maximally supersymmetric Yang--Mills theory corresponds to the special case of~(\ref{LII}) with $n_f=4$, $n_s=6$
and matrices $T_I$ (with $I=1$, \dots, $6$) given by (chiral blocks of) Dirac matrices in six-dimensional Euclidean space,
$T_I T_J^\dagger + T_J T_I^\dagger = \delta_{IJ}$ (see Appendix~B in ref.~\cite{Belitsky:2003sh} for details).

\section{Wilson lines and HQET}
\label{S:HQET}

Wilson lines can be conveniently studied using the heavy quark effective theory.%
\footnote{Methods of calculation of multiloop Feynman diagrams in HQET are reviewed, e.g.,
in \cite{Grozin:2004yc,Grozin:2003ak,Grozin:2008tp,Grozin:2013bva}.}
In fact, the HQET Lagrangian was first introduced as a technical device for this purpose~\cite{Gervais:1979fv,Arefeva:1980zd}.

\begin{figure}[t!bp]
\centering
\psfrag{a}[cc][cc]{(a)}
\psfrag{b}[cc][cc]{(b)}
\psfrag{0}[cc][cc]{$0$}
\psfrag{x}[cc][cc]{$x$}
\psfrag{x1}[cc][cc]{$x_1$}
\psfrag{x2}[cc][cc]{$x_2$}
 \includegraphics[width=0.7\textwidth]{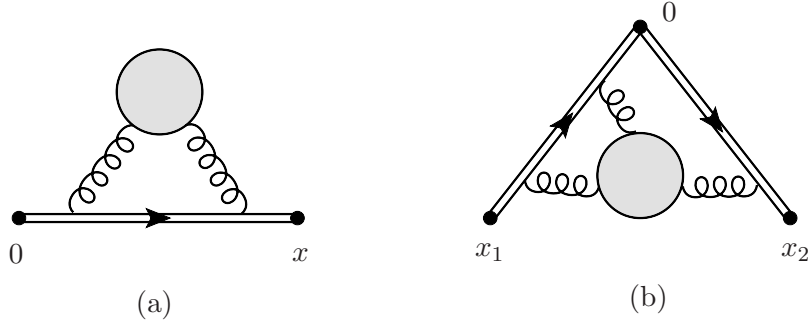} 
\caption{Correlation functions of two HQET fields (a) and of two HQET fields and the current (b).}
\label{F:Green}
\end{figure}

The heavy quark effective fields $h_v(x)$ and $h^\dagger_v(x)$ depend on the unit four-vector $v_\mu^2=1$ which has the meaning
of the heavy quark velocity.  
The correlation function of two HQET fields (figure~\ref{F:Green}(a)) is
\begin{equation}
-i\langle {h_v(x) h_v^\dagger(0)}\rangle = \delta^{(3)}(x_\bot) W(t)\,,
\label{R:Green2}
\end{equation}
where $t=v\cdot x$ and $x_\bot^\mu = (g^{\mu\nu} - v^\mu v^\nu)x_\nu$ is the projection of $x$ onto the subspace orthogonal to $v$.
Also, $W(t)$ is the expectation value of the Wilson line evaluated along the segment of length $t$ oriented along $v^\mu$.

It is convenient to work in momentum space. Introducing the notation for the Fourier transformed HQET field
\begin{align}
\tilde h_v(\omega) =\int_0^\infty dt  \int d^3 x_\perp  \e^{i\omega t} h_v(x) 
\end{align} 
with $\omega$ being the so-called residual energy of heavy quark, we find from (\ref{R:Green2})
\begin{align}\label{S}
\langle {\tilde h_v(\omega)  h_v^\dagger(0)}\rangle = i \int_0^\infty dt \,\e^{i\omega t} \,W(t)\equiv S_v(\omega)\,.
\end{align}
Expanding  $S_v(\omega)$ in powers of the coupling constant yields Feynman diagrams
shown in figure~\ref{HQET-fig} for $\omega=-1/2$.

Within the HQET framework, the cusp anomalous dimension can be identified as anomalous dimension of 
the local gauge-invariant operator
\begin{equation}
J(x) = h^\dagger_{v_2}(x) h_{v_1}(x)\,.
\label{R:ZJ}
\end{equation}
To see this, we consider the correlation function of two HQET fields and the current~(\ref{R:ZJ})
\begin{equation}
(-i)^2\langle h_{v_2}(x_2) J(0) h_{v_1}^\dagger(x_1)\rangle= \delta^{(3)}(x_{1\bot}) \delta^{(3)}(x_{2\bot}) W(t_1,t_2;\phi)\,,
\label{R:Green3}
\end{equation}
where  $x_{i\perp}$ stands for the component of $x_i$ orthogonal to $v_i$
and
$W(t_1,t_2;\phi)$ is the expectation value of Wilson line evaluated along two segments of lengths $t_1$ and $t_2$
separated by a cusp angle $\phi$ (see figure~\ref{F:Green}(b)).  In particular, for $v_1=v_2$, or equivalently $\phi=0$, we have
\begin{equation}
W(t_1,t_2;0) = W(t+t')\,.
\label{R:W0}
\end{equation}
Going to the momentum space, we obtain
\begin{align}\notag
G(\omega_1,\omega_2;\phi)  {}& = \langle \tilde h_{v_2}(\omega_2) J(0) \tilde h_{v_1}^\dagger(\omega_1)\rangle
\\  
 {}& =- \int_0^\infty d t_1\, dt_2 \, \e^{i t_2 \omega_2 + i t_1\omega_1}  W(t_1,t_2;\phi)
= V(\omega_1,\omega_2;\phi) S_{v_1}(\omega_1) S_{v_2}(\omega_2)\,,
\label{R:G}
\end{align}
 where $\omega_1$ and $\omega_2$ are the residual energies,
$S_v(\omega)$ is the propagator of the  HQET field (\ref{S})  
and $V(\omega_1,\omega_2;\phi)$ is the one-particle irreducible vertex function
(without the external-leg propagators). For $\omega_1=\omega_2=-\delta$ the vertex function $V(\omega_1,\omega_2;\phi)$
coincides with $V(\phi)$ in (\ref{lnV}) and is given by Feynman diagrams shown in figure~\ref{HQET-fig1}. 

It is convenient to extract the renormalization  $Z-$factor from
\begin{equation}
\log\left[\frac{G(\omega_1,\omega_2;\phi) }{G(\omega_1,\omega_2;0) }\right] =
\log\left[\frac{V(\omega_1,\omega_2;\phi) }{V(\omega_1,\omega_2;0) }\right]  = \log Z(\phi) + \text{finite}\,.
\label{R:Zcoord}
\end{equation}
Note that $Z(\phi)$ should be gauge invariant and independent on $\omega_1$ and $\omega_2$.
In order to avoid infrared divergences, $\omega_1$ and $\omega_2$ should be different from zero.
It is convenient to choose $\omega_1=\omega_2=-\delta$. Then, the dependence of the HQET integrals
on $\delta$ can be trivially obtained by dimension counting. To simplify the calculation, we can set $\delta=1/2$ and evaluate 
the resulting dimensionless integrals in $D=4-2\epsilon$ dimensions. The cusp anomalous dimension 
can be found by matching $\log Z(\phi)$ into the expected result (\ref{lnZ}).

At $\phi=0$  we find from (\ref{R:W0}) that the vertex function $V(\omega_1,\omega_2;0)$ satisfies the Ward identities
\begin{equation}
V(\omega_1,\omega_2;0) = \frac{S^{-1}(\omega_1) - S^{-1}(\omega_2)}{\omega_1 - \omega_2}\,,\qquad
V(\omega,\omega;0) = \frac{d S^{-1}(\omega)}{d\omega}\,.
\label{R:Ward}
\end{equation}
As a consequence, UV divergences of $V(\omega,\omega;0)$ match those of the heavy quark propagator (\ref{S}) leading to 
\begin{equation}
\log V(\omega,\omega;0) = - \log Z_h + \text{finite}\,,
\label{R:ZhV}
\end{equation}
where $Z_h^{1/2}$ is the renormalization factor for the HQET field $h_v(x)$.
Note that $Z_h$ is not gauge invariant;
at three loops it has been calculated in~\cite{Melnikov:2000zc,Chetyrkin:2003vi}.
We reproduced this result from our three-loop calculation of $V(\omega,\omega;\phi)$
by setting $\phi=0$ and using~(\ref{R:ZhV}).
  
\section{Abelian large-$n_f$ terms}
\label{S:nf}

\begin{figure}[tbp]
\centering
\psfrag{a}[cc][cc]{(a)}
\psfrag{b}[cc][cc]{(b)}
\psfrag{c}[cc][cc]{(c)}
 \includegraphics[width= \textwidth]{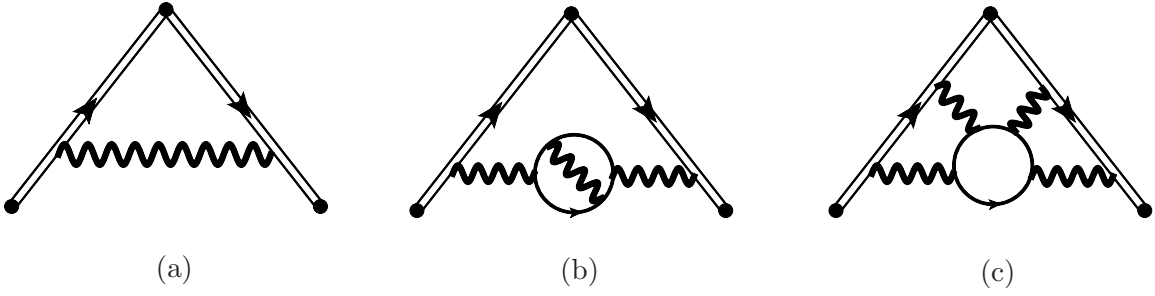} 
\caption{QED like diagrams contributing to the cusp anomalous dimension in the large $n_f$ limit at order  L$\beta_0$ (a),  NL$\beta_0$ (b)
and  NNL$\beta_0$ (c).  Fat wavy line denotes the full photon propagator with the L$\beta_0$ accuracy.}
\label{F:large-nf}
\end{figure}

In this appendix, we compute the special class of QCD corrections to the cusp anomalous dimension (\ref{R:NA})
of the form $(T_F n_f)^{L-1} \alpha_s^L$ and $C_F (T_F n_f)^{L-2} \alpha_s^L$. They originate from QED like diagrams 
which have the form of the one-loop diagram shown in figure \ref{HQET-fig}(a) with a free gluon propagator dressed by fermion loop corrections (see figures~\ref{F:large-nf}(a) and (b)).
\footnote{Starting from $(T_F n_f)^{L-3} \alpha_s^L$ order, we also have to take into account the additional abelian diagrams 
shown in figure~\ref{F:large-nf}(c). They involve the light-by-light scattering and their calculation is more involved compared
to the diagrams shown in figures~\ref{F:large-nf}(a) and (b).}

To compute the contribution of such diagrams it is sufficient to consider QED with $n_f$ massless lepton flavors. In this case, we put
$C_F=T_F=1$, $C_A=0$ and treat the one-loop beta-function
\begin{equation}
\beta_0 = - \frac{4}{3} n_f
\label{nf:beta0}
\end{equation}
as a large parameter. Then, the above mentioned corrections take the form $\beta_0^{L-1}\alpha_s^L$ and $\beta_0^{L-2}\alpha_s^L$. We shall refer to them as the leading (L$\beta_0$) and next-to-leading (NL$\beta_0$) 
large-$\beta_0$ corrections, respectively.

\begin{figure}[tbp]
\centering
\psfrag{a}[cc][cc]{(a)}
\psfrag{b}[cc][cc]{(b)}
\psfrag{+}[cc][cc]{$+$}
\includegraphics[width=\textwidth]{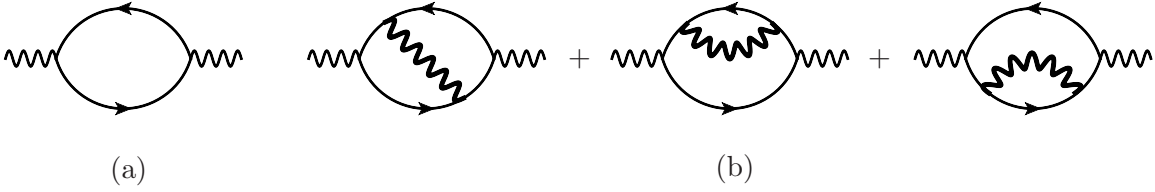}
\caption{Photon self-energy at L$\beta_0$ order (a) and NL$\beta_0$ order (b). Fat wavy line denotes the full photon propagator
with the L$\beta_0$ accuracy.}
\label{F:Pi}
\end{figure}

For our purposes we need the expression for the photon self-energy $(g_{\mu\nu}k^2 -k_\mu k_\nu) \Pi(k^2)$
with the NL$\beta_0$ accuracy 
\begin{equation}
\Pi(k^2) = \Pi_0(k^2) + \frac{\tilde{\Pi}(k^2)}{\beta_0}+ O(1/\beta_0^2)\,.
\label{nf:Pi}
\end{equation}
Here the leading term comes from the diagram shown in figure~\ref{F:Pi}(a)
\begin{align}
{}& \Pi_0(k^2) =   \frac{e_0^2\, \beta_0}{(4\pi)^{2-\epsilon}} \frac{D(\epsilon)}{\epsilon} (-k^2\e^{\gamma_{\text{E}}}/\mu^2)^{-\epsilon}\,,
\nonumber\\
{}& D(\epsilon) =  \e^{\gamma_{\text{E}}\epsilon}
\frac{(1-\epsilon) \Gamma(1+\epsilon) \Gamma^2(1-\epsilon)}{(1-2\epsilon) (1-\frac{2}{3}\epsilon) \Gamma(1-2\epsilon)}
= 1 + \frac{5}{3} \epsilon + \cdots\,,
\label{nf:Pi0}
\end{align}
where $e_0^2$ is a bare QED coupling constant.
The diagrams shown in figure~\ref{F:Pi}(b) produce the next-to-leading correction to (\ref{nf:Pi}). It can be written in the 
form~\cite{PalanquesMestre:1983zy,Broadhurst:1992si}
\begin{equation}
\tilde{\Pi}(k^2) = 3 \epsilon \sum_{L=2}^\infty \frac{F(\epsilon,L\epsilon)}{L} [\Pi_0(k^2)]^L\,,
\label{nf:Pi1}
\end{equation}
where the function $F(\epsilon,u)$ is given by
\begin{align}
 F(\epsilon,u) {}& =
\frac{2 (1-2\epsilon)^2 (3-2\epsilon) \Gamma^2(1-2\epsilon)}%
{9 (1-\epsilon) (1-u) (2-u) \Gamma^2(1-\epsilon) \Gamma^2(1+\epsilon)}\bigg[\frac{2\Gamma(1+u) \Gamma(1-u+\epsilon)}{\Gamma(1-u-\epsilon) \Gamma(1+u-2\epsilon)}
\nonumber\\[2mm]
{}& \times
\frac{2(1+\epsilon)(3-2\epsilon) - (4+11\epsilon-7\epsilon^2)u + \epsilon(8-3\epsilon)u^2 - \epsilon u^3}
{(1-u) (2-u) (1-u-\epsilon) (2-u-\epsilon)} 
\nonumber\\
{}&  - u
\frac{2-3\epsilon-\epsilon^2 + \epsilon(2+\epsilon)u - \epsilon u^2}{\Gamma^2(1-\epsilon)}
I(1+u-2\epsilon)
\biggr],
\label{nf:Feu}
\end{align}
with the Euclidean integral (with $p^2=1$)
\begin{equation*}
I(n) = \frac{1}{\pi^D} \int \frac{d^D k_1\,d^D k_2}%
{k_1^2 k_2^2 (k_1+p)^2 (k_2+p)^2 \left[(k_1-k_2)^2\right]^n}
\end{equation*}
that can be expressed via a hypergeometric ${}_3F_2-$function of unit argument~\cite{Kotikov:1995cw,Broadhurst:1996ur}.

The function $F(\epsilon,u)$ is regular at the origin and admits a double series expansion
\begin{equation}
F(\epsilon,u) = \sum_{n,m=0}^\infty F_{nm} \epsilon^n u^m\,,
\label{nf:Feuexp}
\end{equation}
with the coefficients $F_{nm}$ that can be calculated to any order in terms of multiple $\zeta$ values.
For $u=0$, the function $F(\epsilon,0)$ reduces to Euler  gamma functions~\cite{Broadhurst:1992si}
(the same holds for  $F(\epsilon,2\epsilon)$).
To save space, we do not presentan explicit expression for $F(\epsilon,u)$.

It is convenient to introduce the renormalized coupling constant
\begin{equation}
b = \beta_0 \frac{\alpha(\mu)}{4\pi}\,.
\label{nf:b}
\end{equation}
In the large $\beta_0$ limit, we keep $b$ fixed and use $1/\beta_0$ as an expansion parameter. In the 
\MS{} scheme the renormalized coupling is defined as
\begin{equation}
\beta_0 \frac{e_0^2 }{(4\pi)^{2-\epsilon}} (\mu^2 \e^{\gamma_{\text{E}} })^{-\epsilon} = b Z_\alpha(b)  \,,
\label{nf:MS}
\end{equation}
where $e_0^2$ is a bare coupling constant and the charge renormalization constant is given with the NL$\beta_0$ 
accuracy by
\begin{align}\notag
{}& Z_\alpha(b) = \frac{1}{1+b/\epsilon}
\left(1 + \frac{\tilde{Z}_\alpha (b)}{\beta_0} + O(1/\beta_0^2) \right)\,,\qquad
\\
{}& \tilde{Z}_\alpha (b) = \frac{\tilde{Z}_{\alpha,1}(b)}{\epsilon} + \frac{\tilde{Z}_{\alpha,2}(b)}{\epsilon^2} + \cdots\,.
\label{nf:Za}
\end{align}
Here expansion of $\tilde{Z}_{\alpha,1}(b)$ starts from $b^2$,
that of $\tilde{Z}_{\alpha,2}(b)$ from $b^3$, etc. The charge renormalization constant satisfies the 
renormalization group equation that allows us to express $Z_\alpha(b)$ in terms of beta-function.

In the abelian theory, $Z_\alpha(b)$ is related to the photon self-energy (\ref{nf:Pi}) expressed in terms of renormalized 
coupling constant 
\begin{align}\label{RG}
\log(1-\Pi(k^2)) = \log Z_\alpha(b) + O(\epsilon^0) =  -\int_0^b {d b\, \beta(b)\over b(\epsilon+\beta(b))} + O(\epsilon^0)\,.
\end{align}
Substituting (\ref{nf:Pi})  and (\ref{nf:Za}) into this relation and equating the coefficients in front of $1/(\epsilon\beta_0)$ on the
both sides, we find that $\tilde{Z}_{\alpha,1}(b)$ in (\ref{nf:Za}) is given by the coefficient of $\epsilon^{-1}$
in $-(1+b/\epsilon) \tilde{\Pi}(k^2)$.
It is convenient to choose the renormalization scale as
\begin{equation}
\mu^2 =  (-k^2)  \lim_{\epsilon\to 0}\, [D(\epsilon)]^{-1/\epsilon}   = -k^2  \e^{-\frac{5}{3}}   \,.
\label{nf:mu}
\end{equation}
Then, we use (\ref{nf:Pi1})  to obtain $-(1+b/\epsilon) \tilde{\Pi}(k^2) = - 3 b \sum_{L=2}^\infty  {F(\epsilon,L\epsilon)} 
(b/(\epsilon+b))^{L-1}/{L}$.
Expanding $(b/(\epsilon+b))^{L-1}$ in powers of $b$ and replacing $F(\epsilon,L\epsilon)$ with (\ref{nf:Feuexp}),
we find that all coefficients but $F_{n0}$ cancel leading to
\begin{equation}
\tilde{Z}_{\alpha,1} = - 3 \sum_{n=0}^\infty \frac{F_{n0} (-b)^{n+2}}{(n+1) (n+2)}\,.
\label{nf:Z1}
\end{equation}
We can use this relation to find the $\beta$ function with NL$\beta_0$ accuracy~\cite{PalanquesMestre:1983zy,Broadhurst:1992si}
\begin{align}\notag
{}& \beta(g) =b+\tilde{\beta}(b)/\beta_0+O(1/\beta_0^2)\,,
\\
{}& \tilde{\beta}(b) = - \frac{d\tilde{Z}_{\alpha,1}(b)}{d\log b}
= 3 b^2  +\frac{11}{4}b^3 -\frac{77}{36}b^4+ O(b^5)\,.
\label{nf:beta}
\end{align}
Finally, we substitute $\beta(g)$ into the last relation in (\ref{RG}) 
and obtain $O(1/\beta_0)$ correction to the charge renormalization constant (\ref{nf:Za})
\begin{align}
\tilde{Z}_\alpha (b) ={}& - \epsilon \int_0^b \frac{\tilde{\beta}(b)\,d b}{b (\epsilon+b)^2}
\nonumber\\
={}& - \frac{3}{2} \frac{b^2}{\epsilon}
+ \frac{1}{2} \left(4 + F_{10} \epsilon \right) \frac{b^3}{\epsilon^2}
- \frac{1}{4} \left(9 + 3 F_{10} \epsilon + F_{20} \epsilon^2 \right) \frac{b^4}{\epsilon^3}
+ \cdots
\label{nf:Ztilde}
\end{align}

We are now ready to determine the cusp anomalous dimension at the NL$\beta_0$ order. To this end,
we have to repeat the one-loop calculation  of the vertex function $V(\omega,\omega;\phi)$ (see (\ref{zero1}) for $\omega=\delta$), with  
a free photon propagator modified by self-energy corrections
\begin{align} \label{pro-exp}
{1\over k^2} {}& \to  {1\over k^2(1-\Pi(k^2))}  
={1\over k^2(1-\Pi_0(k^2))} \left[1+ {1\over \beta_0}{\tilde{\Pi}(k^2)\over 1-\Pi_0(k^2)} + O(1/\beta_0^2)\right]
\end{align}
and, then, express the result in terms of the renormalized coupling constant (\ref{nf:b}). Performing the calculation
we obtain
\begin{align}
V(\omega,\omega;\phi) - V(\omega,\omega;0)
{}& = \frac{1}{\beta_0} \sum_{L=1}^\infty \frac{f(\epsilon,L\epsilon;\phi)}{L}
\left(\frac{b}{\epsilon+b}\right)^L
\nonumber\\
&{}\times\left[1 + L \frac{\tilde{Z}_\alpha(b)}{\beta_0}
+ \frac{3\epsilon}{\beta_0} \sum_{L'=2}^{L-1} \frac{L-L'}{L'} F(\epsilon,L'\epsilon) \right]
+ \mathcal{O}\left(\frac{1}{\beta_0^3}\right)\,,
\label{nf:Vf1}
\end{align}
where the coupling constant $b$ is defined at the scale $\mu^2 = \e^{-\frac{5}{3}} (2\omega)^2$, the function $F(\epsilon,L'\epsilon)$ is given by (\ref{nf:Feuexp}) and the notation was introduced
for 
\begin{align}\notag
 f(\epsilon,u;\phi) {}&=
- \frac{(1-\frac{2}{3}\epsilon) \Gamma(2-2\epsilon) \Gamma(1-u) \Gamma(1+2u)}%
{(1-\epsilon) \Gamma^2(1-\epsilon) \Gamma(1+\epsilon) \Gamma(2+u-\epsilon)}\\
&{}\times\left[ \bigl((2+u-2\epsilon)\cos\phi-u\bigr)
\,_2F_1\left(\left.\begin{array}{c}1,1-u\\3/2\end{array}\right|\frac{1-\cos\phi}{2}\right)
-2(1-\epsilon) \right]\,.
\label{nf:fphi}
\end{align}
The function $ f(\epsilon,u;\phi)$ is regular at the origin:
\begin{equation}
f(\epsilon,u;\phi) = \sum_{n,m=0}^\infty f_{nm}(\phi) \epsilon^n u^m\,.
\label{nf:feuexp}
\end{equation}
In particular, for $u=0$ we have 
\begin{align}\notag
{}&  {f}(\epsilon,0;\phi) = - 2  {f}(\epsilon)
(\phi \cot\phi - 1)\,,\quad
\\
{}&  {f}(\epsilon) = \frac{(1-\frac{2}{3}\epsilon) \Gamma (2-2 \epsilon )}{ \Gamma^2 (1-\epsilon ) \Gamma
   (2-\epsilon ) \Gamma (1+\epsilon)}= \sum_{n=0}^\infty  {f}_n \,\epsilon^n\,.
\end{align}
 
The cusp anomalous dimension is related to the residue at the pole $Z_1(b;\phi)/\epsilon$ in the expression (\ref{nf:Vf1})
\begin{equation}
\Gamma_{\rm cusp} (b,\phi) = - 2 \frac{d Z_1(b;\phi)}{d\log b}\,.
\end{equation}
Replacing ${f}(\epsilon,L\epsilon;\phi)$ in (\ref{nf:Vf1}) with its general expression (\ref{nf:feuexp}), we find that the coefficient
in front of $1/\epsilon$ on the right-hand side of (\ref{nf:Vf1}) only depends on the coefficients $f_{n0}(\phi)= -2 (\phi \cot\phi - 1) f_n$.
Then, at the NL$\beta_0$ order the cusp anomalous dimension is given by 
\begin{align}
 \Gamma_{\rm cusp}(b,\phi)   = 4 (\phi \cot\phi - 1){}& \bigg[  \frac{b}{\beta_0} {f}(-b)  
+ \frac{b^3}{\beta_0^2}
 \biggl\{
\frac{3}{2} \bigl(F_{10} + 2 F_{01} - 2 {f}_1 \bigr)
\nonumber\\
{} &
- \bigl(2 F_{20} + 3 (F_{11} + F_{02}) + 3 F_{01} {f}_1 - 6 {f}_2 \bigr) b
\nonumber\\
& 
+ \biggl( \frac{9}{4} F_{30} + 3 (F_{21} + F_{12} + F_{03})
+ (F_{20} + 3 (F_{11} + F_{02})) {f}_1
\nonumber\\
& 
- \frac{3}{2} \bigl(F_{10} - 2 F_{01}\bigr)  {f}_2
- 9  {f}_3 \biggr) b^2
+ O(b^3)\bigg\}
 \biggr]  + \mathcal{O}\left({1}/{\beta_0^3}\right) \,.
\label{nf:NL}
\end{align}
Replacing the coefficients $F_{nm}$ and $f_n$ by their explicit expressions, we finally obtain
\begin{align}
\Gamma_{\rm cusp}(b,\phi) ={}& 4 \left[ \frac{b}{\beta_0} \Gamma_0(b) - \frac{b^3}{\beta_0^2} \Gamma_1(b) \right]
\left( \phi \cot\phi - 1 \right) + \mathcal{O}\left(\frac{1}{\beta_0^3}\right)\,,
\nonumber\\[2mm]
\Gamma_0(b) ={}& \frac{(1+\frac{2}{3}b) \Gamma(2+2b)}{(1+b) \Gamma^3(1+b) \Gamma(1-b)}
\nonumber\\
={}& 1 + \frac{5}{3} b - \frac{1}{3} b^2
- \left( 2 \zeta_3 - \frac{1}{3} \right) b^3
+ \left( \frac{\pi^4}{30} - \frac{10}{3} \zeta_3 - \frac{1}{3} \right) b^4
+ \cdots\,,
\nonumber\\
\Gamma_1(b) ={}& 12 \zeta_3 - \frac{55}{4}
+ \biggl( 40 \zeta_3 - \frac{\pi^4}{5} - \frac{299}{18} \biggr) b
\nonumber\\
&{}+ \biggl( 24 \zeta_5 + \frac{233}{6} \zeta_3 - \frac{2}{3} \pi^4 + \frac{15211}{864} \biggr) b^2
\nonumber\\
&{}+ \biggl( 80 \zeta_5 - 48 \zeta_3^2 + \frac{1168}{15} \zeta_3 - \frac{2}{63} \pi^6 - \frac{167}{225} \pi^4 - \frac{971}{240} \biggr) b^3
+\cdots\,.
\label{nf:res}
\end{align}
This expansion can be extended to any number of loops. The leading term $\Gamma_0(b)$ has been derived in~\cite{Beneke:1995pq}, the
result for $\Gamma_1(b)$ is new. The first term in the expression for $\Gamma_1(b)$ is in agreement with our result for $\gamma_{Ff}$ in (\ref{resultc4}).

In a similar manner, we can use (\ref{R:ZhV}) to compute the anomalous dimension of the HQET field
\begin{equation}
\gamma_{h} (b) = - 2 \frac{d Z_{h,1}(b)}{d\log b}\,.
\end{equation}
where $Z_{h,1}(b)$ denotes the residue at the simple pole $1/\epsilon$ in the expression for $Z_h$.
Performing the calculation, we find in Landau gauge at the NL$\beta_0$ order 
\begin{align}
\gamma_h(b)  {}& =  - 6 \left[ \frac{b}{\beta_0} \gamma_{0}(b) - \frac{b^3}{\beta_0^2} \gamma_{1}(b) \right]
+ \mathcal{O}\left(\frac{1}{\beta_0^3}\right)\,,
\nonumber
\\
\gamma_{0}(b) {}& =
\frac{\left(1+\frac{2}{3}b\right)^2 \Gamma(2+2b)}{(1+b)^2 \Gamma^3(1+b) \Gamma(1-b)}
\nonumber\\
{}& =  1 + \frac{4}{3} b - \frac{5}{9} b^2
- \left( 2 \zeta_3 - \frac{2}{3} \right) b^3
- \left( \frac{8}{3} \zeta_3 - \frac{\pi^4}{30} + \frac{7}{9} \right) b^4
+ \cdots\,,
\nonumber
\\
\gamma_{1}(b) {}& = 3 \biggl( 4 \zeta_3 - \frac{17}{4} \biggr)
+ \biggl( 36 \zeta_3 - \frac{\pi^4}{5} - \frac{103}{9} \biggr) b
\nonumber\\
&{} + \biggl( 24 \zeta_5 + \frac{59}{2} \zeta_3 - \frac{3}{5} \pi^4 + \frac{14579}{864} \biggr) b^2
\nonumber\\
&{} + \biggl( 72 \zeta_5 - 48 \zeta_3^3 + \frac{3229}{45} \zeta_3
- \frac{2}{63} \pi^6 - \frac{44}{75} \pi^4 - \frac{5191}{540} \biggr) b^3
+ \cdots
\label{nf:res0}
\end{align}
The L$\beta_0$ result $\gamma_{0}(b)$ has been derived in~\cite{Broadhurst:1994se}. The first term in the expression for $\gamma_{1}(b)$ matches the $C_F^2 T_F n_f$ term in the three-loop $\gamma_h$~\cite{Melnikov:2000zc,Chetyrkin:2003vi}.
 
\section{Abelian large-$n_f$ terms in the quark-antiquark potential}

We can use the methods of appendix~\ref{S:nf} to compute $(T_F n_f)^{L-1} \alpha_s^L$ and $C_F(T_F n_f)^{L-2} \alpha_s^L$ 
corrections to the quark-antiquark potential (\ref{V-gen}) and, then, to find corrections to the coefficient function
$C(\alpha_s)$ defined in (\ref{anomaly}) and (\ref{C-fun}). 

As before, it is sufficient to perform calculations in QED with $n_f$ lepton flavors. With the  NL$\beta_0$  accuracy,
the potential (\ref{poten-R}) is determined by the full photon propagator in the Coulomb gauge,
\begin{align}\notag\label{V01-pot}
V(\boldsymbol{q}) {}&= - \frac{e_0^2}{\boldsymbol{q}^2} \frac{1}{1 - \Pi(-\boldsymbol{q}^2)} +O(1/\beta_0^3)
\\
{}&=-{(4\pi)^2\over \boldsymbol{q}^2}\bigg[{b\over \beta_0}V_0(b) - {b^3\over \beta_0^2}  V_1(b) \bigg]+O(1/\beta_0^3)\,,
\end{align}
where $\Pi(-\boldsymbol{q}^2)$ is given by (\ref{nf:Pi}) for $q^\mu=(0,\boldsymbol{q})$. Beyond the NL$\beta_0$ order, this relation 
is modified by corrections due to light-by-light scattering.

At the L$\beta_0$ order, we have from (\ref{nf:Pi0}), (\ref{nf:MS}) and (\ref{nf:Za}) 
\begin{equation}\label{V0-pot}
V_0(b) = 
{\epsilon\over b} \sum_{L=1}^\infty \left( \frac{D(\epsilon)\, b}{\epsilon+b}\right)^L=  \frac{1}{1 - \frac{5}{3} b}\,,
\end{equation}
where the coupling constant $b$ is defined at the scale $\mu^2=\boldsymbol{q}^2$.
At the NL$\beta_0$ order we use (\ref{pro-exp}) and (\ref{nf:Pi1}) to get
\begin{align}
V_1(b)  =-{\epsilon\over b^3}
\sum_{L=1}^\infty   \left(\frac{D(\epsilon)\,b}{\epsilon+b}\right)^L
\left[ L {\tilde{Z}_\alpha(b)} +  {3 \epsilon} 
\sum_{L'=2}^{L-1} \frac{L-L'}{L'} F(\epsilon,L'\epsilon) \right]\,.
\end{align}
Replacing $F(\epsilon,L'\epsilon)$ with (\ref{nf:Feuexp}) we find after some algebra
\begin{align}\notag
V_1(b) {}& = -\frac{3}{2} \left( F_{10} + 2 F_{01} + 2 v_{1} \right) 
+ \frac{1}{2} \left[ F_{20} - 6 F_{02} - 6 \left( F_{10} + 3 F_{01} \right) v_{1} - 30 v_{2} \right] b
\\
&{} - \frac{1}{4} \left[ F_{30} + 24 F_{03} - 4 \left( F_{20} + 12 F_{02} \right) v_{1}
+ 36 \left( F_{10} + 4 F_{01} \right) v_{2} + 312 v_{3} \right] b^2 + \cdots\,,
\end{align}
where $v_n=(5/3)^n/n!$ are the expansion coefficients of $[D(\epsilon)]^{u/\epsilon} = \sum_n v_n u^n + O(\epsilon)$. 
We use the known results~\cite{Broadhurst:1992si}  for the coefficients $F_{n0}$ and $F_{0n}$ to obtain
\begin{align}\label{V1-pot}
V_1(b) ={}& 12 \zeta_3 - \frac{55}{4}
+ \left( 78 \zeta_3 - \frac{7001}{72} \right) b 
+ \left( 60 \zeta_5 + \frac{723}{2} \zeta_3 - \frac{147851}{288} \right) b^2
\nonumber\\
&{} + \left( 770 \zeta_5 + \frac{276901}{180} \zeta_3 + \frac{\pi^4}{200} - \frac{70418923}{25920} \right) b^3 +\cdots\,.
\end{align}
Substituting (\ref{V0-pot}) and (\ref{V1-pot}) into (\ref{V01-pot}), we verify that $O(b^3)$ and $O(b^4)$ corrections to $V(\boldsymbol{q})$
are in agreement with the known $C_F (T_F n_f)^2 \alpha_s^3$ and $C_F^2 T_F n_f \alpha_s^3$ terms in the two-loop potential~\cite{Schroder:1998vy},
as well as the $C_F (T_F n_f)^3 \alpha_s^4$ and $C_F^2 (T_F n_f)^2 \alpha_s^4$ terms at three loops~\cite{Smirnov:2008pn}.

We can use  (\ref{nf:res}) together with (\ref{V-cusp}) to determine the function $V_{\rm cusp}$ in the large $n_f$-limit.
Comparing this function with the potential (\ref{V01-pot}) we verify the anomaly relation (\ref{anomaly}) and compute the
corresponding coefficient function 
\begin{align} 
C(b) = C_0(b) - {b^2\over \beta_0} C_1(b) + O(1/\beta_0^2) \,,
\end{align}
 with $C_0(b)=(\Gamma_0(b) - V_0(b))/b$ and $C_1(b)=(\Gamma_1(b) - V_1(b))/b+  {\tilde{\beta}(b) C_0(b)}/{b^3} $ given by 
 \begin{align} \notag
C_0(b) ={}& 
 -\frac{28}{9} b
-  \left(2 \zeta_3 + \frac{116}{27} \right) b^2
-  \left( \frac{10}3 \zeta_3 - \frac{\pi^4}{30} + \frac{652}{81} \right) b^3
+ \cdots\,,
\\
C_1(b) ={}&- \left( 38 \zeta_3 + \frac{\pi^4}{5} - \frac{1711}{24} \right)
- \left( 36 \zeta_5 + \frac{986}{3} \zeta_3 + \frac{2}{3} \pi^4
        - \frac{110059}{216} \right) b 
\nonumber\\
&{} - \left( 690 \zeta_5 + 48 \zeta_3^2 + \frac{53135}{36} \zeta_3
        + \frac{2}{63} \pi^6 + \frac{233}{360} \pi^4
        - \frac{13910875}{5184} \right) b^2+\cdots\,.
\end{align}
Here the expansion can be extended to any desired order.
We verify that the $n_f-$dependent term in 
(\ref{C-fun}) matches the first term in the expression for $C_0(b)$.
Notice that $1/b$ term cancels in $C_1(b)$ leading to the absence of the $b/\beta_0$ term in $C(b)$.
This explains why (\ref{C-fun}) does not contain an abelian color factor $C_F$.   

\bibliographystyle{JHEP} 


\providecommand{\href}[2]{#2}\begingroup\raggedright\endgroup

\end{document}